\documentclass[preprint,12pt,eqsecnum,nofootinbib,amsmath,amssymb]{revtex4-2}
\newcommand{\mydate}{6/5/2024}

\usepackage{graphicx}  
\usepackage{dcolumn}  
\usepackage{bm}             
\usepackage{latexsym}  
\usepackage{fancyhdr}  
\usepackage{wrapfig}
\usepackage{comment}
\usepackage{dsfont}
\usepackage{mathtools}
\usepackage{datetime}
\usepackage{mathrsfs}
\usepackage{amsbsy}
\usepackage{tikz}
\usepackage{bbold}
\usepackage{multirow}
\usepackage[export]{adjustbox}

\newcommand{\smN}{{\scriptscriptstyle N}}
 
\newcommand{\bodyskip}{\baselineskip 18pt plus 1pt minus 1pt}
\newcommand{\bibskip}{\baselineskip16pt plus 1pt minus 1pt}
\newcommand{\tableofcontentsskip}{\baselineskip 14pt plus 1pt minus 1pt}
\newcommand{\footnoteskip}{\baselineskip 12pt plus 1pt minus 1pt}
\newcommand{\abstractskip}{\baselineskip 13pt plus 1pt minus 1pt}

\begin{document}

\baselineskip 20pt plus 1pt minus 1pt

\bodyskip

\baselineskip 15pt plus 1pt minus 1pt

\title{
Truncated Modular Exponentiation Operators: 
\\[-5pt]
A Strategy for Quantum Factoring
}

\author{Robert L. Singleton Jr.$^\dagger$}

\affiliation{
SavantX Research Center\\
Santa Fe,  New Mexico,  USA
}

\date{\mydate}

\begin{abstract}
\abstractskip
\vskip0.3cm 
\noindent

Modular exponentiation (ME) operators are one of the 
fundamental components of Shor's algorithm, and the 
place where most of the quantum resources are deployed. 
These operators are often referred to as the bottleneck 
of the algorithm. I propose a method for constructing 
the ME operators that requires only $3n + 1$ qubits
with no ancillary qubits. 
The method relies upon the simple observation that the 
work register starts in state $\vert 1 \rangle$. Therefore, 
we do not have to create an ME operator $U$ that accepts 
a {\em general} input, but rather, one that takes an input 
from the periodic sequence of states $\vert f(x) \rangle$ 
for $x \in \{0, 1, \cdots, r-1\}$. Here, the ME function with
base $a$ is defined by $f(x) = a^x ~({\rm mod}~N)$ and
has a period  of $r$. For an $n$-bit number $N$, the
operator $U$ can be partitioned into~$r$ levels, where 
the gates in level $x \in \{0, 1, \cdots, r-1\}$ increment 
the state $\vert f(x) \rangle = \vert w_{n-1} \cdots w_1 
w_0 \rangle$ to the state $\vert f(x+1) \rangle = \vert 
w_{n-1}^\prime \cdots w_1^\prime w_0^\prime\rangle$. 
The gates below $x$ do not affect the state $\vert f(x+1)
\rangle$. This amounts to transforming an $n$-bit binary 
number $w_{n-1} \cdots w_1 w_0$ into another binary 
number $w_{n-1}^\prime \cdots w_1^\prime w_0^\prime$, 
without altering the previous states, which can be 
accomplished by a set of formal rules involving 
multi-control-NOT gates  and  single-qubit NOT gates. 
The process of gate construction can therefore be 
automated, which is essential for factoring larger 
numbers. The obvious problem with this method is 
that it is self-defeating: If we knew the operator~$U$, 
then we would know the period $r$ of the ME function, 
and there would be no need for Shor's algorithm. I show, 
however, that the ME operators are very forgiving, and 
truncated approximate forms in which levels have been 
omitted are able to extract factors just as well as the 
exact operators.  I demonstrate this by factoring the 
numbers $N = 21, 33, 35, 143, 247$ by using less than 
half the requisite number of levels in the ME operators. 
This procedure works because the method of continued 
fractions only requires an approximate phase value, 
which suggests that implementing Shor's 
algorithm might not be as difficult as first suspected.  
This is the basis for a factorization strategy in which 
one level at a time is iterated over using an automated
script. In this way, we fill the circuits for the ME 
operators with more and more gates, and the 
correlations between the various composite 
operators $U^p$ (where $p$ is a power of two) 
compensate for the missing levels. 

\vfill
\noindent
$\dagger$
corresponding email: robert.singleton@savantx.com or 
bobs1@comcast.net
\end{abstract}

\maketitle

\vfill


\pagebreak
\tableofcontentsskip
\tableofcontents

\newpage
\bodyskip

\pagebreak
\clearpage

\section{Introduction}

Modern computer security relies upon RSA encryption and 
related methods\,\cite{rsa, df1, df2}. Very large semi-primes 
$N$ are used to create {\em public-private} key pairs based upon 
the factors of~$N$. The number $N$ is called the {\em public
key}, and it can be known to everyone. It is typically thousands 
of bits long; for example, standard RSA can employ keys of
4096-bits and larger. The {\em  private key}, {\em i.e.} the factors 
of $N$, must remain secret. Indeed, if we knew the private 
key, this would break the encryption. Therefore, the security 
of these schemes rests upon the fact (or rather the empirical 
observation) that it is very hard to factor large numbers. A  
classical digital computer must essentially check every
combination of factors in succession, and breaking 
a 4096-bit key would take much longer than the age of the 
universe, even on the largest high performance supercomputers. 
In contrast,  Shor's algorithm\,\cite{shor1, shor2} runs on 
a quantum computer, exploiting the massive parallelism 
inherent in quantum mechanics,  so that all numbers can 
be tested simultaneously rather than sequentially. This 
provides for a {\em polynomial} factorization method. 
Thus, by employing quantum processes,  Shor's  algorithm 
can factor large numbers exponentially faster than a classical
computer, placing the security of RSA in grave danger.\footnote{
\footnoteskip
It is indeed ironic to think that the digital security of 
entire nations  and the world economy rests upon the 
{\em assumption} that factoring is difficult. 
}

Just how close are we to manufacturing an existing quantum 
device that can perform multipurpose factoring of large 
numbers relevant for encryption? To address this question, 
it is necessary to assess the current experimental state of 
quantum factoring. The first factorization was performed 
in 2001 in Ref.~\cite{ibm2001}, which factored $N = 15$ 
using an NMR device with 7 qubits. Over the following 
decade, a number of experiments were fielded that 
employed photonic devices, which factored $N = 15$ and 
$N =21$\,\cite{lu1, lanyon1, politi1, lopez1}. Significantly, 
these experiments observed entanglement, which is 
essential for a quantum speedup. Finally, over the last 
decade, the numbers $N = 15, 21, 35$ were factored 
using a trapped ion device and IBM's superconducting 
qubits\,\cite{monz1, amico1, skosana1}, although $N= 35$ 
had a success rate of only 14\%. All of these experiments 
have relied upon some form of simplification to reduce 
the qubit and gate count, but they successfully provide 
proof-of-principle demonstrations of the feasibility of 
Shor's algorithm. 

These are indeed seminal experiments, and they have 
pushed the limits of existing technology. However, in 
some sense, only modest progress has been made over 
the last 20 years, and it is clear that decoherence remains 
the primary obstacle. It is encouraging, however, that 
the rate of progress seems to be increasing.  The 
experiments have all been confined to small numbers, 
$N =15, 21, 35$,  and small periods such as $r = 2, 3, 4$. 
All but Ref.~\cite{ibm2001} employs some type of 
{\em complied} algorithm, in which knowledge of 
the solution is exploited to decrease the number of 
qubits and gates on the machine. All current machines 
lack a sufficient number of qubits and acceptable 
decoherence times to run a full non-compiled version 
of Shor's algorithm, even for small numbers. The qubit 
count for a full version of Shor is just too demanding. 
For example, for an $n = 4096$ bit number $N$,  the 
method proposed in this manuscript would require 
$3n +1 = 12289$ total qubits. The gate count for the 
modular exponential (ME) operator is also problematic.  
To this point, the general purpose~ME operator of 
Ref.~\cite{preskill1} requires of order $72 n^3$ gates 
for an $n$-bit number. Therefore, the ME operator for 
a 4096-bit key would need $5 \times 10^{12}$ gates! 
Breaking RSA consequently requires tens of thousands
or even trillions of high quality gates, in addition to 
very long decoherence times. Modern quantum 
computers are currently quite far from this domain,  
although future machines will undoubtedly be able 
to handle these requirements.  Quantum computers 
must first emerge from the Noisy Intermediate-Scale 
Quantum (NISQ) regime\,\cite{nisq1} before large 
numbers can be factored using a full implementation 
of Shor's algorithm (rather than complied versions 
with semi-classical Fourier transforms). But there is 
every reason to believe that at some point in the not 
too distant future, quantum computers will be able 
to break RSA keys of length 4096 and higher. 

The ME operators are the most critical component of 
Shor's algorithm. These operators are also responsible 
for the entanglement of the quantum system, which is 
required for a quantum speedup. Shor's algorithm is 
based on the observation that factoring a semi-prime 
number $N$ is equivalent to finding the period $r$ of 
the ME function $f(x) = a^x ~({\rm mod}~N)$. The number 
$a$ is called the {\em base}, and it is chosen randomly 
such that $1 < a < N$ and \hbox{${\rm gcd}(a, N) = 1$}. 
The number of bits required to represent a number $N$ 
is given by \hbox{$n = \lceil \log_2 N \rceil$}, and the ME 
operator $U$ acts on an $n$-bit {\em work register}
to transform the state $\vert f(x) \rangle$ to the
next state $\vert f(x + 1) \rangle$ for $x \in \{0, 1, 2,
\cdots \}$. 
This procedure lies at the heart of Shor's algorithm, and
consumes the most quantum resources. One can fairly
say that implementing the ME operator is the primary 
bottleneck of the algorithm. The method introduced in 
Ref.~\cite{sing1} (and presented in this manuscript) is
designed to reduce the gate count dramatically, and 
relies on the following observation. Since the work 
register starts in state $\vert 1 \rangle$, we do not have 
to create an ME operator $U$ that takes a {\em general} 
input, as in Ref.~\cite{preskill1}. Rather, we can use 
only the inputs from the periodic sequence of states 
$\vert f(x) \rangle$ for $x \in \{0, 1, \cdots, r-1\}$. An ME 
operator $U$ with period $r$ can therefore be partitioned 
into $r$ segments indexed by the integers $x \in \{0, 1, 
\cdots, r-1\}$. The gates in segment $x$ transform the 
work-state $\vert f(x) \rangle \equiv \vert w_{n-1} 
\cdots w_0 \rangle$ into $\vert f(x+1) \rangle \equiv 
\vert w_{n-1}^\prime \cdots w_0^\prime \rangle$,
while the levels below $x$ have no effect on the
state $\vert f(x + 1) \rangle$. 
Transforming a binary state $w_{n-1} \cdots w_0$ 
into another binary state $w_{n-1}^\prime \cdots 
w_0^\prime$, without altering the  preceding states, 
can be turned into a set of formal rules involving 
multi-control-NOT gates and single-qubit NOT gates. 
This procedure can therefore be automated, which will
be essential for factoring large numbers.  Indeed, this 
is how most of the gates in this manuscript and  in 
Ref.~\cite{sing1} were constructed. 

There is an obvious problem with this method, however. 
Knowing the operator $U$ is equivalent to knowing the 
period of the modular exponentiation function $f(x)$, 
which means that Shor's algorithm is not needed. 
However,  it turns out that the entire sequence of 
transitions that define the $U$ operator is {\em not}
required.  We can instead use {\em truncated} approximate 
versions 
of the ME operators in which levels are omitted. The 
corresponding phase histograms show that employing 
this truncation strategy still permits us to extract the 
appropriate phases, and therefore the correct factors. 
This is because the method of continued fractions uses 
only an {\em approximate} phase in order to extract  
the period of $f(x)$, and omitting levels in $U$ still 
provides an adequate approximation for the measured 
phase that determines the period. I will perform a 
number of systematic truncation studies, with the 
general result that well over half the requisite number 
of levels can be omitted without affecting the ability 
to factor. This makes factorization much easier, and 
provides for a practical factorization strategy in which 
the ME operators are constructed one level at a time
from the bottom up. 

This paper is organized as follows. To establish notation
and context,  Section~\ref{sec_review} presents a brief 
review of Shor's algorithm. The theory of ME operators
and the proposed method for constructing them are
introduced in Section~\ref{sec_truncate}. I use this
truncated formalism to factor the numbers $N = 21, 
33, 35, 143, 247$,  carrying out corresponding systematic 
truncation studies. It should be emphasized that 
factorization can still be performed when well over 
half of the levels of the ME operators are omitted. 
Finally, Section~\ref{sec_conclusions} provides some 
concluding remarks and potential paths for future 
research. 

\vfill
\pagebreak

\section{Review of Shor's Algorithm}
\label{sec_review}

I shall now present  a review of Shor's algorithm. To establish
some notation, we first look at the qubit ordering convention 
employed by OpenQASM/Qiskit\,\cite{OpenQASM_ref}.
For an $m$-qubit system, the computational basis states 
start with qubit $0$ in the upper position, working their 
way down to the last qubit labeled by $m - 1$ (in both the
quantum circuit and the tensor product representation).
The basis states can therefore be expressed by 
\begin{eqnarray}
   \vert k \rangle 
  &=&
  \vert k_0 \rangle  \otimes \vert k_1 \rangle  \otimes \cdots \otimes
 \vert k_{m-2} \rangle  \otimes \vert k_{m-1} \rangle 
  \ ,
\end{eqnarray}
where $k_\ell \in \{0, 1\}$. 
\begin{figure}[b!]
\includegraphics[scale=0.45]{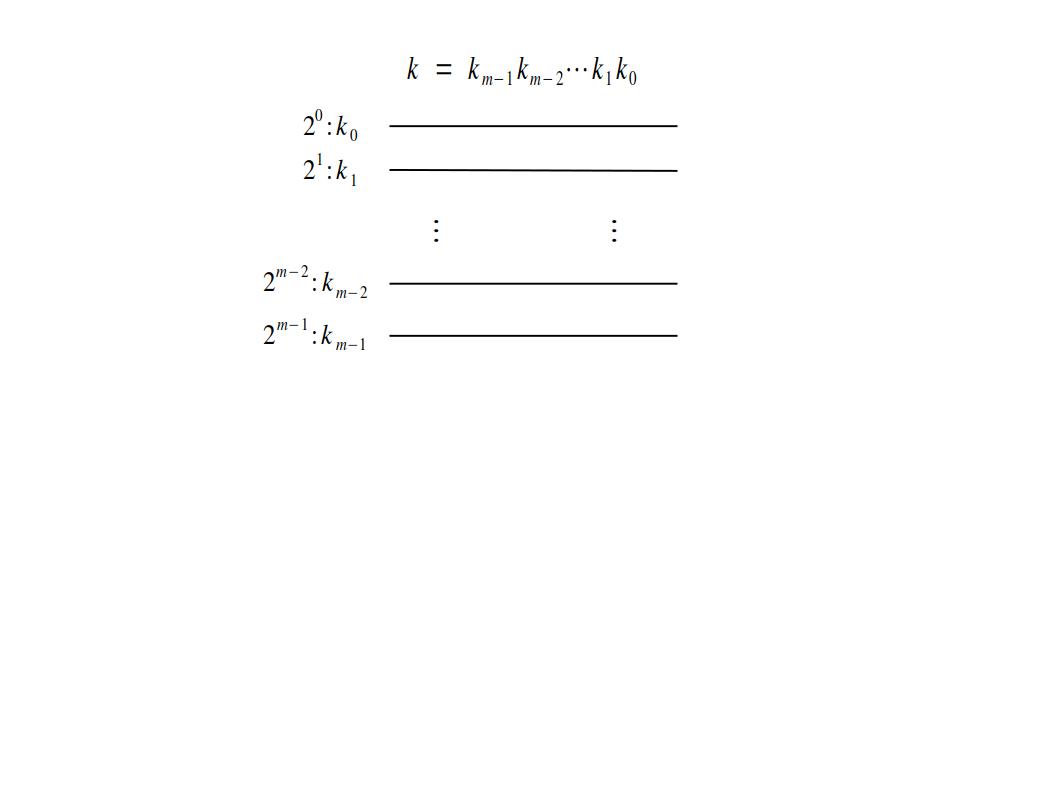} 
\vskip-5.3cm 
\caption{\footnoteskip  
An $m$-bit binary integer $k$ can be encoded on a gated 
quantum computer using a number of bit-conventions. 
The OpenQASM/Qiskit convention starts at the top with
qubit 0, and works its way down to the last qubit labeled 
by $m-1$.  Binary integers are expressed by the standard 
bit encoding $k = k_{m-1} \cdots k_1 k_0$,  which places the 
lowest order bit $k_0$ at the top of the circuit.   
}
\label{fig_conv}
\end{figure}
%
There are $M = 2^m$ quantum states in the system, with 
each state being indexed by an \hbox{$m$-bit} integer 
$k \in \{0, 1,  \cdots, M-1\}$. In binary form, these integers
are represented by
\begin{eqnarray}
  k &=& 
  k_{m-1} k_{m-2} \cdots k_{1} k_{0}
\\[5pt]
  &=&
  2^{m-1} k_{m-1} +  2^{m-2} k_{m-2} +
  \cdots  + 2^1 \, k_{1} +   2^0 \, k_{0}
  \ ,
\end{eqnarray}
where $k_0$ is the least significant bit. The OpenQASM/Qiskit 
bit-convention is shown in Fig.~\ref{fig_conv}, and we will use 
this convention throughout the manuscript. This convention 
is more in line with computer science, and differs slightly from 
the convention employed in physics. Note that the relation
\begin{eqnarray}
 \theta_k
 \equiv
  \frac{k}{M} 
  &=&
  \frac{k_{m-1}}{2^1} +  \frac{k_{m-2}}{2^2} +
  \cdots + \frac{k_{1}}{2^{m-1}} +   \frac{k_{0}}{2^m}
\\[5pt]
  &=&
  0.k_{m-1} \cdots k_1 k_0
\end{eqnarray}
\noindent
is an $m$-bit phase angle between 0 and 1. Given any $m$-bit 
phase $\theta_k$, we will often use the fact that $M \theta_k 
= k$ is an $m$-bit integer between 0 and $M - 1$.

Now that the qubit ordering conventions have been defined,
I am ready to summarize Shor's algorithm, which rests 
upon two fundamental quantum algorithms: the quantum 
Fourier transform (QFT), and the quantum phase estimation 
(QPE).  The QFT implements the discrete Fourier transform 
on a gated quantum computer,
and like the classical Fourier transform,  it extracts frequency 
signals from an input source. The QPE algorithm, in contrast,  
finds the complex phases or the eigenvalues of an arbitrary 
{\em unitary} linear operator.  Shor's algorithm elegantly 
combines the QFT and QPE to construct a powerful quantum 
algorithm for factoring very large integers.   
The mathematics behind Shor's algorithm is based on a 
simple but profound result from number theory,  which 
maps the factoring 
problem onto another mathematical problem that finds
the period of the {\em modular exponential function}.  The 
period of this function is directly related to the factors 
of the number in question,  and the QPE extracts this 
period using the method of continued fractions,  thereby 
providing the sought after factors. 

\begin{figure}[b!]
\begin{centering}
\includegraphics[scale=0.45]{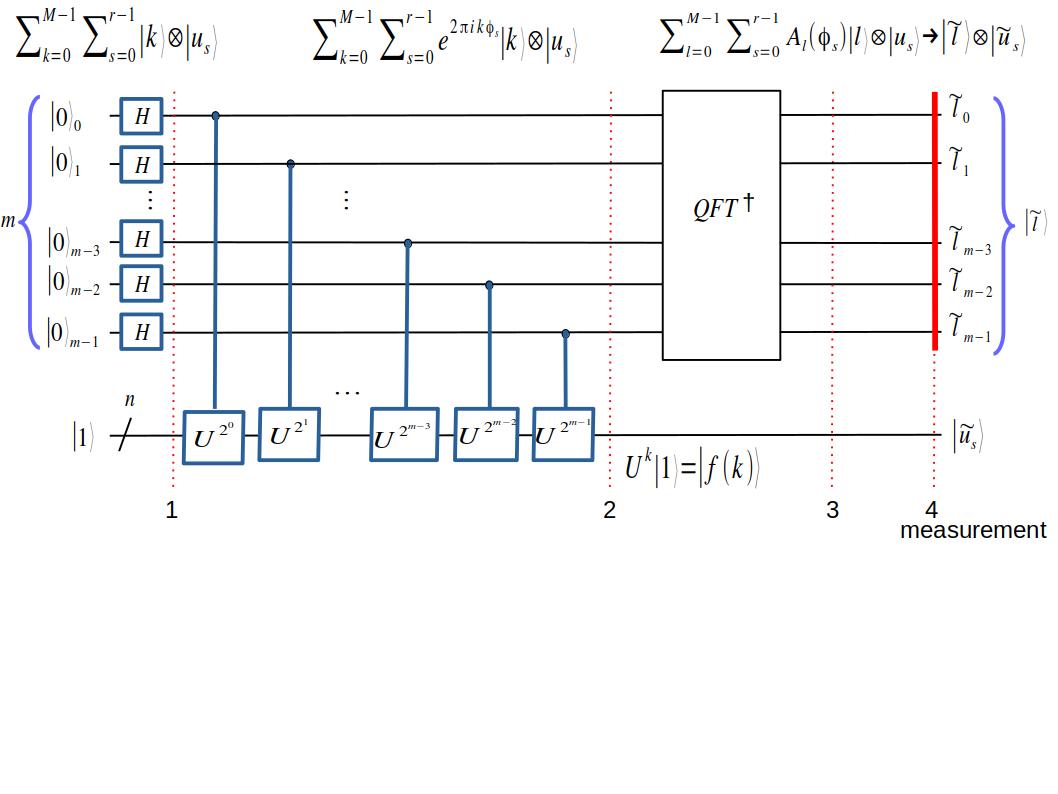}
\par
\end{centering}
\vskip-3.0cm
\caption{\footnoteskip
Shor's algorithm for factoring a semi-prime $N$ in the 
OpenQASM/Qiskit convention.  There are two registers,
a control register containing $m$ qubits, and a work
register containing $n$ qubits, with $n = \lceil \log_2 
N \rceil$. Measuring the control register projects the 
system into a state $\vert\tilde\ell \rangle \otimes 
\vert \tilde u_s \rangle$ for some index $\tilde\ell = 
\tilde\ell_{m-1}\cdots \tilde\ell_0 \tilde\ell_0 \in 
\{0, 1, \cdots, M-1\}$ and some ME eigenstate $\vert 
\tilde u_s \rangle$, where $M = 2^m$ is the number 
of quantum states in the control register. This gives 
the \hbox{$m$-bit} measured phase $\tilde\phi = 
\tilde\ell/M = 0.\tilde\ell_{m-1} \cdots \tilde\ell_1
\tilde\ell_0$, which lies close to  the actual phase 
$\phi_s$ with high probability. Upon knowing $\tilde
\phi$, we can extract the exact phase $\phi_s = s/r$ 
using the method of continued fractions.
}
\label{fig_shor_4}
\end{figure}

The general circuit for Shor's algorithm is illustrated in 
Fig.~\ref{fig_shor_4}. Two registers are required: (i) 
a control register of $m$ qubits, and (ii) a work register 
of $n$ qubits. The work register stores all possible values 
of the modular exponentiation (ME) function 
\begin{eqnarray} 
  f(x) = a^x ~({\rm mod}~N)
  \ ,
\label{eq_fx_def}
\end{eqnarray}
where $a$ is called the {\em base}, and it satisfies the
conditions $1 < a < N$ and ${\rm gcd}(a, N) = 1$. The
factors of $N$ are related to the period $r$ of the ME
function by ${\rm gcd}(a^{r/2} \pm 1, N)$. Since the values 
of $f(x)$ lie between 0 and $N-1$, we choose the 
number of work qubits to be $n = \lceil \log_2 N \rceil$.  
The control register, which might also be called the period
register or the {\em phase register}, stores all possible
$m$-bit approximations to the eigenphases of the 
corresponding ME operator. Note that there are $M = 
2^m$ quantum states in the control register. The continued 
fractions algorithm dictations that the number of qubits 
should be set to $m = 2 n + 1$ for sufficient phase 
resolution\,\cite{qcqi}. 
The work-space states are indexed by an $n$-bit binary 
integer $w = w_{n-1} \cdots w_1 w_0$, where $w 
\in\{0, 1, \cdots, 2^n-1\}$. We will denote states in the 
work register by either their binary representation 
$\vert w_{n-1} \cdots w_1 w_0  \rangle_{\rm w}$,  or 
by their numerical form $\vert w \rangle_{\rm w}$ 
(usually the later). These states can be viewed as the 
computational basis for the work register, and therefore 
linear operators are specified by their action on these 
states. 

The modular exponentiation (ME) operator for Shor's  
algorithm is defined by 
\begin{eqnarray} 
  U_{a,  \smN}\, \vert w \rangle_{\rm w}
  =
  \vert a \times w ~({\rm mod}~N) \rangle_{\rm w}
  \ ,
\label{eq_Uw_def}
\end{eqnarray}
and therefore
\begin{eqnarray}
  U_{a,  \smN} \,\vert f(x) \rangle_{\rm w} &=& \vert f(x+1) \rangle_{\rm w} \ ,
  ~~\text{or equivalently}~~
\\[5pt]
   U_{a,  \smN}^x \, \vert 1 \rangle_{\rm w}
  &=&
  \big\vert f(x)  \big\rangle_{\rm w}
   ~~\text{for any}~ x \in \{0, 1, 2,  \cdots \} 
   \ .
\end{eqnarray}
We will use the short-hand notation $U = U_{a, \smN}$ in 
the future, with the base $a$ and the number $N$ being 
left implicit. The eigenvalue problem for the ME operator
$U$ can be expressed by 
\begin{eqnarray}
  U \vert u_s \rangle_{\rm w}
  &=&
  e^{2\pi i \phi_s}\, \vert u_s \rangle_{\rm w}
  ~~~\text{where}~~ \phi_s = \frac{s}{r}
  ~~\text{with}~~ s \in \{0, 1, \cdots, r-1\}
  \ ,
\label{eq_us_def1}
\end{eqnarray}
and the eigenstates are given by 
\begin{eqnarray}
  \vert u_s \rangle_{\rm w}
  &=&
  \frac{1}{\sqrt{r}}\sum_{k=0}^{r-1}\, e^{-2\pi i k \, \phi_s}\,
  \vert f(k) \rangle_{\rm w}
\label{eq_us_def2}
  \ .
\end{eqnarray}
By using the relation
\begin{eqnarray}
 \sum_{s = 0}^{r-1} e^{2\pi i \, k \phi_s}
 =
 r \, \delta_{k, 0}
\label{eq_sum_exps}
\end{eqnarray}
for any integer $k$, we can invert (\ref{eq_us_def2}) 
to obtain
\begin{eqnarray}
 \vert f(k) \rangle_{\rm w}
  &=&
  \frac{1}{\sqrt{r}} \sum_{s=0}^{r-1}\, e^{2\pi i k \, \phi_s}\,
  \vert u_s \rangle_{\rm w}
\ .
\label{eq_ak_again}
\end{eqnarray}
It is clear from (\ref{eq_ak_again}) that $\vert f(k) \rangle_{\rm w}$
is periodic in $k$ with period $r$. Note that for $k = 0$,  we find
the interesting expression
\begin{eqnarray}
  \vert 1 \rangle_{\rm w}
  =
  \vert 0 \cdots 01 \rangle_{\rm w}
  =
  \frac{1}{\sqrt{r}}\sum_{s=0}^{r-1} \vert u_s \rangle_{\rm w}
  \ .
\label{eq_one}
\end{eqnarray}
Thus, populating the work register with state $\vert 1 
\rangle_{\rm w}$ is equivalent to filling that register 
with a uniform linear sum of the eigenstates $\vert u_s 
\rangle_{\rm w}$ for all values of $s $. This is the origin
of the entanglement between the work and control
registers. We are now ready to commence with Shor's 
algorithm.

\vskip0.2cm
\noindent
- (i) To factor a semi-prime number $N$, randomly choose 
a base $a$  such that $1 < a <N$ and ${\rm gcd}(a, N) = 1$ 
(if there {\em were} non-trivial common factors of $a$ and 
$N$, then we have found a factor of $N$, which was the
objective). The ME function $f(x)$ is now defined as in 
(\ref{eq_fx_def}), and we seek the period $r$ of this 
function,  that is to say, the smallest positive integer $r$ 
such that $a^r~({\rm mod}~N) = 1$. If ${\rm gcd}(a, N) 
= 1$, such a period $r \le N$ always exists.

\noindent
- (ii)
The control and work registers are initialized to the state 
\begin{eqnarray}
  \vert \psi_0 \rangle
  = 
  \vert\underbrace{ \,00 \cdots 0\,}_{m}  \rangle_{\rm c}
  \,\otimes \, 
  \vert\underbrace{ 0 \cdots 01 \hskip0.05cm}_{n} \rangle_{\rm w}
  =
  \vert 0 \rangle_{\rm c} \otimes \vert 1 \rangle_{\rm w}
  \ .
\label{eq_psi0}
\end{eqnarray}

\noindent
- (iii) Hadamard gates $H^{\otimes m}$ act on the control 
register to create the uniform distribution 
\begin{eqnarray}
  \vert \psi_1 \rangle 
  &=&
  \frac{1}{\sqrt{M}}\sum_{k=0}^{M-1}\vert k \rangle_{\rm c} 
  \otimes \vert 1 \rangle_{\rm w}
\label{eq_psi1}
\\[5pt]
  &=&
    \frac{1}{\sqrt{r M}}\sum_{k=0}^{M-1}\sum_{s=0}^{r-1}
    \vert k \rangle_{\rm c} \otimes \vert u_s \rangle_{\rm w}
  \ ,
\label{eq_psi1_b}
\end{eqnarray}
where the number of quantum states in the control register 
is $M = 2^m$.  I have used (\ref{eq_one}) in passing from 
the first form of state-1 to the second form, which is more 
convenient when taking the inverse QFT later in the circuit. 
The Hadamard gates ensure that all possible combinations 
of $k$ can be processed simultaneously via the massive 
parallelism of quantum mechanics. 

\begin{figure}[b!]
\includegraphics[scale=0.40, center]{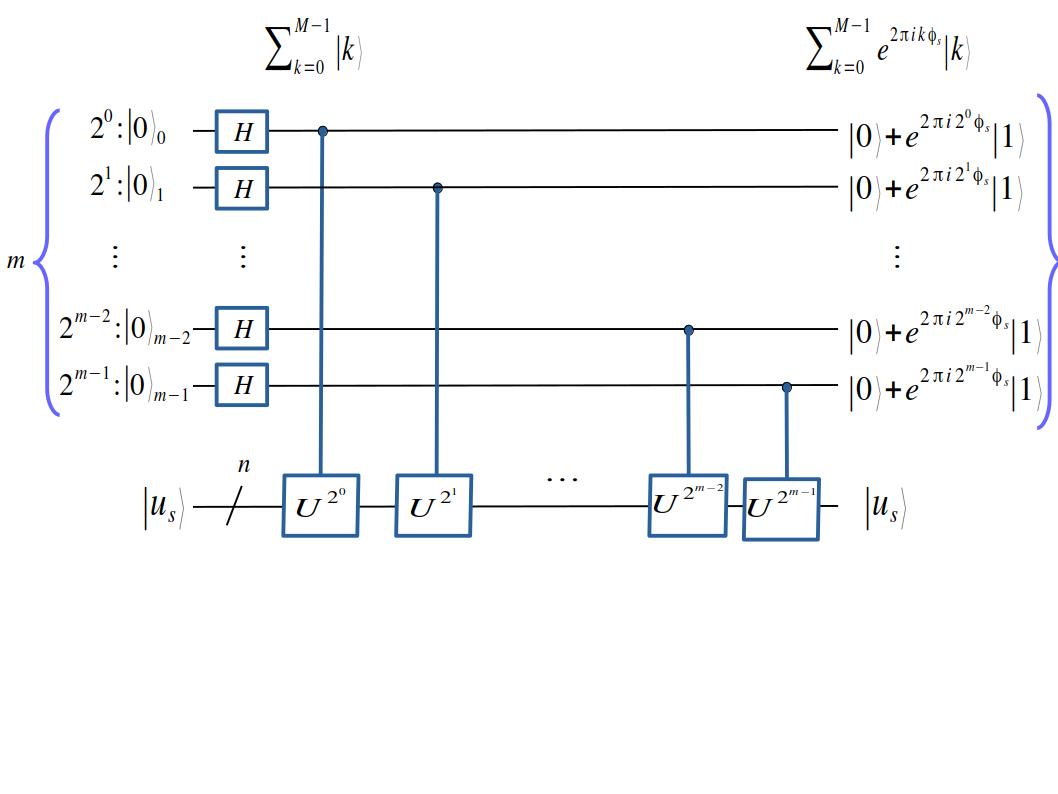} 
\vskip-2.5cm 
\caption{\footnoteskip  
  Quantum phase estimation front-end in the OpenQASM/Qiskit
  convention. 
  The work register has been populated by the eigenstate
  $\vert u_s \rangle$. 
}
\label{fig_qpe_1_conv1}
\end{figure}

\noindent
- (iv) Controlled modular exponentiation operators $CU^p$ 
for $p \in \{2^0, 2^1, \cdots, 2^{m-1}\}$ 
act between the control and work registers to form the highly
entangled state
\begin{eqnarray}
  \vert \psi_2 \rangle 
  &=&
  \frac{1}{\sqrt{r M}}\sum_{k=0}^{M-1}\sum_{s=0}^{r-1}\, 
  e^{2\pi i k \, \phi_s} \, \vert k \rangle_{\rm c} \otimes 
  \vert u_s \rangle_{\rm w}
  \label{eq_psi2}
\\[5pt]
  &=&
      \frac{1}{\sqrt{M}}\sum_{k=0}^{M-1}\vert k \rangle_{\rm c} 
  \otimes \vert f(k) \rangle_{\rm w}
  \ ,
\label{eq_psi2_again}
\end{eqnarray}
where I have used (\ref{eq_ak_again}) in passing from
the first form to the second form of state-2. These two
equivalent forms express different ways of thinking about 
the state after the controlled ME operators $CU^p$ have 
acted. Figure~\ref{fig_shor_4} emphasizes forms 
(\ref{eq_psi2}) and (\ref{eq_psi1_b}) expressed in terms
of the eigenphases $\vert u_s \rangle$. 
See Fig.~\ref{fig_qpe_1_conv1} and the surrounding 
discussion for an explanation of the role of phase 
kickback and entanglement in expression (\ref{eq_psi2}).

\noindent
- (v) The previous state involves the Fourier transform of the
phases $\phi_s$. These phases can therefore be recovered 
with an inverse Fourier transform applied to the control register,
which takes the form 
\begin{eqnarray}
   QFT^\dagger_{\rm c}
   &=&
   \frac{1}{\sqrt{M}}\sum_{\ell = 0}^{M-1}\sum_{k=0}^{M-1}
   e^{-2\pi i\, k \ell /M}\, 
   \vert \ell \rangle \langle k \vert_{\rm c}
    \ .
\label{eq_qft_dagger}
\end{eqnarray}
The final state therefore becomes
\begin{eqnarray}
  \vert \psi_3 \rangle 
  &=& 
  QFT^\dagger_{\rm c} \, \vert \psi_2\rangle
  =
  \sum_{\ell=0}^{M-1} \sum_{s =0}^{r-1}
  A_\ell(\phi_s)\,
  \vert \ell \rangle_{\rm c} \otimes \vert u_s \rangle_{\rm w}
  \ ,
\label{eq_psi3}
\end{eqnarray}
where the amplitudes are given by 
\begin{eqnarray}
   A_\ell(\phi_s)
   &=&
   \frac{1}{\sqrt{r} M} 
   \sum_{k = 0}^{M-1} \, e^{2\pi i k (\phi_s - \ell / M)}
 \\[3pt]
  &=&
  \frac{1}{\sqrt{r} M} \,
  \frac{1 - e^{2 \pi i \, (\phi_s - \ell/M)M}}{1 - e^{2 \pi i \, 
  (\phi_s - \ell/M)}}
  \ .
\label{eq_Als}
\end{eqnarray}

\noindent %
- (vi) We now measure the control register to obtain 
$\tilde\ell = \tilde\ell_{m-1} \cdots \tilde\ell_1\, \tilde\ell_0$. 
Dominant peaks occur at the phase values $\tilde\phi = 
\tilde\ell/M = 0.\tilde\ell_{m-1} \cdots \tilde\ell_1\, 
\tilde\ell_0$  near the exact ME  phases $\phi_s=s/r$ 
for $s \in \{0, 1, \cdots, r - 1\}$. The continued fractions
algorithm is then applied to the control register to 
extract the integers $s$ and $r$ of the exact phase 
$\phi_s = s/r$. As a practical matter, for the continued 
fractions algorithm to work, the integers $s$ and $r$ 
can have no non-trivial common factors. This  slightly 
reduces the number of reliable peaks from which $r$ 
can be obtained. The number of control qubits is 
required to be $m = 2 n + 1$, so that
$\tilde\phi$ has enough resolution to reliably extract 
the phase $\phi_s = s/r$ using the continued fractions
method\,\cite{qcqi}. If the period turns out to be odd, 
then return to step (i) and choose another base~$a$ 
(unless $a$ is a perfect square). Otherwise, check that 
$a^{r/2} \ne -1 ~({\rm mod}~N)$ and that $a^r = 1 
~({\rm mod}~N)$. Note that $a^{r/2} \ne 1 ~({\rm mod}
~N)$ holds automatically (otherwise the period would
be $r/2$ rather than $r$). If these conditions are not 
met, then return to (ii) and try again. The probability 
of success is quite high after only a few iterations.
Once a solution for $r$ is obtained, proceed to the 
next and final step. 

\noindent
- (vii) The factors are given by ${\rm gcd}\big( a^{r/2} 
\pm 1, N \big)$. The greatest common divisor can be 
calculated quickly using Euler's algorithm on a classical 
computer. 

The modular exponentiation operator of (iv) is the most 
crucial component of Shor's algorithm, and it is useful to 
review it further, with an eye on entanglement. Let us start 
with a single eigenstate $\vert u_s \rangle$ in the work 
register. The control and work subscripts have been dropped 
for readability. A series of controlled operators $C U^p$ 
with control qubit \hbox{$q \in \{0, 1, \cdots, m - 1\}$} and 
power $p \in \{2^0, 2^1,\cdots, 2^{m-1}\}$ act on the work 
register. Each power $p$ corresponds to a power of 2 in 
the binary representation of an $m$-bit number $k = k_{m-1} 
\cdots k_1 k_0$ in the control register. After the action 
of the controlled ME operator $CU^p$ (position-2 in 
Fig.~\ref{fig_shor_4}), phase kickback produces the state
\begin{eqnarray}
   CU^p H \vert 0 \rangle  \otimes \vert u_s \rangle
   &=&
   \frac{1}{\sqrt{2}}\, \vert 0 \rangle \otimes \vert u_s \rangle
   +
   \frac{1}{\sqrt{2}}\, \vert 1 \rangle \otimes U^p \vert u_s \rangle
\\[3pt]
  &=&
   \frac{1}{\sqrt{2}}\, \vert 0 \rangle\otimes \vert u_s \rangle
   +
   \frac{1}{\sqrt{2}}\, \vert 1 \rangle \otimes e^{2\pi i p\, \phi_s} 
   \vert u_s \rangle
\\[3pt]
  &=&
     \frac{1}{\sqrt{2}}\, \Big(\vert 0 \rangle
   +
   e^{2\pi i p\, \phi_s}  \, \vert 1 \rangle
   \Big) 
   \otimes \vert u_s \rangle
   \ .
\end{eqnarray}

\noindent
This procedure is represented in Fig.~\ref{fig_qpe_1_conv1} 
for the state $\vert u_s \rangle$ in the work register. The 
system is now in the state
\begin{eqnarray}
  \vert \psi_2^s \rangle
  &=&
  \frac{1}{2^{m/2}}\,
  \big( \vert 0 \rangle + e^{2\pi i\, 2^0 \phi_s} \,\vert 1 \rangle \big)_0 
  \otimes \cdots \otimes
  \big( \vert 0 \rangle + e^{2\pi i\, 2^{m-1} \phi_s} \,\vert 1 \rangle \big)_{m-1} 
  \otimes
  \vert u_s \rangle
\\[4pt]
  &=&
    \frac{1}{2^{m/2}}\,
    \sum_{k_0 = 0, 1} e^{2\pi i \, 2^0 \cdot k_0} \vert k_0 \rangle
    \otimes \cdots \otimes  \!\!
    \sum_{k_{m-1} = 0, 1} e^{2\pi i \, 2^{m - 1}\cdot k_{m-1}} \vert k_{m-1}
    \rangle
    \otimes \vert u_s \rangle
\\[5pt]
  &=&
  \frac{1}{2^{m/2}}
  \sum_{k_{0}=0,1} \cdots \sum_{k_{m-1}=0,1} 
  e^{2\pi i \phi_s (2^0 k_0 \,+ \,\cdots \,+ \,2^{m-1} k_{m-1})} 
  \vert k_0 \rangle \otimes \cdots \otimes \vert k_{m-1} \rangle
  \otimes
  \vert u_s \rangle \hskip1.2cm
\\[5pt]
  &=&
  \frac{1}{\sqrt{M}} \sum_{k=0}^{M-1} e^{2\pi i k \phi_s}\,
  \vert k \rangle \otimes \vert u_s \rangle
  \equiv
  \vert {\rm control} \rangle \otimes \vert {\rm work} \rangle
  \ .
\label{eq_psi2s}
\end{eqnarray}
\noindent
Note that the control register for state (\ref{eq_psi2s}) is entangled
with itself (and indeed, this entanglement will produce the quantum 
speedup). However, the control and work registers are not entangled, 
as $\vert \psi_2^s \rangle$ is a simple tensor product of a control 
state and a work state. However, if the work register is in state 
$\vert 1 \rangle$, then we must sum over all $s$ with equal weights, 
so that
\begin{eqnarray}
  \vert \psi_2 \rangle
  =
  \frac{1}{\sqrt{r}} \sum_{s=0}^{r-1}\, \vert \psi_2^s \rangle
  =
  \frac{1}{\sqrt{r M}} \sum_{k=0}^{M-1} \sum_{s=0}^{r-1}
  e^{2\pi i k \phi_s}\,
  \vert k \rangle \otimes \vert u_s \rangle
  \ ,
\end{eqnarray}
from which state (\ref{eq_psi2}) is obtained. Thus, by starting 
the work register in state $\vert 1 \rangle$, entanglement 
between  the control and work registers is introduced. 
This is why a measurement on the control register in 
Fig.~\ref{fig_shor_4} also projects the work register into 
an eigenstate of $U$. Although the work state is not
measured, it might be interesting to do so for diagnostic
purposes, but that is beyond the scope of this work. 

It is informative to calculate $\vert \psi_2 \rangle$ in yet another 
way. Suppose the  control register is in the state~$\vert k \rangle$, 
where $k = k_{m-1} \cdots k_1 k_0$, and take the work register 
to be in state $\vert 1 \rangle$. Then the action of the controlled 
ME operators $CU^p$ produces the work-state
\begin{eqnarray}
   U^{2^0 \cdot k_0} \,{\scriptstyle\times}\, U^{2^1 \cdot k_1} 
   \,{\scriptstyle\times}\,\cdots \,{\scriptstyle\times}\,
   U^{2^{m-1} \cdot k_{m-1}} \vert 1 \rangle
 =
 U^k \vert 1 \rangle 
 =
 \vert f(k) \rangle  
\label{eq_worstate_k}
  \ ,
\end{eqnarray}
and state-2 becomes
\begin{eqnarray}
  \vert \psi_2 \rangle 
  &=&
  \frac{1}{\sqrt{M}}\sum_{k=0}^{M-1}\vert k \rangle
  \otimes \vert f(k) \rangle
  \ ,
\end{eqnarray}
in agreement with (\ref{eq_psi2_again}). Upon employing
(\ref{eq_ak_again}), state-2 can be written in the first form  
(\ref{eq_psi2}). This is more convenient when taking the 
inverse QFT. 

There is a  further simplification when the $\phi_s$ 
are $m$-bit phases. In this case $\ell_s = M \phi_s$
is an integer between 0 and $M -1$, and  $A_\ell(\phi_s) 
= \delta_{\ell, \ell_s}/\sqrt{r}$\,\,\cite{sing1}. 
Most amplitudes vanish, and there are only $r$
non-zero  values at $\ell = \ell_s$ for $s \in \{0, 1, 
\cdots, r-1\}$, so that 
\begin{eqnarray}
  \vert \psi_3 \rangle 
  &=&
  \frac{1}{\sqrt{r}} \sum_{s=0}^{r-1} \, 
 \vert \ell_s \rangle \otimes \vert u_s \rangle
\label{eq_psi3_mbit}
  \ .
\end{eqnarray}
There is a more revealing way of obtaining this result. Since
$M \phi_s = \ell_s$ is an $m$-bit integer, note that (\ref{eq_psi2}) 
can be expressed in terms of an exact QFT, 
\begin{eqnarray}
  \vert \psi_2 \rangle 
&=&
  \frac{1}{\sqrt{r}}\,   \frac{1}{\sqrt{M}}
  \sum_{k=0}^{M-1}\sum_{s=0}^{r-1}\, 
  e^{2\pi i \, k  \ell_s/M} \, \vert k \rangle \otimes 
  \vert u_s \rangle 
  =
  QFT \cdot
  \frac{1}{\sqrt{r}} \sum_{s=0}^{r-1} \,
  \vert \ell_s \rangle \otimes \vert u_s \rangle
\label{eq_psi2_mbit}
  \ .
\end{eqnarray}
Upon taking the inverse $QFT^\dagger$, the final state is
given by (\ref{eq_psi3_mbit}). This is the case for $N = 15$ 
where $r = 2, 4$ for all bases $a$. This does not occur, for 
example, with a period that contains the factor 3, since $1/3$ 
does not terminate. This explains the richer phase-structure 
for $N =21$ (where $a=2$ gives $r = 6$),  and  for larger 
numbers $N$.

\vfill
\pagebreak
\section{Truncated Modular Exponentiation Operators}
\label{sec_truncate}

Recall that the Shor factoring circuit of Fig.~\ref{fig_shor_4} 
possesses two registers: a control or phase register containing 
$m$ qubits, and a work register containing $n = \lceil \log_2 N 
\rceil$ qubits, where $N$ is the semi-prime number we wish 
to factor. For sufficient phase resolution in the continued 
fractions routine, the control register should contain $m =2 
n + 1$ qubits, although in practice we can sometimes get by 
with fewer qubits. The modular exponentiation (ME) operators 
$U^p$ for $p \in \{2^0, 2^1, \cdots, 2^{m-1}\}$ act on the work 
register, and are the most crucial component of Shor's 
algorithm. They are often called the {\em bottleneck} of the 
algorithm. I would also refer to them as the {\em workhorse} 
of the procedure, as most of the quantum resources are deployed 
here. Furthermore, the ME operators are responsible for the 
entanglement of the control register, which is required for a 
quantum speedup. Shor's algorithm is based on the observation 
that the factors of a semi-prime number $N$ are determined 
by the period $r$ of the ME function $f(x) = a^x ~({\rm mod}~N)$,
and the algorithm 
is designed to extract this period. The ME function $f(x)$ is 
implemented by an ME operator $U$ that performs the 
transitions $\vert f(x) \rangle \to \vert f(x + 1) \rangle$
from one work-state to the next. As previously emphasized, 
the ME operators are indeed both the bottleneck and the 
workhorse of the procedure, as they require the largest 
expenditure of quantum resources. For example, 
constructing a general $U$ operator can employ 
thousands of gates\,\cite{preskill1}, each of which 
must maintain coherence with the others. 

In Ref.~\cite{sing1},  I recently proposed a method for 
reducing the gate count of the ME operators $U^p$ 
with a total of $3n + 1$ qubits in the Shor circuit. This 
reference also serves as a thorough introduction to Shor's 
algorithm. I employed the Qiskit simulator to factor 
a variety of smaller numbers ($N = 15, 21, 33, 35$), as well as 
the larger semi-primes $N = 143, 247$. That smaller values of 
$N$ could be factored in this way is not very remarkable, and 
serves to prove that the method is sound. But to my 
knowledge, numbers in the hundreds had never been factored 
before, and indeed pushed the limits of the simulator, even with 
the reduced gate count. Soon thereafter, Ref.~\cite{tomcala} also 
factored $N = 143, 247$ on the Qiskit simulator using quite similar 
methods. However, the author used much smaller periods than
Ref.~\cite{sing1}. For example, in factoring $N = 143$ and $N = 
247$, Ref.~\cite{tomcala} used the periods $r = 4$  ($a = 21)$ 
and $r = 12$ ($a = 8$), respectively. This should be compared 
to Ref.~\cite{sing1}, which factored these numbers using the
much larger periods of $r = 20$ ($a = 5$)  for $N = 143$ and  
$r = 36$ ($a = 2$) for $N = 247$. I used these larger periods 
in an effort to push the simulator and the continued fractions 
routine to their limits. Using such large periods was facilitated 
by the automation of the gate construction procedure, as 
making an ME operator with a period of $r = 36$ is just too 
time consuming and tedious to perform manually. 

Reference~\cite{tomcala} goes on to survey a large 
number of implementation strategies for  ME operators. 
The author categorizes ME operators into three primary 
types: (i) those that employ basic arithmetic operations, 
(ii) those that employ Fourier rotations, and (iii) specially 
designed circuits.  Examples of method (i) are given in 
Refs.~\cite{preskill1} and \cite{vanmeter}. Most notably, 
the general routine of Ref.~\cite{preskill1} takes of order 
$72 n^3$ gates for an $n$-bit ME operator. 
While the number of gates is {\em only} polynomial 
in $n$, the cubic growth renders the gate count far too 
large for current machines. Even the smallest number 
$N = 15$ for which Shor's algorithm is applicable would 
require 9000 gates with this procedure. An example of 
method (ii) is Ref.~\cite{pavlidis1}, which requires of 
order $2000 n^2$ gates, and is therefore also out of 
current experimental reach. After a thorough review, 
Ref.~\cite{tomcala} concludes that only method (iii) 
involving specially designed circuits is feasible for existing 
and near-term devices. The method of Ref.~\cite{sing1}
falls into category (iii) of this taxonomy. 
 
The proposed method is not really profound, and lies 
on the simple observation that
\begin{eqnarray}
  U \vert f(x) \rangle &=& \vert f(x+1) \rangle 
  \ ,
\label{eq_U_fx}
\end{eqnarray}
or equivalently, 
\begin{eqnarray}
   U^x \vert 1 \rangle 
  &=&
  \big\vert f(x)  \big\rangle 
   ~~\text{for any}~ x \in \{0, 1, 2,  \cdots \} 
   \ ,
\end{eqnarray}
where $f(x)$ is the ME function defined by (\ref{eq_fx_def}). 
Since the work register starts in state $\vert 1 \rangle$, 
we do not have to create an ME operator $U$ that 
takes a {\em general} input, but rather, one that takes an input 
from the periodic sequence of states $\vert f(x) \rangle$ for 
$x \in \{0, 1, 2, \cdots\}$. An ME operator $U$ with period $r$ 
can therefore be partitioned into $r$ segments indexed by 
the integers $x \in \{0, 1, \cdots, r-1\}$. The gates in segment 
$x$ transform the work-state $\vert f(x) \rangle \equiv \vert 
w_{n-1} \cdots w_0 \rangle$ into $U \vert f(x) \rangle = \vert 
f(x+1) \rangle \equiv \vert w_{n-1}^\prime \cdots w_0^\prime 
\rangle$, while the levels below $x$ have no effect on the
state $\vert f(x + 1) \rangle$. Transforming a binary number 
$w_{n-1} \cdots w_1 w_0$ into another binary number 
$w_{n-1}^\prime \cdots w_1^\prime w_0^\prime$, without 
altering the  preceding states, can be turned into a set of 
formal rules involving multi-control-NOT gates $CC \cdots 
C X$ and single-qubit NOT gates $X$.  This procedure can 
therefore be automated. Indeed, this is how most of the 
gates in this manuscript and in Ref.~\cite{sing1} were 
constructed: The factorization script essentially wrote 
itself, including the Qiskit gate output for the next level 
of the ME operator $U$. The same script can be used 
for the composite operators $U^p$ for $p \in \{2^0, 2^1, 
\cdots, 2^{m-1}\}$. This method consequently requires 
${\cal O}(m \times n r) \sim {\cal O}(n^2 r)$ gates, as each 
of the $m =2 n +1$ operators $U^p$  requires of order 
$n r$ gates.

Let us make the following observation before continuing
with an example. We shall denote the $2^n$-dimensional 
space for the  work-register by ${\cal W}_n$, and a general 
ME operator~$U$ can act on this entire Hilbert space. 
Consider now the $r$-dimensional subspace defined by 
\begin{eqnarray}
  {\cal U}_r 
  \equiv
  {\rm Span}\,\Big\{ \,\big\vert f(x) \big\rangle\, \,\Big\vert\, 
  x \in \{0, 1,  \cdots,  r-1\}  \Big\} \subseteq {\cal W}_n
  \ .
\end{eqnarray}
As discussed above, the  $U$ operator transforms one basis 
element of ${\cal U}_r$ into another basis element,  that is to 
say, the ME operator $U$ leaves the subspace ${\cal U}_r$ 
invariant, so that \hbox{$U[{\cal U}_r] = {\cal U}_r$}. 
Thus, as the $U$ operator acts successively, the states in the 
work register only vary over the $r$-dimensional subspace 
${\cal U}_r$.  This means that the subspace ${\cal U}_r$ is 
the only portion of ${\cal W}_r$ that is entangled with the 
control register. 

\subsection{$\bm{N=21 = 3 \times 7}$, $\bm{a=2}$, $\bm{r=6}$}

\begin{figure}[b!]
\hskip-1.0cm
\begin{minipage}[c]{0.4\linewidth}
\includegraphics[scale=0.48]{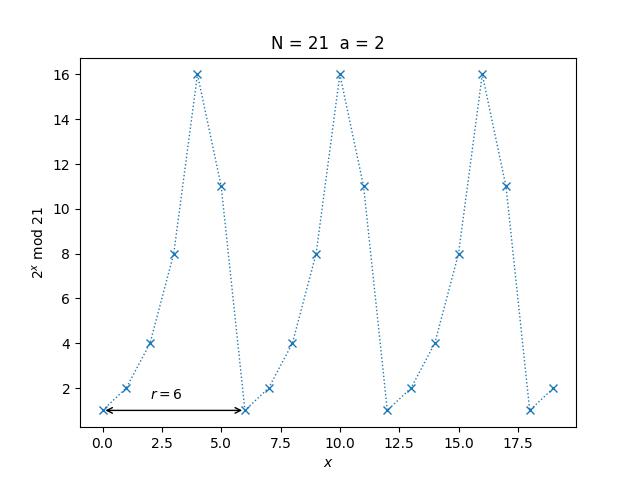}
\end{minipage}
\hskip1.6cm
\begin{minipage}[c]{0.4\linewidth}
\begin{tabular}{|c|c|} \hline
 \multicolumn{2}{|c|}{~$U\vert w \rangle = 
\big\vert 2 \times w ~({\rm mod}~21) \big\rangle$~}  
\\\hline
$~~~U\vert 1 \rangle ~= \vert 2 \rangle$~~~~~&
~~$U\vert 00001 \rangle = \vert 00010 \rangle$~~~~~\\[-5pt]
$~~~U\vert 2 \rangle ~= \vert 4 \rangle$~~~~~&
~~$U\vert 00010 \rangle = \vert 00100 \rangle$~~~~~\\[-5pt]
$~~~U\vert 4 \rangle ~= \vert 8 \rangle$~~~~~&
~~$U\vert 00100 \rangle = \vert 01000 \rangle$~~~~~\\[-5pt]
$~~~~U\vert 8 \rangle ~= \vert 16 \rangle$~~~~~&
~~$U\vert 01000 \rangle = \vert 10000 \rangle$~~~~~\\[-5pt]
$~~~U\vert 16 \rangle ~= \vert 11 \rangle$~~~~~&
~~$U\vert 10000 \rangle = \vert 01011 \rangle$~~~~~\\[-5pt]
$~~~U\vert 11 \rangle~ = \vert 1  \rangle$~~~~~~&
~~$U\vert 01011 \rangle = \vert 00001 \rangle$~~~~~\\\hline
\end{tabular} 
\end{minipage}
\caption{\footnoteskip
$N = 21$, $a = 2$, $r = 6$:
The left panel illustrates the modular exponential function 
$f_{2, 21}(x) = 2^x ~({\rm mod}~21)$, while the right panel 
shows the action of the modular exponentiation operator 
$U_{2, 21}$ on the closed sequence $[1, 2, 4, 8, 16, 11, 1 ]$.  
The circuit requires $n = \lceil \log_2 21 \rceil = 5$ qubits 
in the work register. 
}
\label{fig_fN21a2x}
\end{figure}

\begin{figure}[t!]
\begin{centering}
\includegraphics[width=\textwidth]{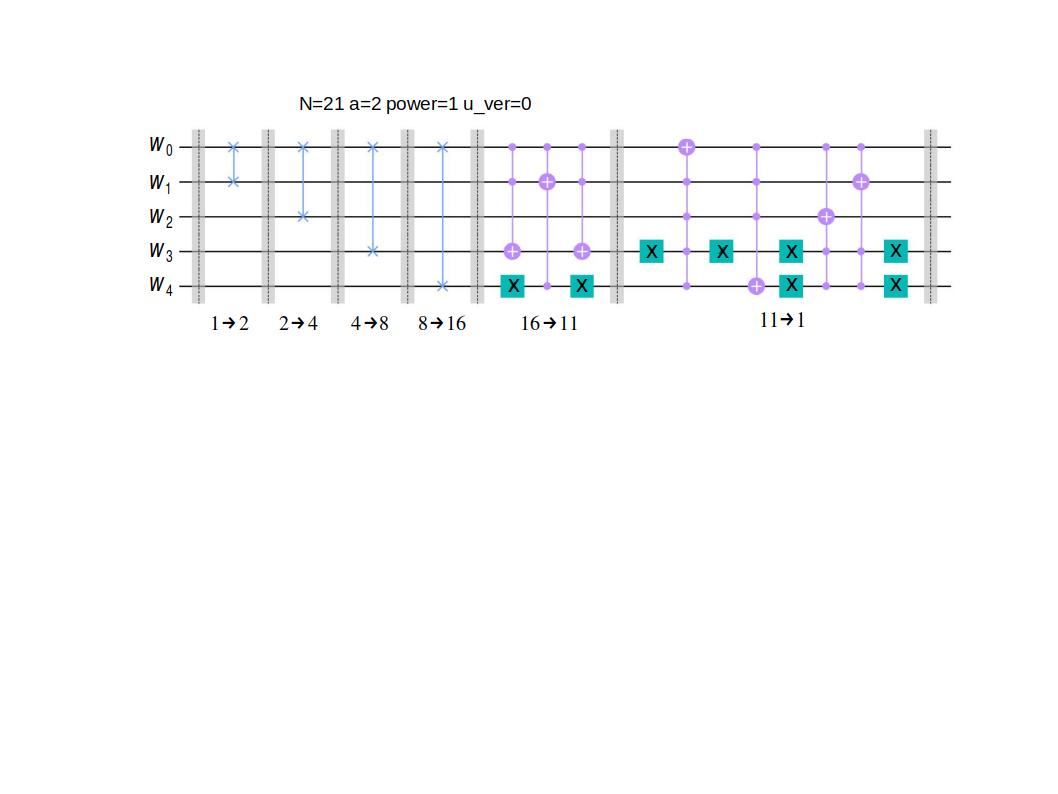} 
\par\end{centering}
\vskip-7.2cm
\caption{\footnoteskip  
$N = 21$, $a = 2$, $r = 6$, $n=5$:
The quantum circuit for the modular exponentiation (ME) operator  
$U_{2 , 21}$ with $n = \lceil \log_2 21 \rceil = 5$ work qubits. The 
quantum gates between the barriers transform the work-states
from one value of $f_{2 , 21}(x)$ to the next in the closed sequence 
$[1, 2, 4, 8, 16, 11, 1]$. This version of the ME operator will be called 
${\tt{u\_ver}}=0$. Note that this circuit requires the 
multi-control gates $CX$, $CCX$, $CCCX$ and $CCCCX$.
}
\label{fig_U21a2_sub}
\end{figure}

As an example of the method,  consider the number $N = 21$ 
with the base $a = 2$. The work register must contain 
$n = \lceil \log_2 21 \rceil = 5$ qubits, and therefore takes 
the form $\vert w \rangle = \vert w_4 \cdots w_0 \rangle$, 
where \hbox{$w = w_4 w_3 w_2 w_1 w_0 \in  \{0, 1, \cdots, 31\}$} 
is a \hbox{$5$-bit} indexing integer. Note that the ME function 
$f(x) = 2^x ~({\rm mod}~ 21)$ has a period of $r = 6$, as illustrated 
in Fig.~\ref{fig_fN21a2x}. Also note that the state transitions 
are given by $\vert 1 \rangle \to  \vert 2 \rangle \to \vert 4 
\rangle\to \vert 8\rangle \to \vert 16 \rangle \to \vert 11 
\rangle \to \vert 1 \rangle$.   Figure~\ref{fig_U21a2_sub} 
shows the corresponding 
ME operator $U_{2, 21}$, which is partitioned into six levels 
delimited by the barriers in the circuit. The first partition in 
the Figure gives $U \vert 1 \rangle = \vert 2 \rangle$, as the 
first SWAP gate changes the initial state \hbox{$\vert 1 
\rangle = \vert 00001 \rangle$} into the next state $\vert 2 
\rangle = \vert 00010 \rangle$.  The remaining sequence of 
gates have no effect on the state $\vert 2 \rangle$. In the 
second partition, the next SWAP operation transforms 
$\vert 2 \rangle = \vert 00010 \rangle$ into $\vert 4 \rangle 
= \vert 00100 \rangle$,  and so on.  More precisely, because the 
gates in each level act successively, the second transition
$\vert 2 \rangle  \to \vert 4\rangle$ is actually accomplished 
by the first two partitions: the SWAP gate in $x=0$ is followed 
by the SWAP gate in $x=1$, which performs the operation 
$\vert 2 \rangle \to \vert 1 \rangle \to \vert 4 \rangle$. The 
remaining partitions have no effect on state $\vert 4 \rangle$. 
Finally, the last partition returns the state $\vert 11 \rangle$ 
back into the state $\vert 1 \rangle$. We can then concatenate 
the operators $U_{2, 21}$ to form the composite operators 
$U_{2, 21}^p$ for $p \in \{2^0, 2^1, \cdots, 2^{m-1}\}$. Simple 
concatenation is the fastest way of confirming that $U$ is 
correct, and that the Shor circuit does indeed perform the 
factorization. We will soon abandon the concatenation strategy
since it requires ${\cal O}(n^3 r)$ gates, and we will build the 
individual operators $U^p$ for every $p$. As previously noted, 
this procedure uses ${\cal O}(n^2 r)$ multi-control-NOT gates and 
single-qubit NOT gates (recall that SWAP gates are just three 
interwoven $CX$ gates). Note that the operator $U_{2,21}$ 
would require a machine that could implement the gates 
$CX$, $CCX$, $CCCX$, and $CCCCX$, which is currently 
beyond the scope of existing machines. 

\begin{figure}[t!]
\begin{centering}
\includegraphics[scale=0.40]{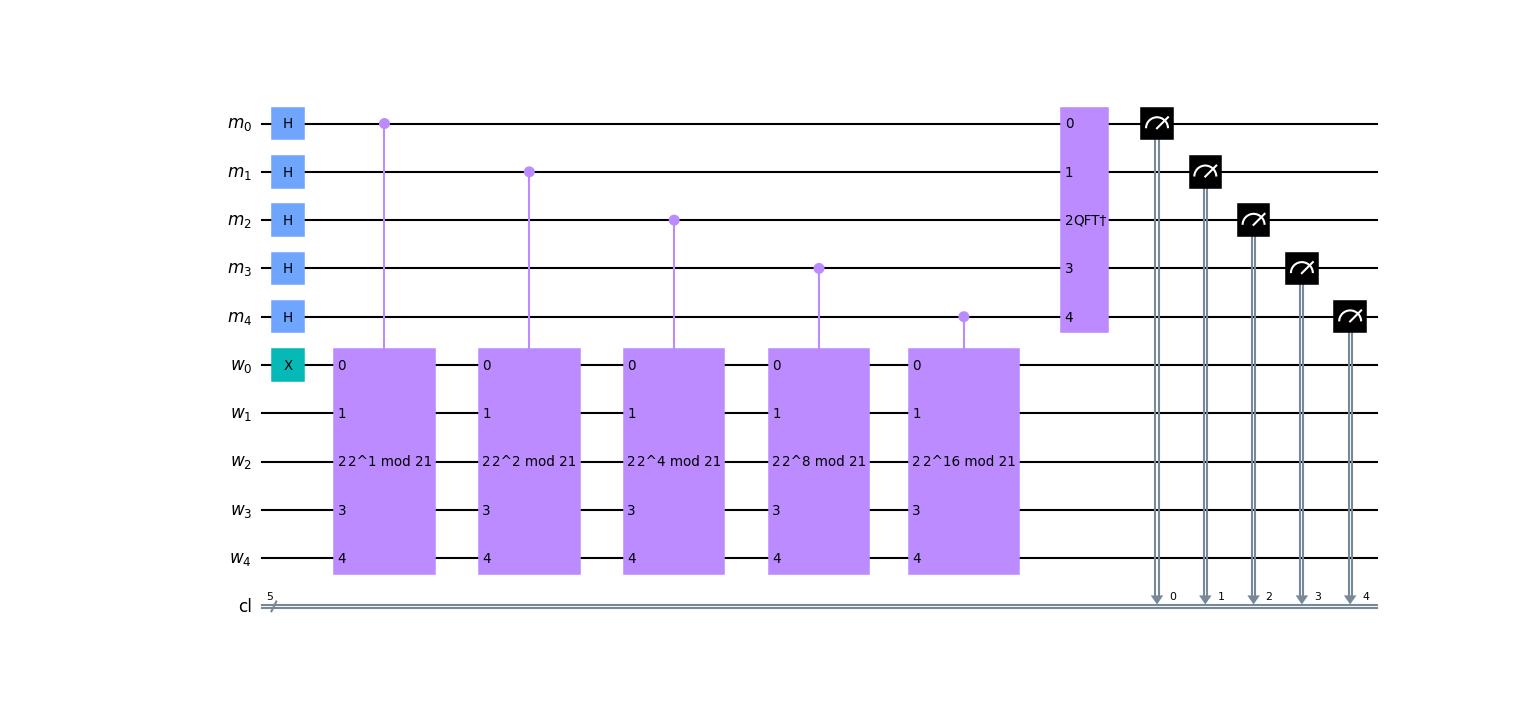} 
\par\end{centering}
\vskip-1.5cm 
\caption{\footnoteskip  
The Shor factoring circuit in Qiskit for $N=21$, $a=2$, $n = 5$, 
and $m=5$. 
}
\label{fig_N21a2_shor_circuit}
\end{figure}

\begin{figure}[t!]
\begin{centering}
\includegraphics[scale=0.45]{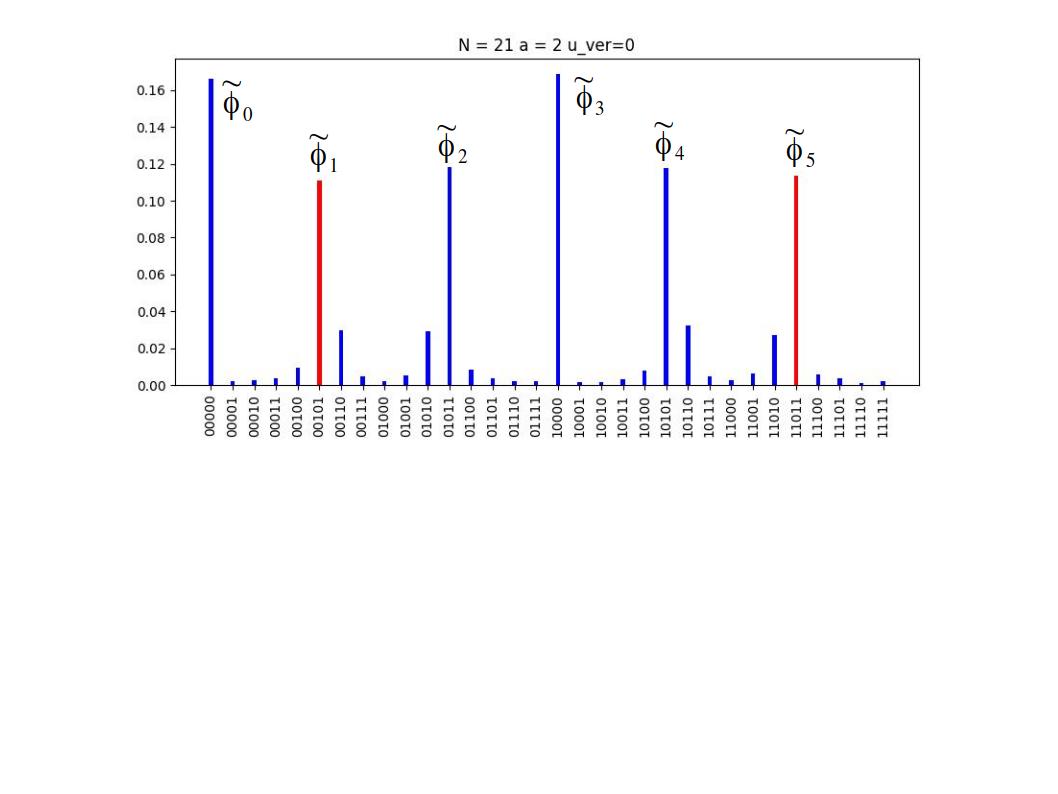} 
\par\end{centering}
\vskip-4.3cm 
\caption{\footnoteskip  
The phase histogram for $N=21$, $a=2$, $n = 5$, and $m=5$ 
from a Qiskit simulation with 4096 runs for the concatenated 
version ${\tt{u\_ver}} = 0$. The abscissa indexes the possible phases, 
and the ordinate provides their corresponding probabilities. The 
six dominant peaks of the histogram occur very close to the six 
eigenphases $\phi_s  = s/6$ of the ME operator $U_{2,21}$ for 
$s \in \{0, 1, \cdots, 5\}$. The phases shown in red corresponding 
to $s=1$ and $s = 5$ are those that produce the factors of 3
and~7, and they occur at the (binary) values $\tilde\phi_1 =
0.00101 \approx \phi_1 = 1/6$ and $\tilde\phi_5 =
0.11011 \approx \phi_5  = 5/6$. 
}
\label{fig_N21a2_shor_hist}
\end{figure}

The Qiskit circuit that factors $N = 21$ with base $a=2$  is 
shown in Fig.~\ref{fig_N21a2_shor_circuit}. The work register 
has $n = 5$ qubits, and the control register has $m = 5$ qubits. 
The NOT gate on qubit $w_0$ places the work register in the 
initial state $\vert 1 \rangle$. Since the period $r = 6$ is so 
small, we are able to employ fewer control qubits than 
the recommended number $m = 2 n + 1 = 11$ set by the 
continued fractions resolution requirement. The operators 
$U^p$ were formed via concatenation with $U = U_{2, 21}$, 
which will be  referred to as version ${\tt{u\_ver}} = 0$. 
Figure~\ref{fig_N21a2_shor_hist} illustrates the corresponding 
output phase-histogram for 4096 shots. 
The abscissa of the Figure gives the possible phase values, 
and the ordinate provides their corresponding probabilities. 
Shor's algorithm is designed so that the six most dominant 
phases of $U_{2, 21}$ (denoted $\tilde\phi_0, \cdots, \tilde
\phi_5$ in the Figure) occur near the six eigenphases $\phi_s 
= s/6$ for \hbox{$s \in \{0, 1, \cdots, 5\}$}. The phases for 
which ${\rm gcd}(s, 6) =1$, namely $s=1$ and $s=5$, lead  
to factors of $N=21$, and these peaks are plotted in red. 
The output of the analysis script is detailed in 
Table~\ref{table_N21a2_shor_table}, where only the 
phase values that produce factors are listed. The variable 
${\tt{l\_measured}}$ corresponds to the measured value
of the control register indexed by the binary integer 
$\tilde\ell = \tilde\ell_4\cdots \tilde\ell_0$, while 
${\tt{phi\_phase\_bin}}$ corresponds to the \hbox{$5$-bit} 
binary phase $\tilde\phi = \tilde\ell/2^5 = 0.\tilde\ell_4 
\cdots \tilde\ell_0$, where $\tilde\ell_k \in \{0, 1\}$ is 
the measured value of the \hbox{$k$-th} qubit of the 
 control register. The Table shows that the phases 
that yield factors are $\tilde\phi_1= 0.00101 \approx 
\phi_1 = 1/6 $ and $\tilde\phi_5 = 0.11011 \approx 
\phi_5 = 5/6$, and the continued fractions algorithm
therefore returns the period $r = 6$.  The decimal 
representation of the phase is also provided for 
convenience, as well as the fractional representation 
${\tt{phi\_phase\_frc}}$. The continued fraction of 
$\tilde \phi$ and its associated convergents are also 
listed, where fractions $s/r$ are denoted by ordered 
pairs $(s, r)$. The analysis script loops over the hand 
full of  convergents, testing the period $r$ to check 
that (i) $r$ is even, (ii) $a^{r/2} \ne -1 ~({\rm mod}~ N)$, 
and (iii) $a^r = 1 ~({\rm mod}~ N) $. If $r$ is a solution, 
then the factors are determined by ${\rm gcd}(a^{r/2} 
\pm 1, N) $.

\begin{table}[t!]
\caption{
\footnoteskip
The output of Shor's algorithm for $N=21$, $a=2$, $n=5$, 
and $m=5$ for $4096$ shots. 
Only the two  phase values that produced 
factors are listed. 
}
\baselineskip 10pt
\begin{verbatim}
l_measured   : 00101 5 frequency: 466
phi_phase_bin: 0.00101
phi_phase_dec: 0.15625
phi_phase_frc: (5, 32)
cont frc of phi  : [0, 6, 2, 2]
convergents of phi: [(0, 1), (1, 6), (2, 13), (5, 32)]
conv: (0, 1) r = 1 : no factors found
conv: (1, 6) r = 6 : factors
factor1: 7
factor2: 3
conv: (2, 13) r = 13 : no factors found
conv: (5, 32) r = 32 : no factors found

l_measured   : 11011 27 frequency: 458
phi_phase_bin: 0.11011
phi_phase_dec: 0.84375
phi_phase_frc: (27, 32)
cont frc of phi  : [0, 1, 5, 2, 2]
convergents of phi: [(0, 1), (1, 1), (5, 6), (11, 13), (27, 32)]
conv: (0, 1) r = 1 : no factors found
conv: (1, 1) r = 1 : no factors found
conv: (5, 6) r = 6 : factors
factor1: 7
factor2: 3
conv: (11, 13) r = 13 : no factors found
conv: (27, 32) r = 32 : no factors found
\end{verbatim}
\label{table_N21a2_shor_table}
\end{table}

\begin{figure}[t!]
\includegraphics[scale=0.40]{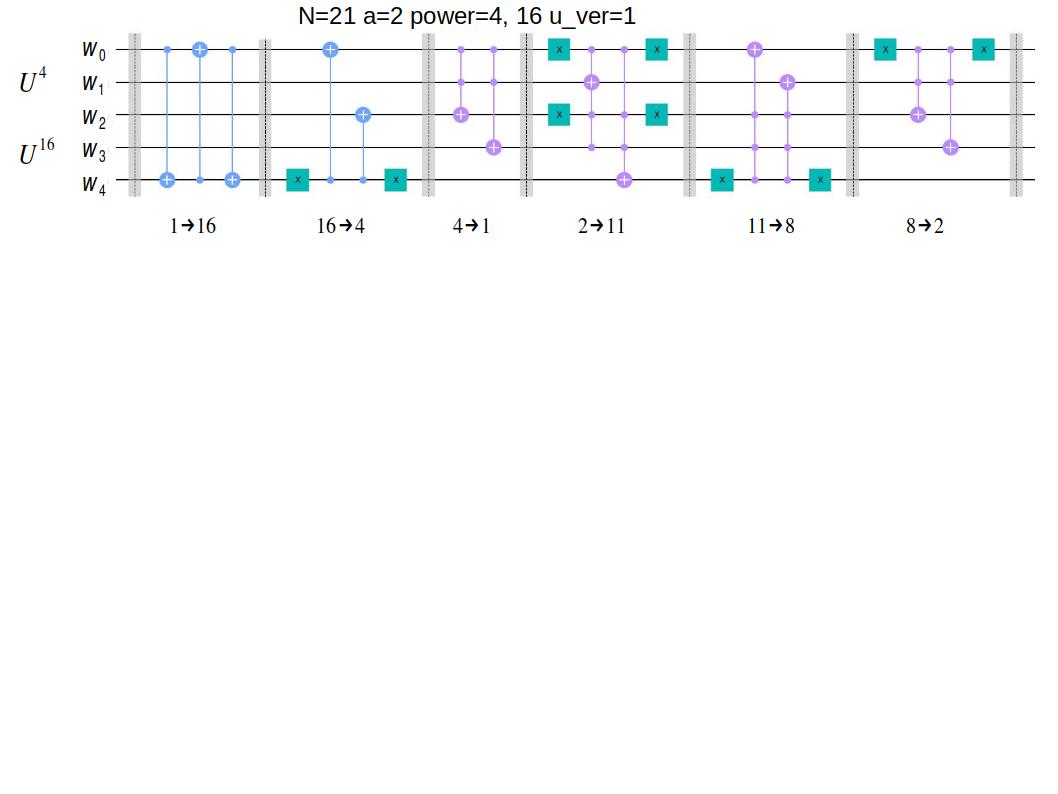}
\vskip-5.5cm
\includegraphics[scale=0.40]{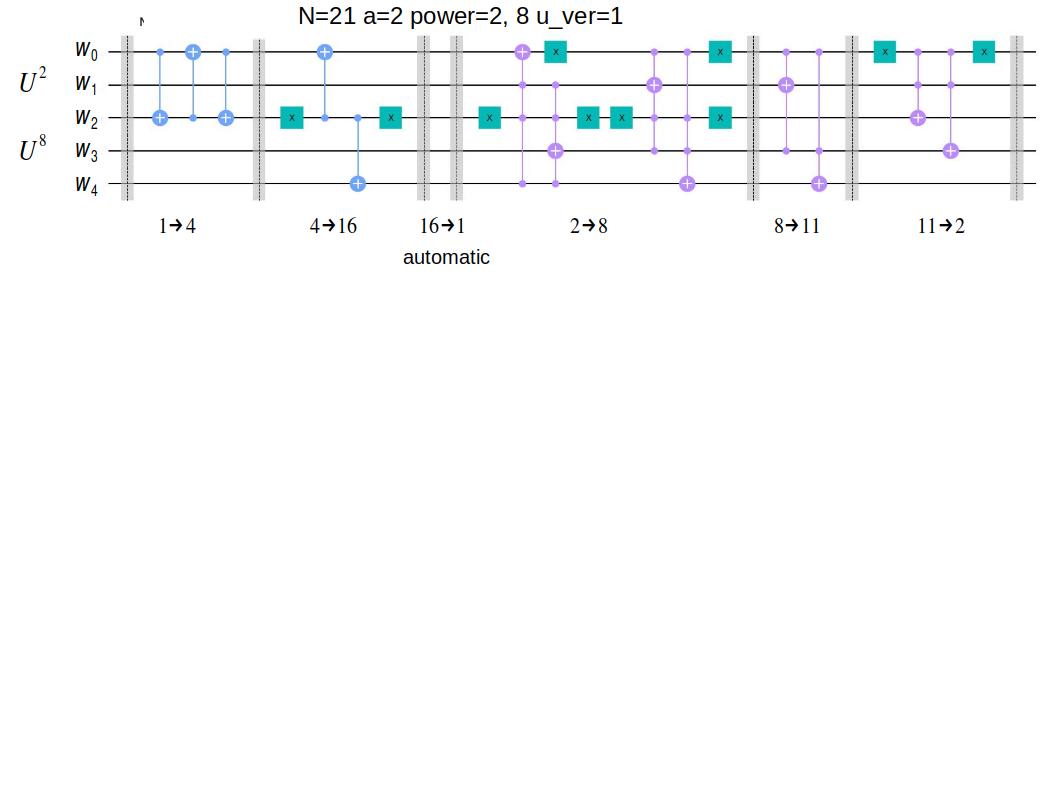}
\vskip-5.5cm
\caption{\footnoteskip
$N = 21$, $a = 2$, $r = 6$, $n =5$, $m=5$: The corresponding composite
operators are $U^2,  U^4 , U^8$, and $U^{16}$. Since $U^2 = U^8$ 
and $U^4 = U^{16}$, we only need to provide two circuits, which 
defines version number ${\tt{u\_ver} }= 1$. The blank barrier in the 
third level of $U^2$ and $U^8$ indicates that the third transition 
$\vert 16 \rangle \to\vert 1 \rangle$ is automatically performed 
by the first two gates. 
}
\label{fig_UpN21a2}
\end{figure}

\begin{figure}[t!]
\includegraphics[scale=0.50]{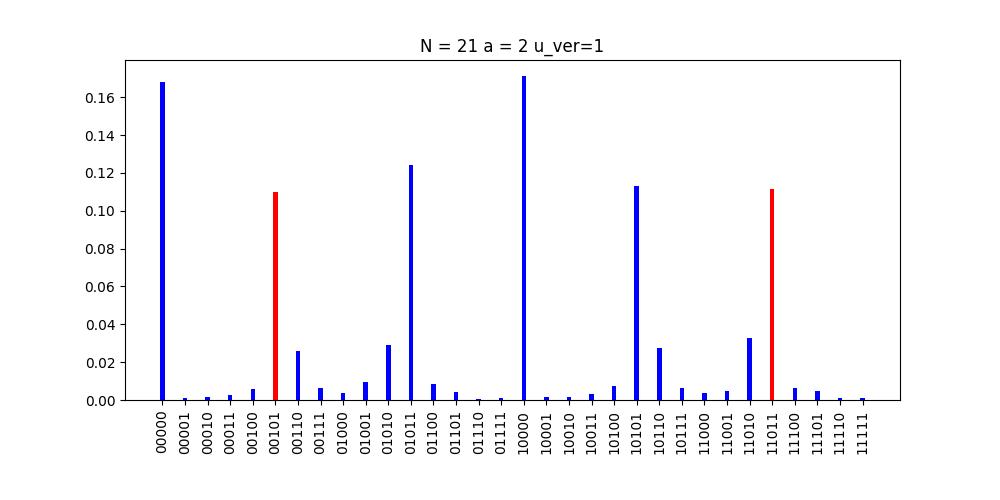}
\vskip-0.5cm
\caption{\footnoteskip
$N = 21$, $a = 2$, $r = 6$, $n=5$, $m=5$: The phase histogram 
for ${\tt u\_ver} = 1$ agrees exactly with the previous phase histogram of 
Fig.~\ref{fig_N21a2_shor_hist} for ${\tt u\_ver} = 0$, as it should.  As before, 
the phases that produce the factors of 3 and 7 are in red. 
}
\label{fig_phase_uver1}
\end{figure}

We are now ready to address the concatenation issue.  
For $m = 5$ qubits in the control register, we require 
the operators $U^2,  U^4,  U^8$ and $U^{16}$. 
We have been constructing the composite operators 
$U^p$ by simply concatenating the $U$ operator the 
appropriate number of times; however, this is an extreme 
waste of quantum gates (although it is a good test 
to ensure that the $U$ operator is in fact correct). 
We should nonetheless construct the operators $U^p$ 
in the same manner (and with the same script) that 
we originally used for $U$. As before, the operators 
$U^p$ are required to act only on the \hbox{6-dimensional} 
subspace ${\cal U}_{r=6}$ of the full Hilbert space 
${\cal W}_{n=5}$. 
Note that $U^2$ acts on every other 
element of the sequence $[1, 2, 4, 8, 16, 11, 1]$, producing 
two closed sub-sequences $[1, 4, 16, 1]$ and $[2, 8, 11, 2]$. 
Consequently, this operator loops over the states $\vert 
f(2 x) \rangle$ and $\vert f(2 x + 1) \rangle$ for $x \in
\{0, 1, 2, \cdots\}$. Similarly,  the operator $U^4$ chooses 
every 4-th element of the sequence, and so on. Therefore, 
the operators $U^2$,  $U^4$, $U^8$, $U^{16}$ act on the 
following closed sub-sequences:
\vskip-0.5cm
\begin{eqnarray}
  U_{2, 21} && ~:~~~ [1, 2, 4, 8, 16, 11, 1]
\nonumber\\[-3pt]
  U^2_{2, 21} && ~:~~~ [1, 4, 16, 1] ~~+~~ [2, 8, 11, 2]
\nonumber\\[-3pt]
  U^4_{2, 21} && ~:~~~  [1, 16, 4, 1] ~~+~~ [2, 11, 8, 2]
\label{eq_Up_N21a2_seq}
\\[-3pt]
  U^8_{2, 21} && ~:~~~ [1, 4, 16, 1]  ~~+~~ [2, 8, 11, 2]
\nonumber\\[-3pt]
  U^{16}_{2, 21} && ~:~~~ [1, 16, 4, 1] ~~+~~ [2, 11, 8, 2]
\nonumber
  \ .
\end{eqnarray}
I have restored the $N=21$ and $a=2$ subscripts on the ME 
operator $U = U_{2, 21}$ for clarity. Note that $U_{2, 21}^2 =
U_{2, 21}^8$ and $U_{2, 21}^4 = U_{2, 21}^{16}$ (the second
relation follows from squaring the first). The corresponding 
circuits that produce these sequences are given in Fig.~\ref{fig_UpN21a2}, 
which will be called version number ${\tt{u\_ver}} = 1$.  Note
that the third partition is blank for the operators $U_{2, 21}^2$ 
and $U_{2,21}^8$. This means that the first two levels also
perform the third transition $\vert 16 \rangle \to \vert 1\rangle$, 
that is to say, they execute the sequence $\vert 4 \rangle \to 
\vert 16 \rangle \to \vert 1\rangle$. This occasionally happens 
when constructing ME operators, and in the future, such blank 
partitions will signify that they are not required to perform the 
requisite sequence of transitions (we keep them only to identify 
the total number of levels in the operator). Finally, 
Fig.~\ref{fig_phase_uver1} illustrates the phase histogram 
from the Shor circuit of Fig.~\ref{fig_N21a2_shor_circuit} using 
these ME operators.  The histogram is identical to that of 
Fig.~\ref{fig_N21a2_shor_hist}, as it should be.


\begin{figure}[b!]
\includegraphics[scale=0.55]{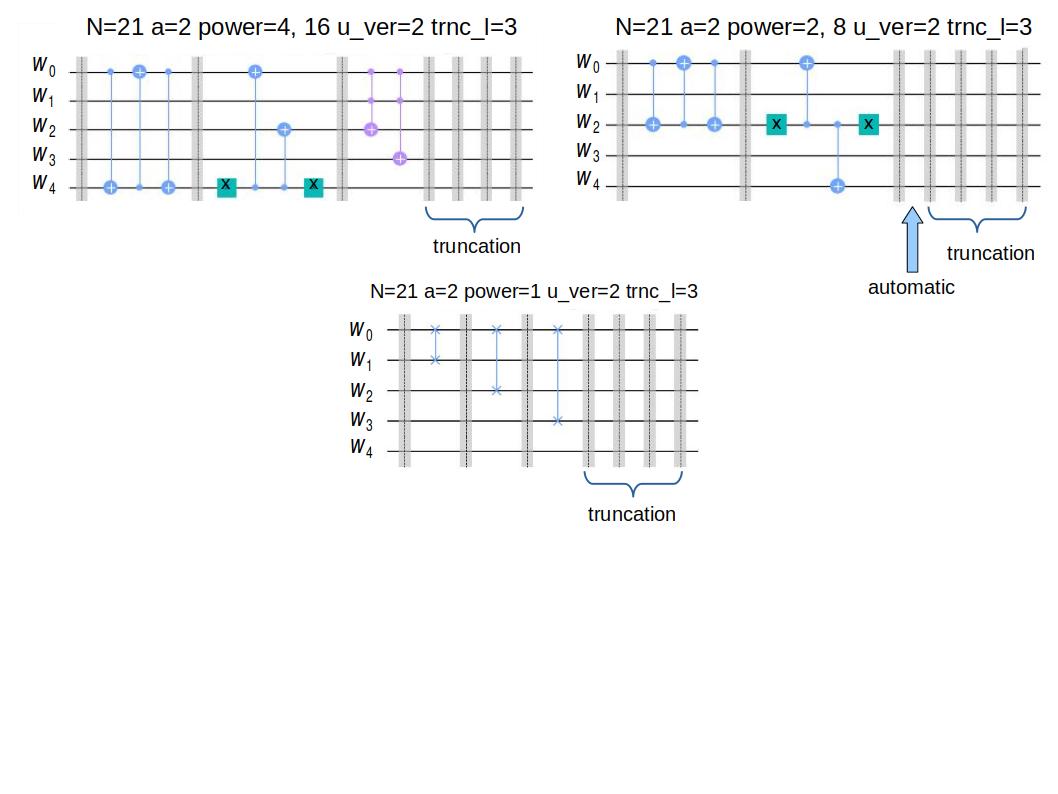}
\vskip-4.0cm
\caption{\footnoteskip
$N = 21$, $a = 2$, $r = 6$, $n = 5$, $m=5$:
Truncated ME operators $U,  U^2, U^4,  U^8$ and $U^{16}$
for version ${\tt{u\_ver}}=2$ and truncation
level ${\tt{trnc\_lv}}=3$. This means that the last 
three levels of the operators have been omitted,
which is indicated by the trivial barriers. Note that $U^2
= U^8$ and $U^4 = U^{16}$, and that the first two have 
an automatic transition at level three. 
}
\label{fig_UpN21a2_trnc}
\end{figure}

\begin{figure}[h!]
\includegraphics[scale=0.45]{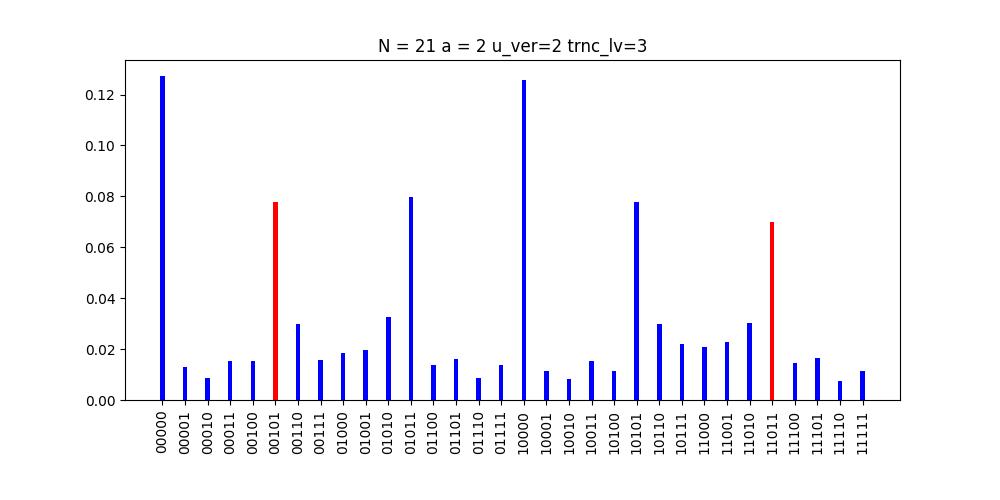}
\vskip-0.5cm
\caption{\footnoteskip
$N = 21$, $a = 2$, $r = 6$, $n = 5$, $m=5$:
Phase histogram for truncation level ${\tt{trnc\_lv}}= 3$ in which
the last three levels have been removed. 
The signal agrees with the previous two versions, with only slightly 
more noise, and the peaks in red correspond to phases that produce
the factors of 3 and 7. 
}
\label{fig_uver2_trnclv3_1}
\end{figure}

At this point, one should levy a serious charge against this 
procedure: We have used the {\em entire} cycle $[1, 2, 4, 8, 
16, 11, 1]$ for the ME operator $U_{2, 21}$,  which means 
we know \hbox{{\em a priori}} that the period of the ME function 
$f_{2, 21}(x)$ is $r=6$.  In other words, if we knew the complete 
closed sequence  for a general number $N$ and base $a$,  
then this is equivalent 
to knowing the period~$r$, and there would be no need for 
Shor's algorithm. However,  it turns out that we do {\em  not} 
require the complete sequence.  This is because the method
of continued fractions needs only an {\em approximate} phase 
in order to extract the corresponding convergents $s/r$, and 
omitting levels in $U$ still provides an adequate approximation 
for the measured phase $\tilde\phi$ (provided we do not 
omit too many levels). Figure~\ref{fig_UpN21a2_trnc} illustrates 
a {\em truncated} version of the operators $U,  U^2,  U^4, U^8$, 
$U^{16}$  in which  the last 3 levels from each operator have 
been omitted. The corresponding phase histogram is given in 
Fig.~\ref{fig_uver2_trnclv3_1}, and shows that employing 
this truncation strategy still permits us to extract the appropriate 
phases, and therefore the correct factors. Not surprisingly, the 
phase histogram has somewhat more noise, but this does not 
overwhelm the signal. 
\begin{figure}[b!]
\includegraphics[scale=0.40]{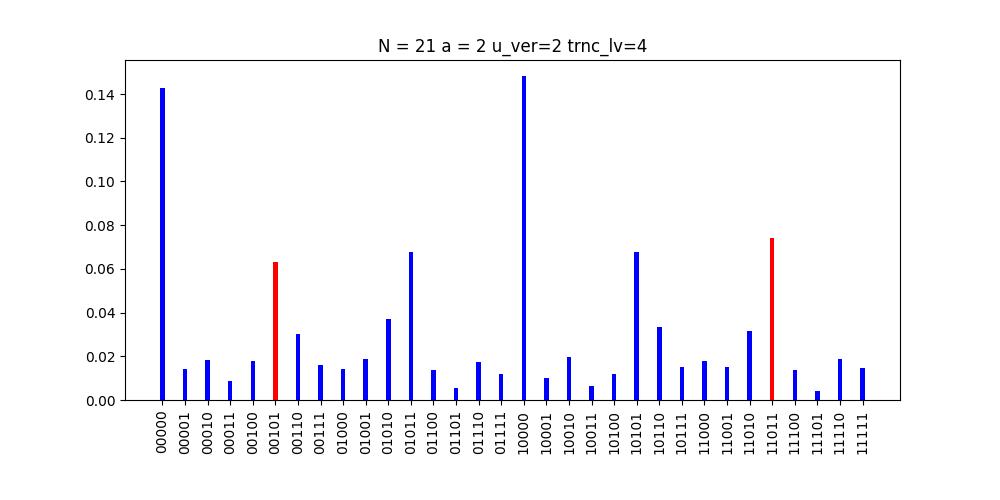}
\includegraphics[scale=0.40]{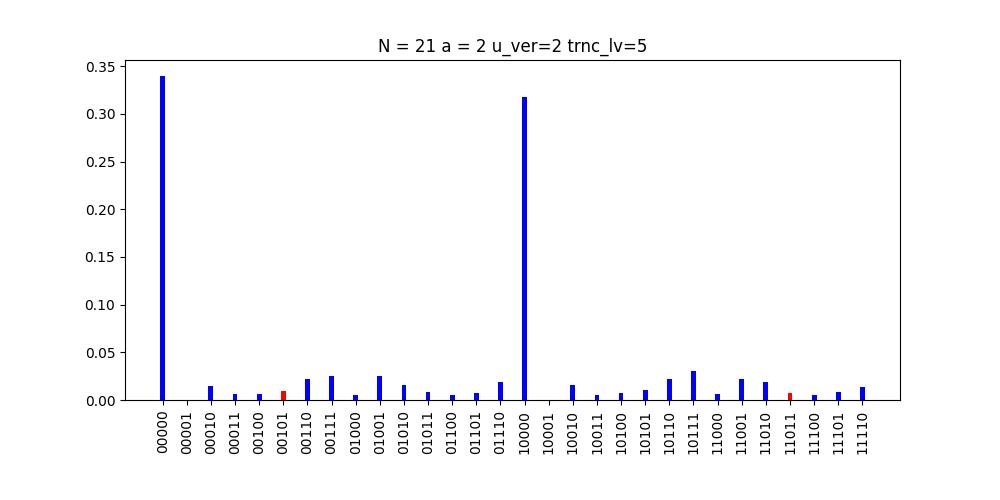}
\vskip-0.5cm
\caption{\footnoteskip
$N = 21$, $a = 2$, $r = 6$, $n = 5$, $m=5$: 
Phase histograms for truncation levels ${\tt{trnc\_lv}}= 4, 5$. 
The top panel shows that level 4 truncation, which performs 
only the first $6 - 4 = 2$ transitions, still permits factorization. 
In the bottom panel, the signal vanishes at level 5 truncation, 
where only the first transition is kept. 
}
\label{fig_uver2_trnclv3_2}
\end{figure}

\begin{figure}[t!]
\includegraphics[scale=0.55]{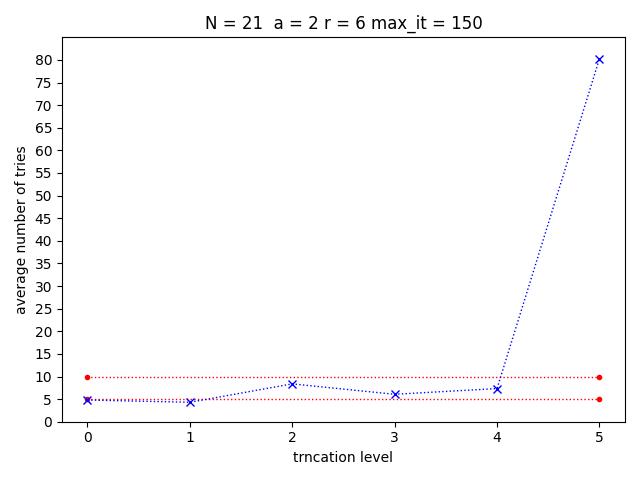} 
\vskip-0.8cm 
\caption{\footnoteskip  
$N = 21$, $a = 2$, $r = 6$, $n = 5$, $m=5$: 
  Average number of tries vs. truncation level. 
}
\label{fig_tries_N21}
\end{figure}

I shall refer to truncated operators as version ${\tt{u\_ver}}
= 2$, and the truncation level will be specified by the flag 
${\tt{trnc\_lv}} =0, 1, 2, \cdots, r-1$. The value 0 means 
that there is no truncation, and all levels are  maintained; 
a value of $k \in \{1, \cdots, r-1\}$ means that the final $k$ 
segments  of the $U^p$ operators have been omitted,
leaving only the first $r - k$ levels. For the case $N =21$ 
and $a = 2$ with period $r = 6$, there are five nonzero 
truncation levels, and we have seen that ${\tt trnc\_lv}
= 3$ still permits factoring (as do truncation levels 
${\tt trnc\_lv} =1, 2$). Figure~\ref{fig_uver2_trnclv3_2} 
illustrates the phase histogram for the remaining truncation 
levels ${\tt trnc\_lv}= 4, 5$.  Even at level 4, 
in which there are only $6 - 4 = 2$ stages of the ME 
operators $U^p$, we are still able to extract factors. 
It is only for the last truncation level, which retains 
just a single stage in the ME operators, that the signal 
recedes into the noise.  These results are summarized
in Fig.~\ref{fig_tries_N21} in a quantitative fashion, 
which plots the average number of tries that are
required to find a factor as a function of the truncation 
level. The average was performed with an ensemble 
of ${\tt num\_it} = 150$ iterations for every truncation 
level. We can factor $N = 21$ with $a = 2$ for truncation 
levels ${\tt trnc\_lv} = 0, 1, \cdots, 4$ using only 5 to 
10 tries. Note the sharp cliff after level 4, in which level 
5 requires of order 80 tries. This is actually worse than 
random, as the phase histogram in bottom panel of 
Fig.~\ref{fig_uver2_trnclv3_2} biasses the phase
toward two major peaks at $\tilde\phi=0.00000$
and $\tilde\phi=0.10000$, neither of which produce
factors. 

That we can continue to factor when well over half
the levels have been omitted  is quite remarkable,
and if this result continues to hold for larger numbers, 
it can be exploited for the factorization process. The 
strategy would be to build the operators $U^p$ one 
level at a time, checking for factors at each level. 
For example, in the case of $N = 21$ with $a = 2$, 
we would start 
with the first level, finding no factors (${\tt trnc\_lv}
= 5)$. We would then proceed to the second level 
(${\tt trnc\_lv} = 4$), after which we would find the 
requisite factors, and there would be no need to 
continue with higher levels. Apparently, there are 
enough correlations in the set of truncated operator 
$U^p$ for $p=1, 2, 4, 8, 16$ to maintain a sufficiently 
correlated control register to permit factorization. 
I will henceforth refer to the concatenated operators 
by $\tt{u\_ver} =0$, and the individually constructed 
composite operators $U^p$ will be denoted by 
$\tt{u\_ver} =1$. The truncated operators will be 
denoted  by version number  $\tt{u\_ver} =2$, with 
the truncation level specified by the flag 
${\tt trnc\_lv} = 0, 1, 2, \cdots, r-1$. 

\vfill
\pagebreak
\subsection{$\bm{N=33 = 3 \times 11}$, $\bm{a=7}$, $\bm{r=6}$}

\begin{figure}[b!]
\begin{minipage}[c]{0.3\linewidth}
\includegraphics[scale=0.50]{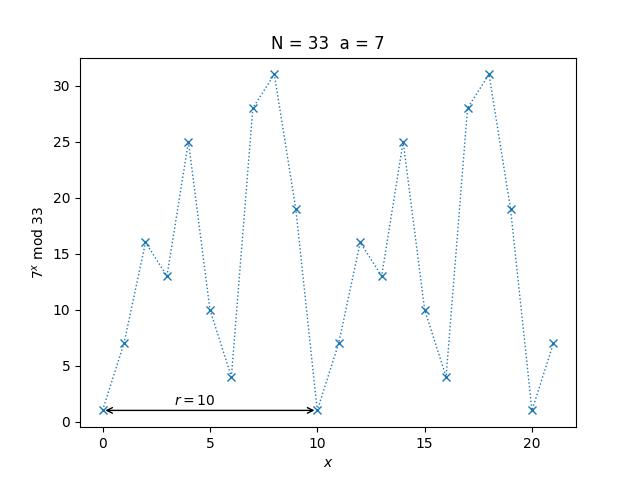}
\end{minipage}
\hskip3.0cm
\begin{minipage}[c]{0.5\linewidth}
\begin{tabular}{|c|c|} \hline
 \multicolumn{2}{|c|}{~$U\vert w \rangle = 
\big\vert 7 \times w ~({\rm mod}~33) \big\rangle$~}  
\\\hline
$~~~U\vert 1 \rangle = \vert 7 \rangle$~~~~~&
~~$U\vert 00001 \rangle = \vert 000111 \rangle$~~~~~\\[-5pt]
$~~~~U\vert 7 \rangle = \vert 16 \rangle$~~~~~&
~~$U\vert 000111 \rangle = \vert 010000 \rangle$~~~~~\\[-5pt]
$~~~U\vert 16 \rangle = \vert 13 \rangle$~~~~~&
~~$U\vert 010000 \rangle = \vert 001101 \rangle$~~~~~\\[-5pt]
$~~~~U\vert 13 \rangle = \vert 25 \rangle$~~~~~&
~~$U\vert 001101 \rangle = \vert 011001 \rangle$~~~~~\\[-5pt]
$~~~U\vert 25 \rangle = \vert 10 \rangle$~~~~~&
~~$U\vert 011001 \rangle = \vert 001010 \rangle$~~~~~\\[-5pt]
$~~~~U\vert 10 \rangle = \vert 4  \rangle$~~~~~~&
~~$U\vert 001010 \rangle = \vert 000100 \rangle$~~~~~\\[-5pt]
$~~~~~U\vert 4 \rangle ~= \vert 28  \rangle$~~~~~~&
~~$U\vert 000100 \rangle = \vert 011100 \rangle$~~~~~\\[-5pt]
$~~~~U\vert 28 \rangle ~= \vert 31  \rangle$~~~~~~&
~~$U\vert 011100 \rangle = \vert 011111 \rangle$~~~~~\\[-5pt]
$~~~~U\vert 31 \rangle ~= \vert 19  \rangle$~~~~~~&
~~$U\vert 011111 \rangle = \vert 010011 \rangle$~~~~~\\[-5pt]
$~~~~U\vert 19 \rangle ~= \vert 1  \rangle$~~~~~~&
~~$U\vert 010011 \rangle = \vert 000001 \rangle$~~~~~\\\hline
\end{tabular} 
\end{minipage}
\caption{\footnoteskip
$N=33$, $a=7$, $r=10$, $n=6$: 
The left panel gives the modular exponential function $f_{7,  33}(x) 
= 7^x ~ ({\rm mod}~33)$,  and the right panel shows the action of 
the ME operator $U_{7, 33}$ on the closed sequence $[1,  7, 16, 13, 
25, 10, 4, 28, 31, 19, 1]$.  The circuit requires $n = \lceil \log_2 33 
\rceil = 6$ qubits in the work register. 
}
\label{fig_fxN33a7_b}
\end{figure}

\begin{figure}[h!]
\includegraphics[scale=0.38]{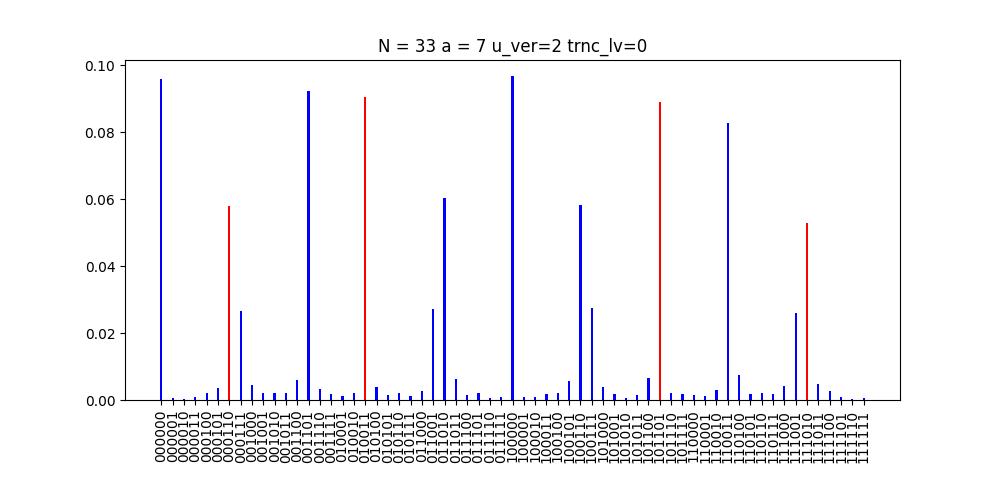}
\vskip-0.9cm
\includegraphics[scale=0.38]{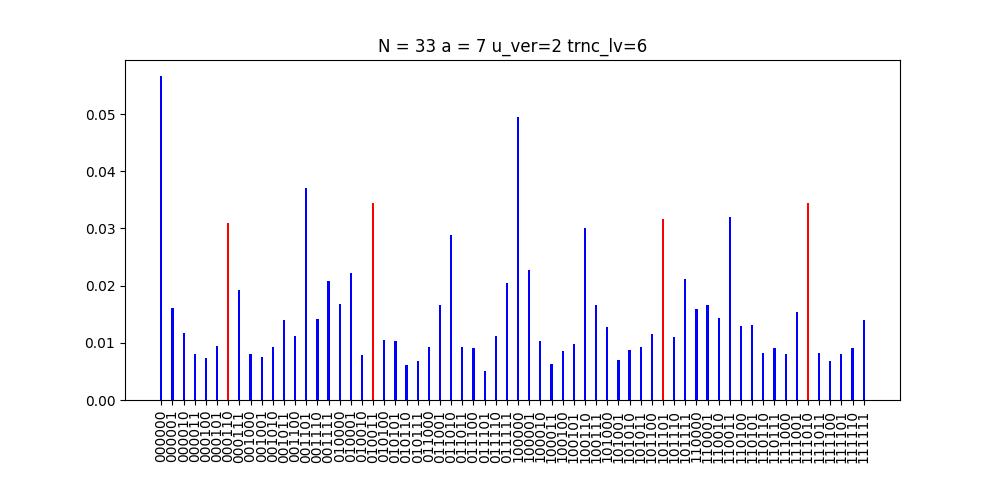}
\vskip-0.9cm
\includegraphics[scale=0.38]{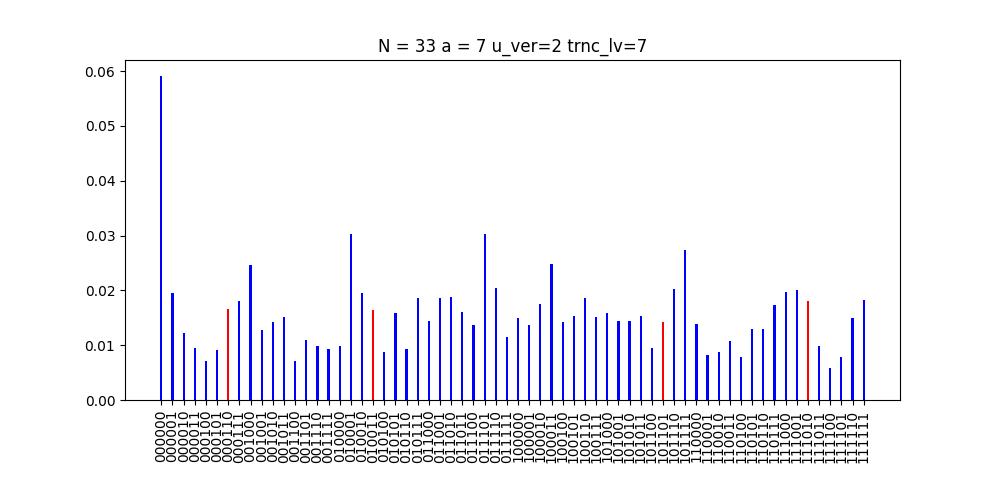}
\vskip-0.5cm
\caption{\footnoteskip
$N = 33$, $a = 7$, $r = 10$, $n = 6$, $m=6$: Truncation level 
study for ${\tt trnc\_lv} = 0, 6, 7$. 
}
\label{fig_trnc_N33_study}
\end{figure}

\begin{figure}[h!]
\includegraphics[scale=0.55]{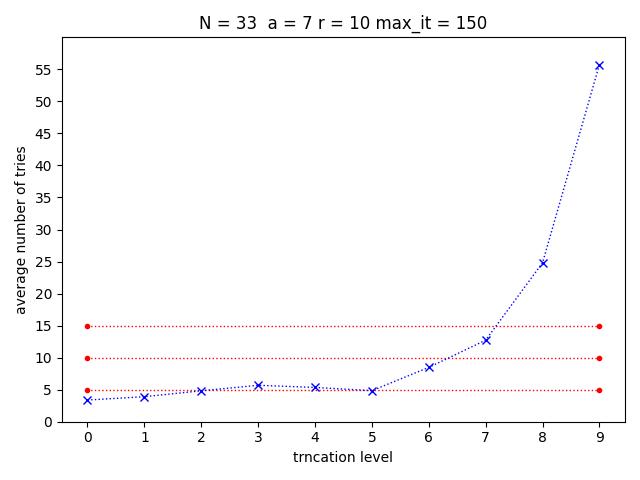} 
\vskip-0.8cm 
\caption{\footnoteskip  
$N = 33$, $a = 7$, $r = 10$, $n = 6$, $m=6$: 
  Average number of tries vs. truncation level. 
}
\label{fig_tries_N33}
\end{figure}

\begin{figure}[h!] 
\includegraphics[scale=0.20, center]{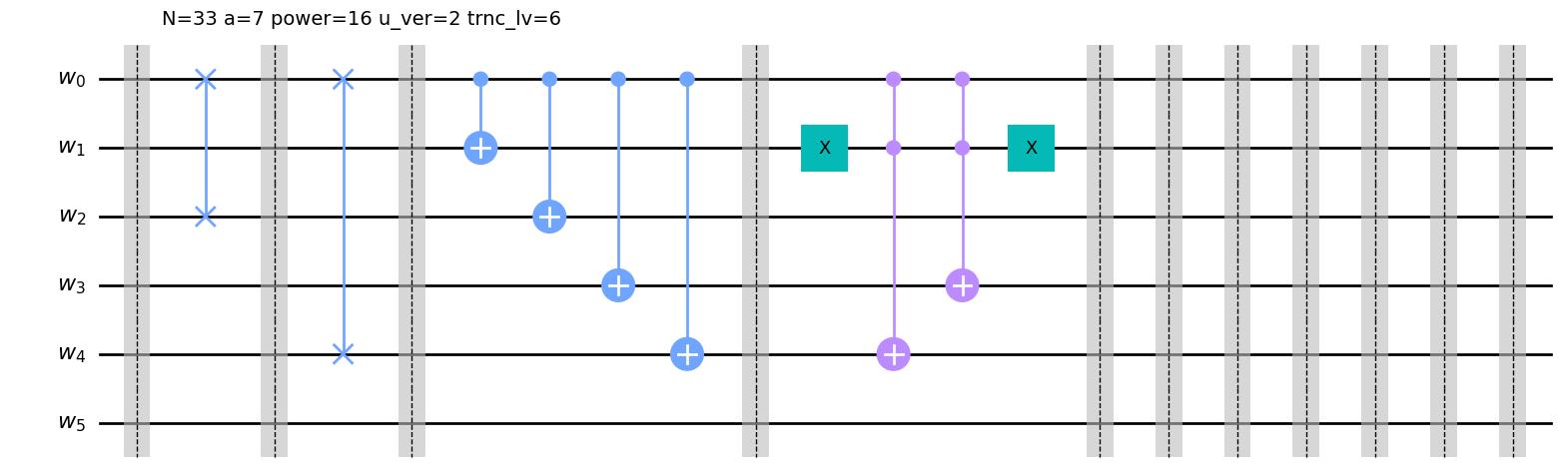}
\includegraphics[scale=0.27, center]{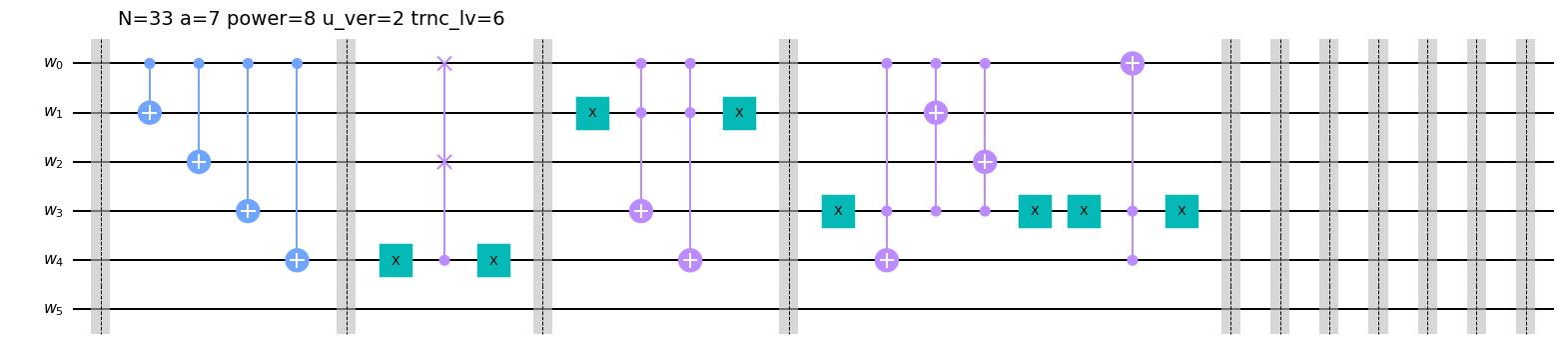}
\includegraphics[scale=0.23, center]{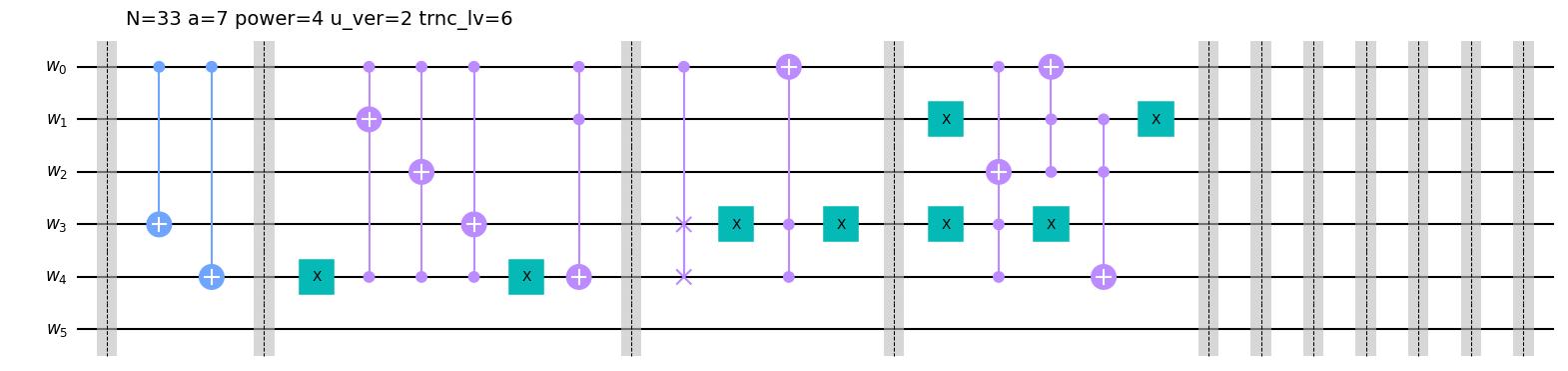}
\includegraphics[scale=0.23, center]{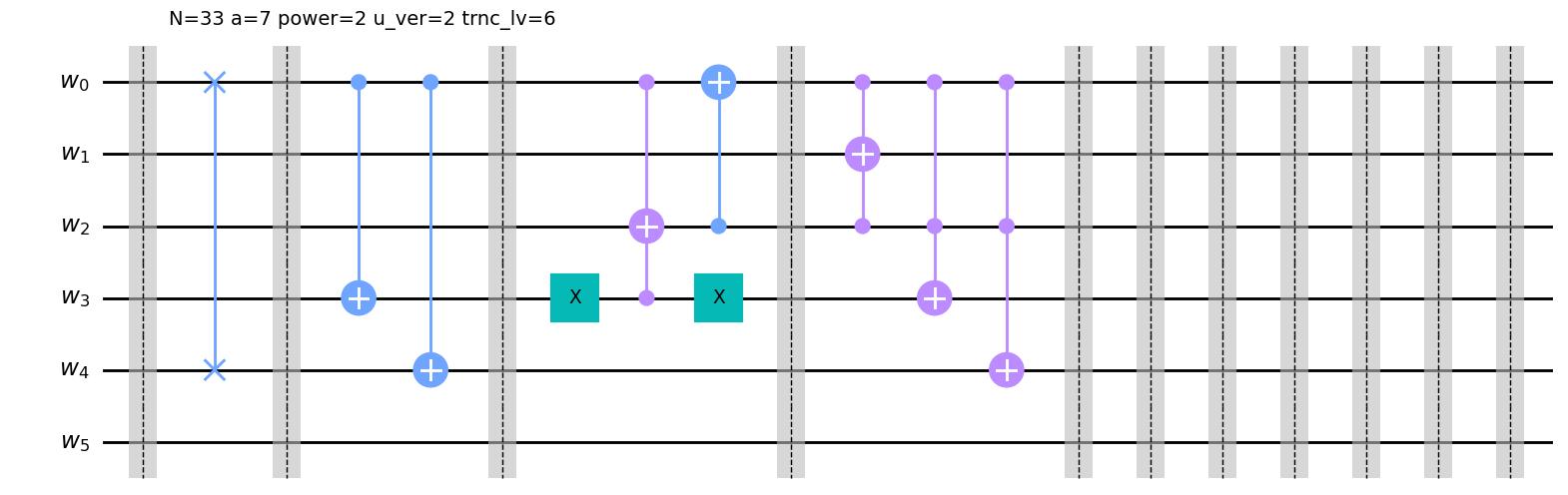}
\includegraphics[scale=0.23, center]{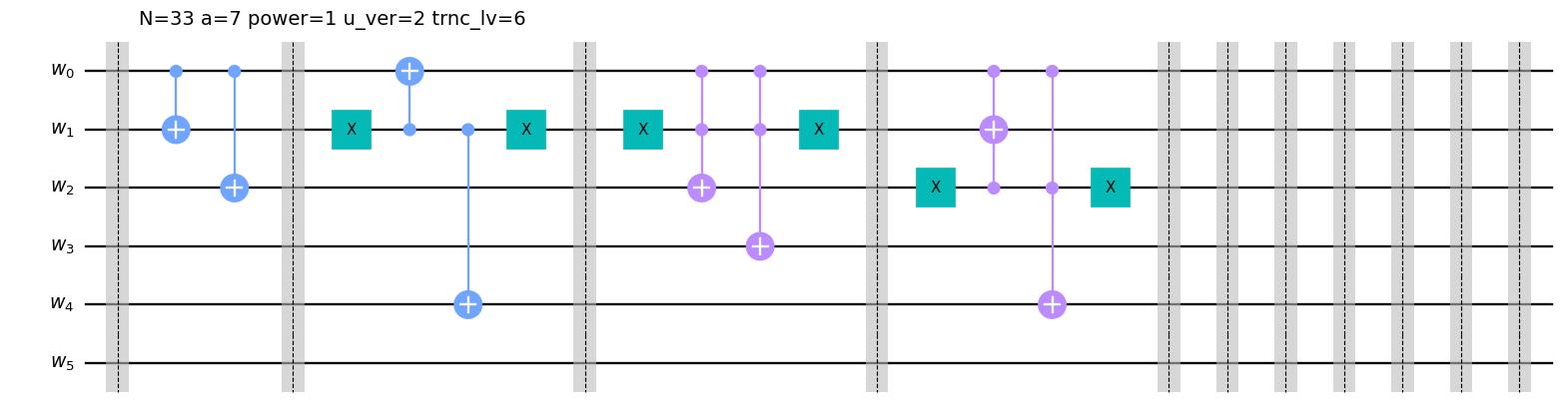}
\caption{\footnoteskip
$N=33$, $a=7$, $r=10$: 
The truncated ME operators $U^p$ for $p = 1, 2, 4, 8, 16$ 
for version $\tt{u\_ver}=2$ and ${\tt trnc\_lv} = 6$. Since 
$U^{32} = U^2$, we do not bother to list the $p=32$ 
operator.
}
\label{fig_UpN33a7_truncate1_a}
\end{figure}

Recall that the difficulty in factoring a specific number 
$N$ lies not in the size of the number itself, but in the 
length of the period $r$ of the modular exponentiation
function\,\cite{pretend}. The ME function for $N = 35$ 
with base $a = 4$ has a period of $r = 6$, and a systematic 
truncation-level study reveals that $N = 35$ can 
be factored with only two levels, just like the previous 
case for $N= 21$ and $a = 2$. It is more interesting 
therefore to examine numbers with larger periods. 
Figure~\ref{fig_fxN33a7_b} shows that $N =33$ 
with base $a = 7$ has a period of $r = 10$. Since the 
period is larger than the previous case, we require more 
resolution in the control register, and so we take $m = 6$ 
qubits. We must therefore construct the operators 
$U, U^2, U^4, \cdots, U^{32}$. As before, the composite 
operators $U^p$ loop over closed sub-cycles, which 
in this case become:
\begin{eqnarray}
  U_{7, 33} && ~:~ [1, 7, 16, 13, 25, 10, 4, 28, 31, 19, 1]
\nonumber\\[-3pt]
  U^2_{7, 33}  && ~:~ [1, 16, 25, 4, 31, 1]
  ~~+~~  [7, 13, 10, 28, 19, 7]
\nonumber\\[-3pt]
  U^4_{7, 33} && ~:~  [1, 25, 31, 16, 4, 1]
  ~~+~~ [7, 10, 19, 13, 28, 7]
\label{eq_Up_N33a7_seq}
\\[-3pt]
  U^8_{7, 33} && ~:~ [1, 31, 4, 25, 16, 1]
  ~~+~~  [7, 19, 28, 10, 13, 7]
\nonumber\\[-3pt]
  U^{16}_{7, 33}  && ~:~ [1, 4, 16, 31, 25, 1]
  ~~+~~ [7, 28, 13, 19, 10, 7]
\nonumber\\[-3pt]
  U^{32}_{7, 33}  && ~:~ [1, 16, 25, 4, 31, 1]
  ~~+~~  [7, 13, 10, 28, 19, 7]
    \ .
\nonumber
\end{eqnarray}
Note that $U^2 = U^{32}$, and therefore $U, U^2, \cdots,
U^{16}$ form a complete set of operators (as any other
power of $p$ can be obtained by squaring). 
\begin{figure}[h!] 
\includegraphics[scale=0.34, center]{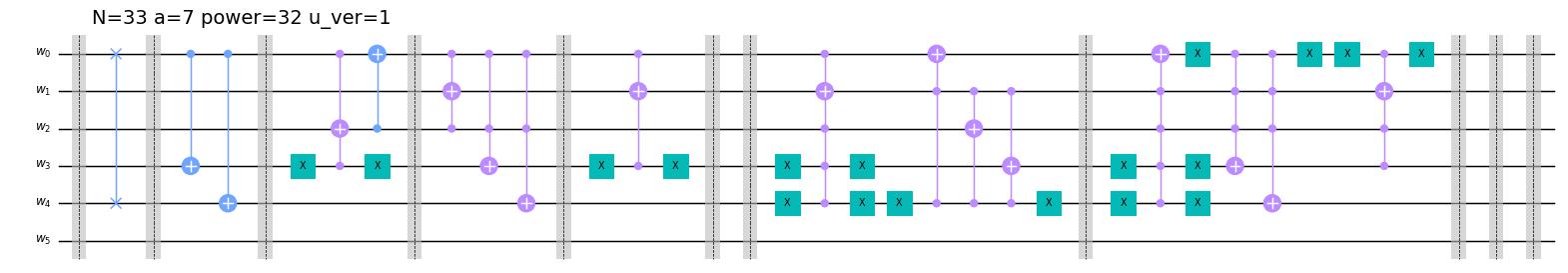}
\includegraphics[scale=0.39, center]{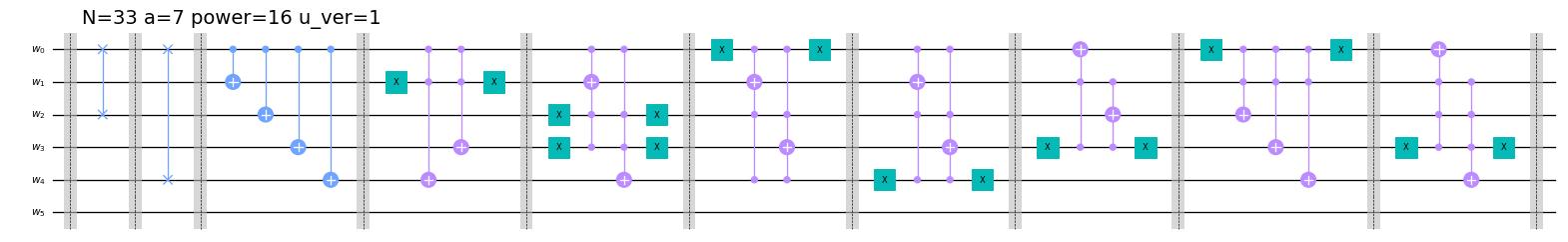}
\includegraphics[scale=0.40, center]{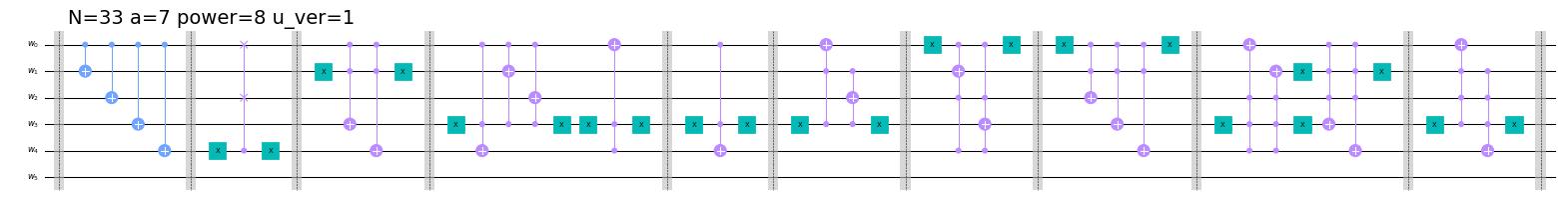}
\includegraphics[scale=0.45, center]{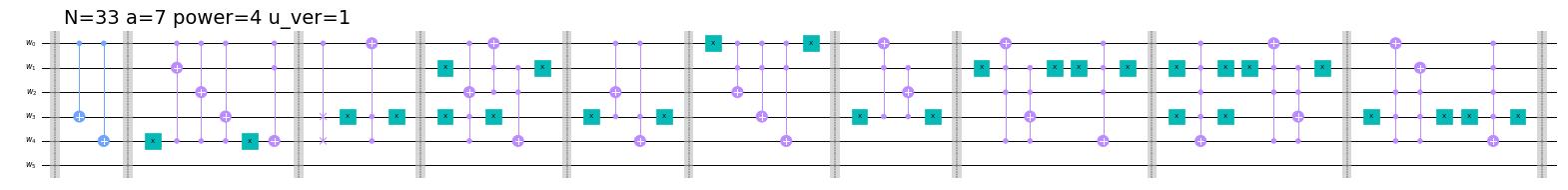}
\includegraphics[scale=0.34, center]{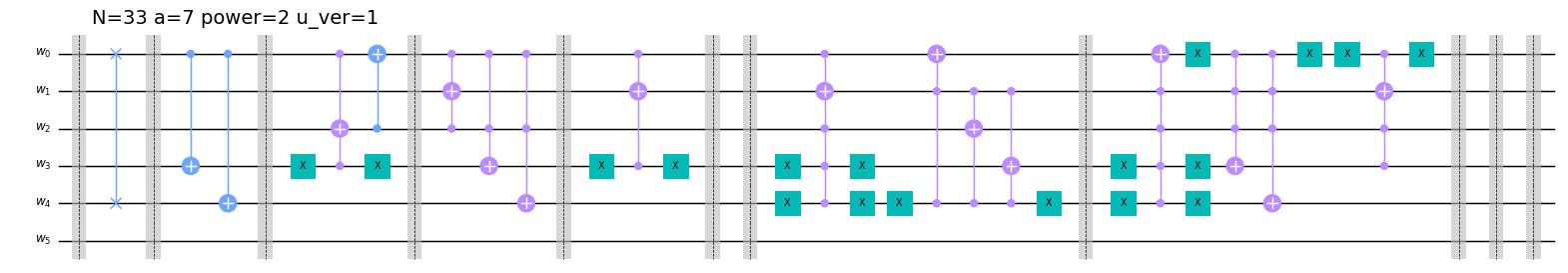}
\includegraphics[scale=0.44, center]{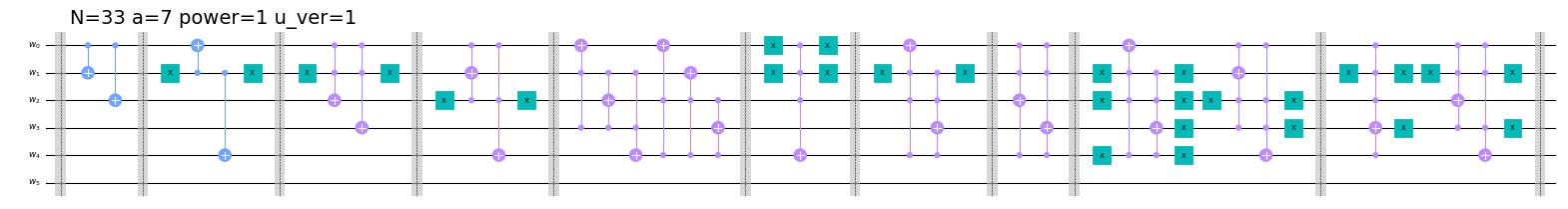}
\caption{\footnoteskip
$N=33$, $a=7$, $r=10$, $n=6$, $m=6$: The ME operators 
$U^p$ for $p=1, 2, 4, 8, 16, 32$ for version ${\tt u\_ver} = 1$ from
bottom to top.  Note that $U^2 = U^{32}$. 
}
\label{fig_UpN33a7_a}
\end{figure}
The composite operators $U^p$ that produce these 
sequences are illustrated in Fig.~\ref{fig_UpN33a7_a}.
Figure~\ref{fig_trnc_N33_study} 
shows the results from a systematic truncation level 
study. The top panel  gives the phase histogram for 
${\tt trnc\_lv} = 0$ (no truncation), and the phase peaks 
are quite distinct, with little or no noise. The middle panel 
show the phase histogram for truncation level ${\tt trnc\_lv} 
= 6$, in which we have kept the first $10 - 6 = 4$ levels. 
While there is more noise, the signal is still large enough 
to extract factors with high probability. The lower panel
illustrates ${\tt trnc\_lv} = 7$, and we see that the signal 
has finally receded into the noise. Figure~\ref{fig_tries_N33}
illustrates the average number of tries required to find a
factor as a function of the truncation level. 
For comparison, 
Fig.~\ref{fig_UpN33a7_truncate1_a} illustrates the 
operators $U^p$ for ${\tt trnc\_lv} = 6$. We see that 
more than half of the levels can be dropped with enough 
signal to extract the factors of $N = 33$ with base $a=7$ 
and period $r = 10$.  
%


\vfill
\clearpage
\subsection{$\bm{N=143 = 11 \times 13}$, $\bm{a=5}$, $\bm{r=20}$}

\begin{figure}[b!]
\begin{minipage}[c]{0.4\linewidth}
\includegraphics[scale=0.50]{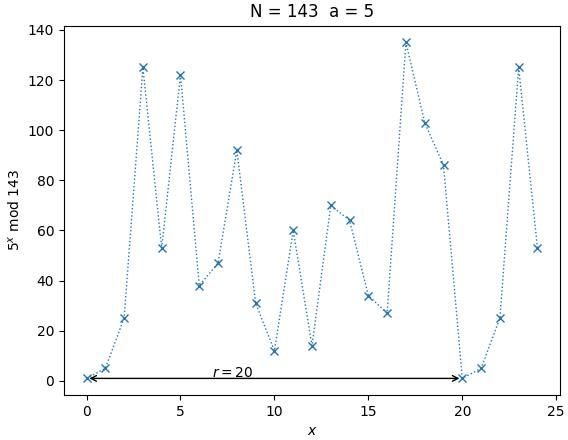}
\end{minipage}
\hskip1.5cm
\begin{minipage}[c]{0.5\linewidth}
\begin{tabular}{|c|c|} \hline
 \multicolumn{2}{|c|}{~$U\vert w \rangle = 
\big\vert 5 \times w ~({\rm mod}~143) \big\rangle$~}  \\\hline
$~~~U\vert 1 \rangle = \vert 5 \rangle$~~~~~&
~~$U\vert 00000001 \rangle = \vert 00000101 \rangle$~~~~~\\[-5pt]
$~~~~U\vert 5 \rangle = \vert 25 \rangle$~~~~~&
~~$U\vert 00000101 \rangle = \vert 00011001 \rangle$~~~~~\\[-5pt]
$~~~U\vert 25 \rangle = \vert 125 \rangle$~~~~~&
~~$U\vert 00011001 \rangle = \vert 01111101 \rangle$~~~~~\\[-5pt]
$~~~U\vert 125 \rangle = \vert 53 \rangle$~~~~~&
~~$U\vert 01111101 \rangle = \vert 00110101 \rangle$~~~~~\\[-5pt]
$~~~~U\vert 53 \rangle = \vert 122 \rangle$~~~~~&
~~$U\vert 00110101 \rangle = \vert 01111010 \rangle$~~~~~\\[-5pt]
$~~~U\vert 122 \rangle = \vert 38 \rangle$~~~~~&
~~$U\vert 01111010 \rangle = \vert 00100110 \rangle$~~~~~\\[-5pt]
$~~~~U\vert 38 \rangle = \vert 47  \rangle$~~~~~~&
~~$U\vert 00100110 \rangle = \vert 00101111 \rangle$~~~~~\\[-5pt]
$~~~~U\vert 47 \rangle ~= \vert 92  \rangle$~~~~~~&
~~$U\vert 00101111 \rangle = \vert 01011100 \rangle$~~~~~\\[-5pt]
$~~~~U\vert 92 \rangle ~= \vert 31  \rangle$~~~~~~&
~~$U\vert 01011100 \rangle = \vert 00011111 \rangle$~~~~~\\[-5pt]
$~~~~U\vert 31 \rangle ~= \vert 12  \rangle$~~~~~~&
~~$U\vert 00011111 \rangle = \vert 00001100 \rangle$~~~~~\\[-5pt]
$~~~~U\vert 12 \rangle ~= \vert 60  \rangle$~~~~~~&
~~$U\vert 00001100 \rangle = \vert 00111100 \rangle$~~~~~\\[-5pt]
$~~~U\vert 60 \rangle = \vert 14 \rangle$~~~~~&
~~$U\vert 00111100 \rangle = \vert 00001110 \rangle$~~~~~\\[-5pt]
$~~~~U\vert 14 \rangle = \vert 70 \rangle$~~~~~&
~~$U\vert 00001110 \rangle = \vert 01000110 \rangle$~~~~~\\[-5pt]
$~~~U\vert 70 \rangle = \vert 64 \rangle$~~~~~&
~~$U\vert 01000110 \rangle = \vert 01000000 \rangle$~~~~~\\[-5pt]
$~~~~U\vert 64 \rangle = \vert 34 \rangle$~~~~~&
~~$U\vert 01000000 \rangle = \vert 00100010 \rangle$~~~~~\\[-5pt]
$~~~U\vert 34 \rangle = \vert 27 \rangle$~~~~~&
~~$U\vert 00100010 \rangle = \vert 00011011 \rangle$~~~~~\\[-5pt]
$~~~~U\vert 27 \rangle = \vert 135  \rangle$~~~~~~&
~~$U\vert 00011011 \rangle = \vert 10000111 \rangle$~~~~~\\[-5pt]
$~~~~U\vert 135 \rangle ~= \vert 103  \rangle$~~~~~~&
~~$U\vert 10000111 \rangle = \vert 01100111 \rangle$~~~~~\\[-5pt]
$~~~~U\vert 103 \rangle ~= \vert 86  \rangle$~~~~~~&
~~$U\vert 01100111 \rangle = \vert 01010110 \rangle$~~~~~\\[-5pt]
$~~~~U\vert 86 \rangle ~= \vert 1  \rangle$~~~~~~&
~~$U\vert 01010110 \rangle = \vert 00000001 \rangle$~~~~~\\\hline
\end{tabular} 
\end{minipage}
\caption{\footnoteskip
$N=143$, $a=5$, $r=20$, $n = 8$: 
The left panel gives the modular exponential function $f_{5,  143}(x) = 
5^x ~ ({\rm mod}~143)$, and the right gives the action of the ME operator
$U_{5, 143}$ on the closed sequence $[1,  5, 25, 125, 53, 122,  38, 47, 92, 31, 
12, 60, 14, 70, 64, 34, 27, 135, 103, 86, 1]$.   The Shor circuit requires $n = \lceil 
\log_2 143 \rceil = 8$ qubits in the work register. 
~~\\
~~\\
~~\\
}
\label{fig_fxN143a5}
\end{figure}

\begin{figure}[h!] 
\includegraphics[scale=0.40, center]{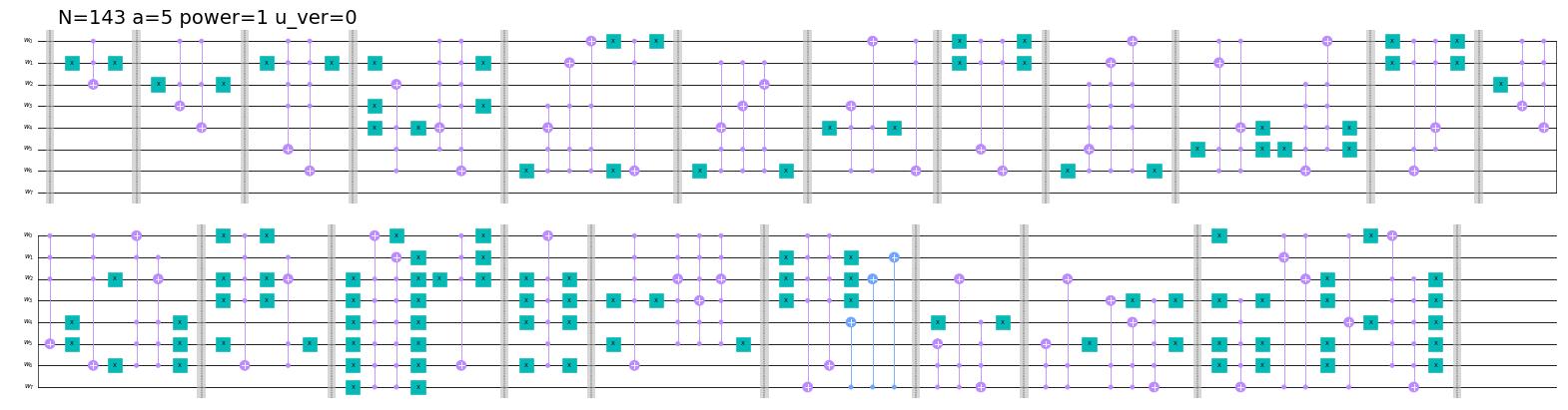}
\caption{\footnoteskip
$N=143$, $a=5$, $r=20$, $n = 8$: 
The modular exponentiation operator $U_{5, 143}$.  
}
\label{fig_fxN143a5_U}
\end{figure}

I will now show that this method continues to work for 
even larger values of~$N$. Let us consider $N = 143$ with 
$a = 5$, which gives  $n = \lceil \log_2 143 \rceil = 8$ work 
qubits. As illustrated in Fig.~\ref{fig_fxN143a5}, 
the modular exponential function $f_{5, 143}(x)$ has 
a  period of $r = 20$, and the corresponding ME operator
$U_{5, 143}$ is given in Fig.~\ref{fig_fxN143a5_U}.  Ideally, 
we need $m = 2 n + 1 = 17$ control qubits; however, since 
the period is not too large, we can get by with only $m = 8$. 
We must therefore implement the operators $U_{5, 143}, 
U_{5, 143}^2, U_{5, 143}^4, U_{5, 143}^8, U_{5, 143}^{16}, 
U_{5, 143}^{32}, U_{5, 143}^{64}$ and $U_{5, 143}^{128}$.  
We will also perform a resolution study on the control
register, taking $m =10$, which requires $U_{5, 143}^{256}$ 
and  $U_{5, 143}^{512}$.


\begin{figure}[t!]
\includegraphics[width=5.9in]{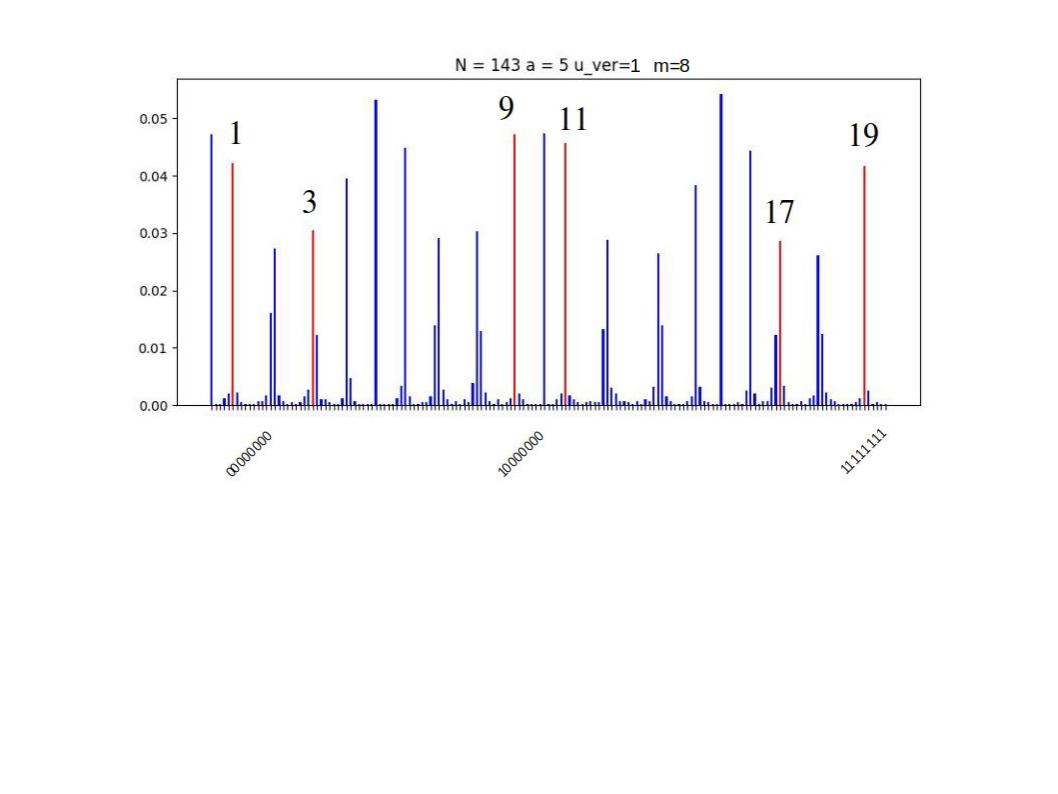} 
\vskip-5.4cm
\includegraphics[width=5.9in]{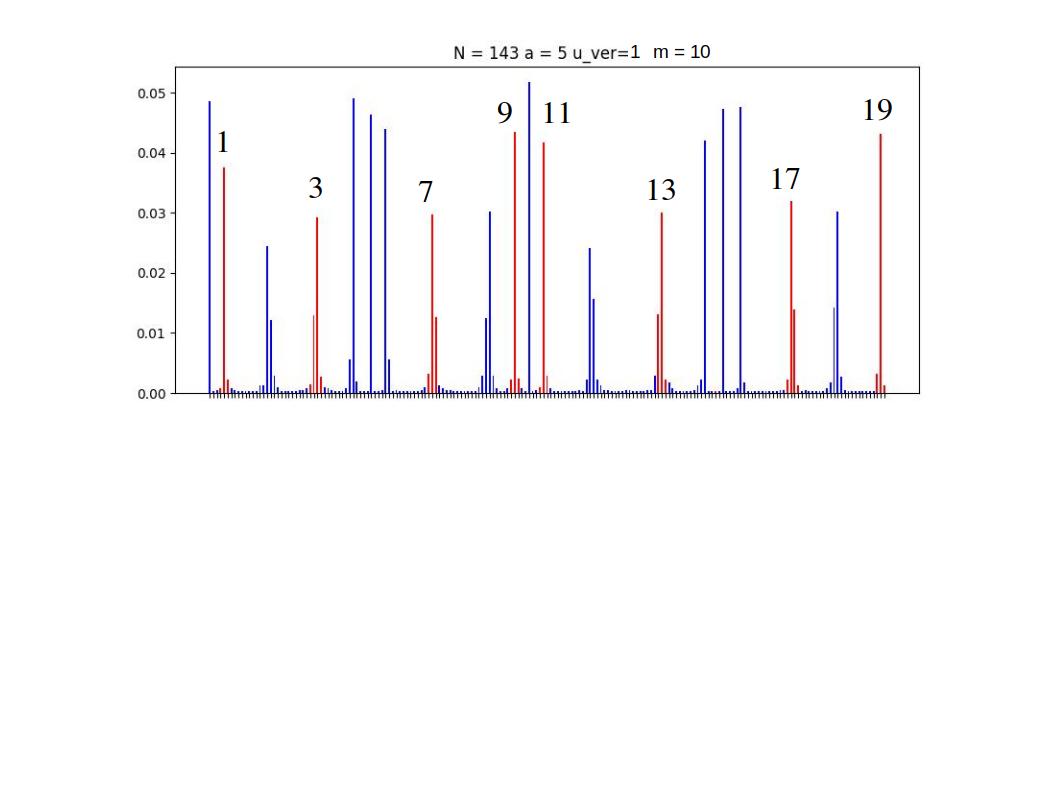}  
\vskip-5.3cm 
\caption{\footnoteskip  
 $N=143$, $a=5$, $r=20$, $n = 8$, $m=8, 10$: 
  Phase histograms for ME operator version $\tt{u\_ver} = 1$.
  The top panel shows $m = 8$, and the bottom panel illustrates
  the higher resolution $m = 10$. 
  The dominant peaks in red correspond to the ME phases 
  $\phi_s = s/20$ with $s \in \{0, 1, \cdots, 19\}$ and ${\rm gcd}
  (s, 20) = 1$. Thus the eight phases $s = 1, 3, 7, 9, 11, 13, 17, 19$ 
  provide the factors of 11 and 13. Note, however, that the peaks 
  for $s=7, 13$ are missing from the top panel.  This is because 
  $m=8$ does not provide sufficient resolution; however, all
  peaks are present for $m = 10$. 
}
\label{fig_fxN143a5m8_9_10_hist}
\end{figure}
\begin{figure}[t!]
\hskip-3.0cm
\begin{minipage}[c]{0.5\linewidth}
\vskip0.3cm
\includegraphics[scale=0.33]{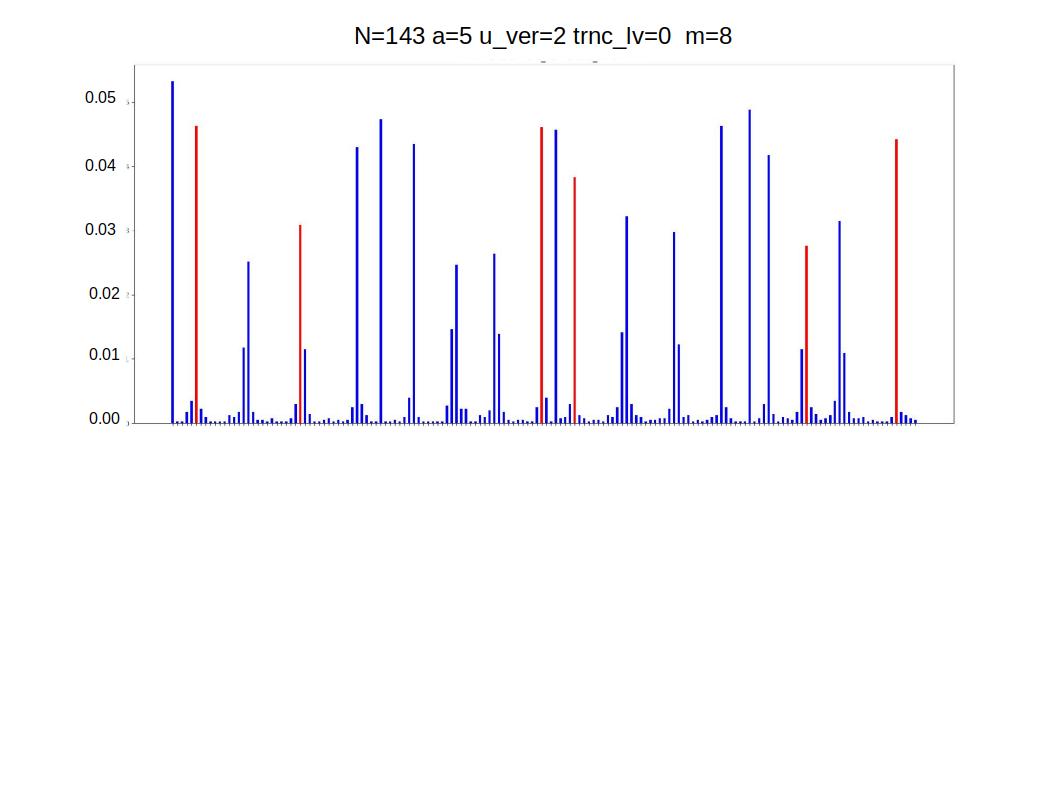}
\vskip-2.6cm
\includegraphics[scale=0.33]{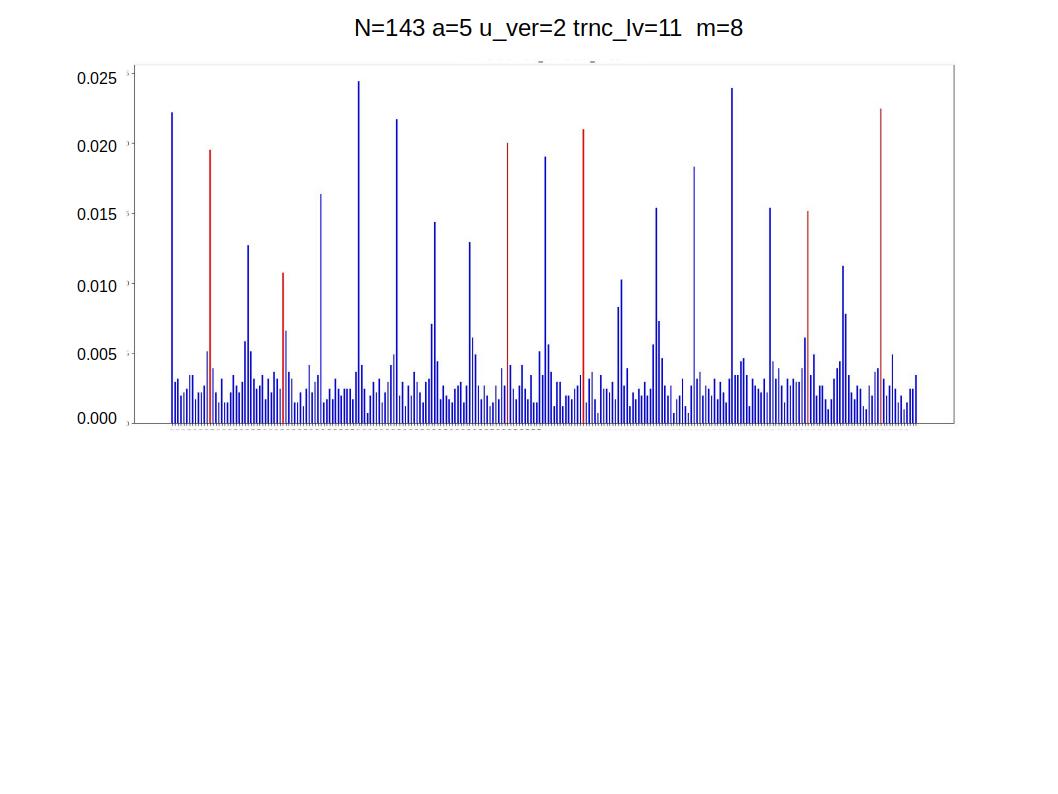}
\vskip-2.7cm
\includegraphics[scale=0.33]{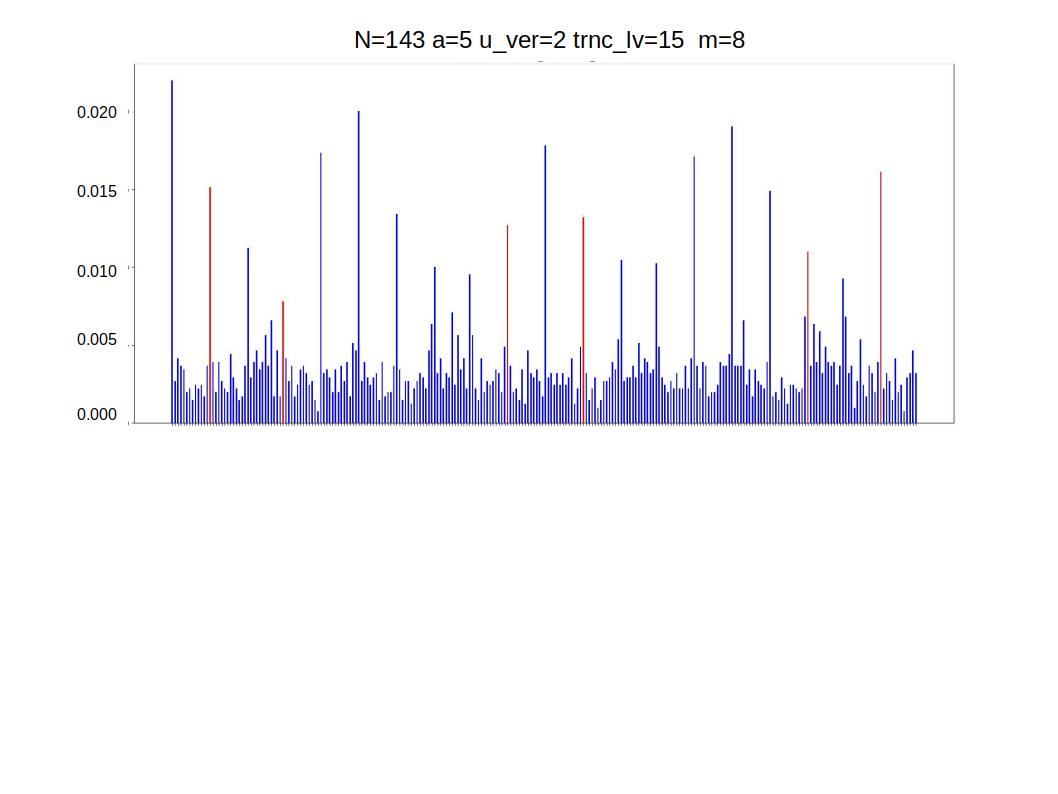}
\vskip-2.7cm
\includegraphics[scale=0.33]{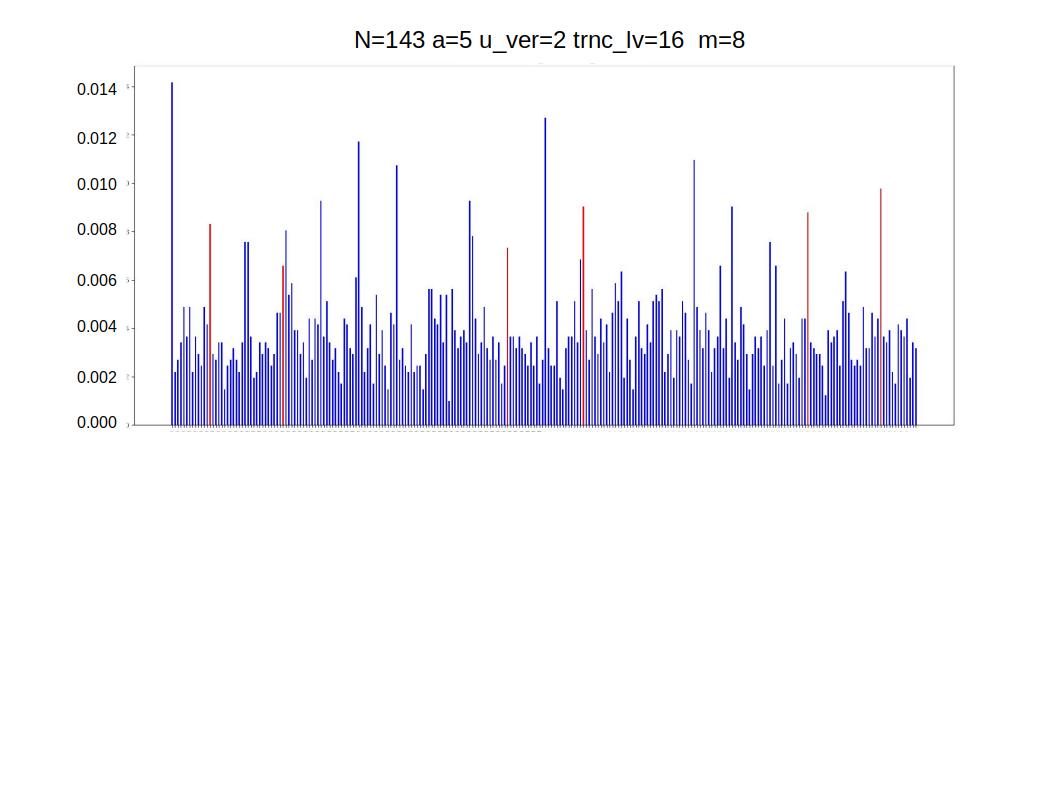}
\vskip-2.7cm
\includegraphics[scale=0.33]{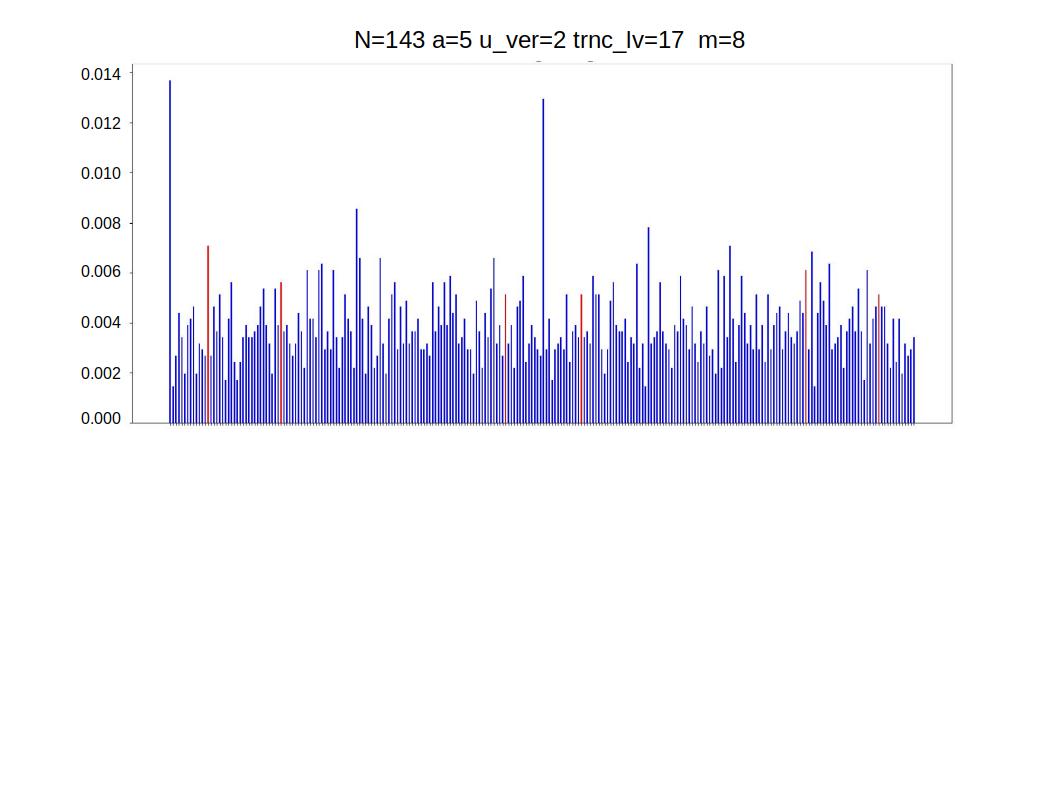}
\end{minipage}
\begin{minipage}[c]{0.4\linewidth}
\vskip0.4cm
\includegraphics[scale=0.33]{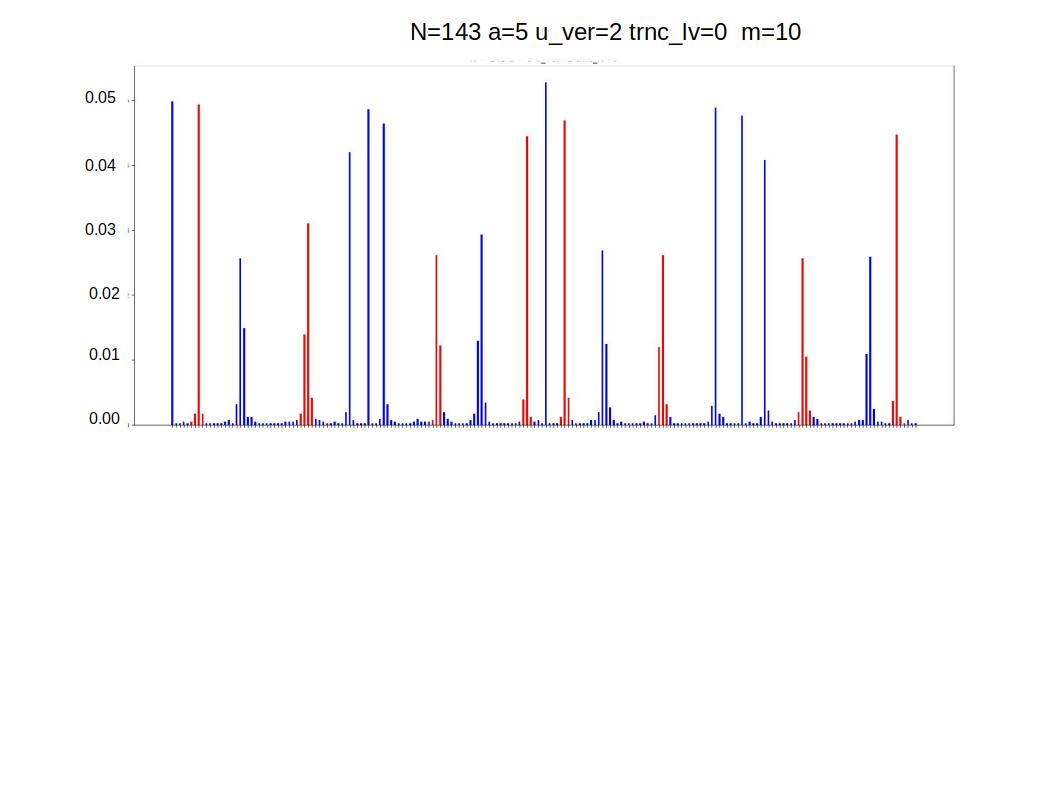}
\vskip-2.7cm
\includegraphics[scale=0.33]{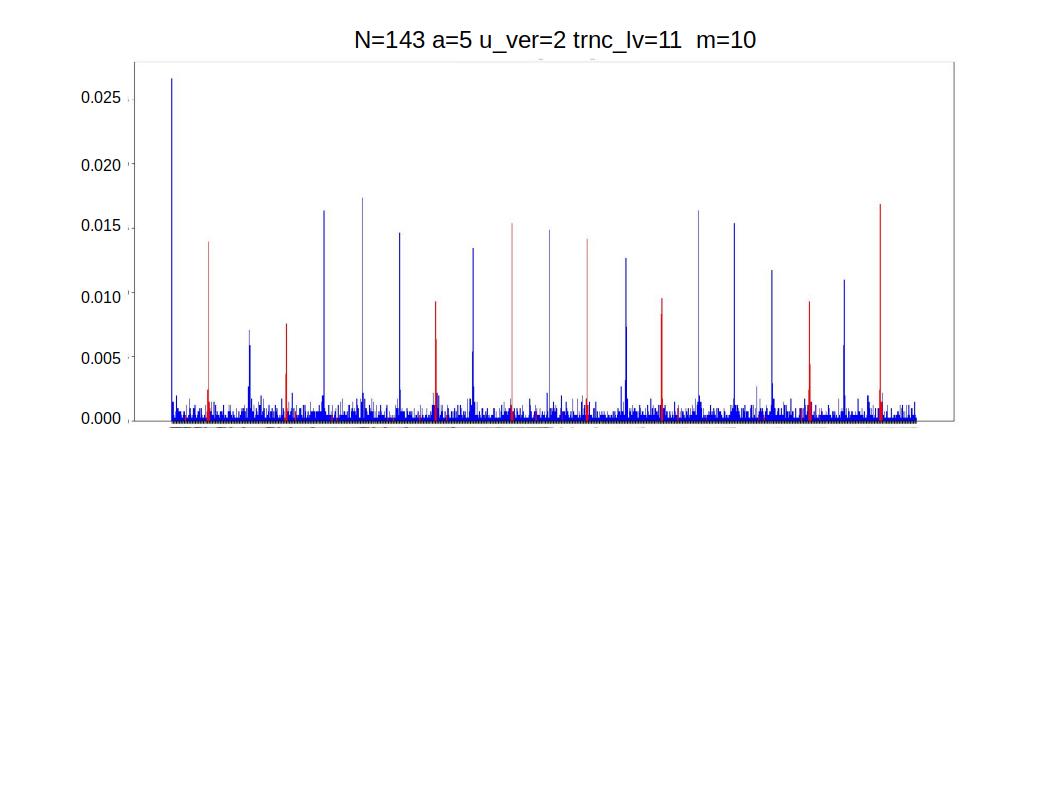}
\vskip-2.7cm
\includegraphics[scale=0.33]{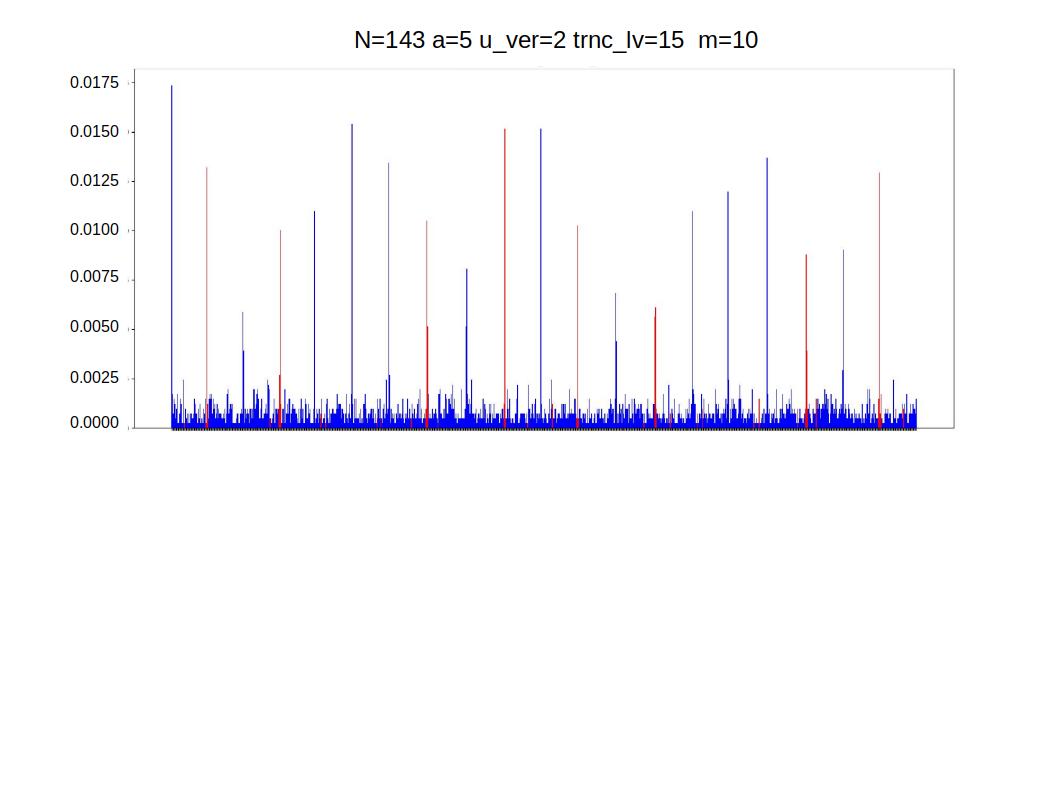}
\vskip-2.7cm
\includegraphics[scale=0.33]{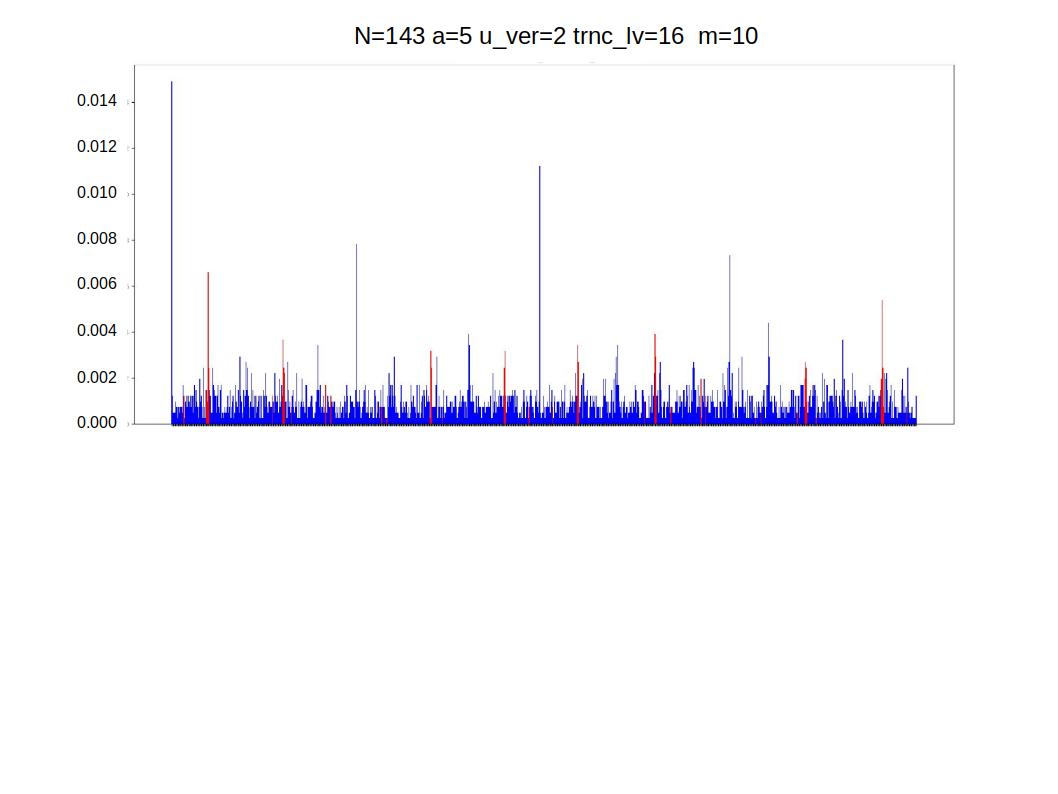}
\vskip-2.7cm
\includegraphics[scale=0.33]{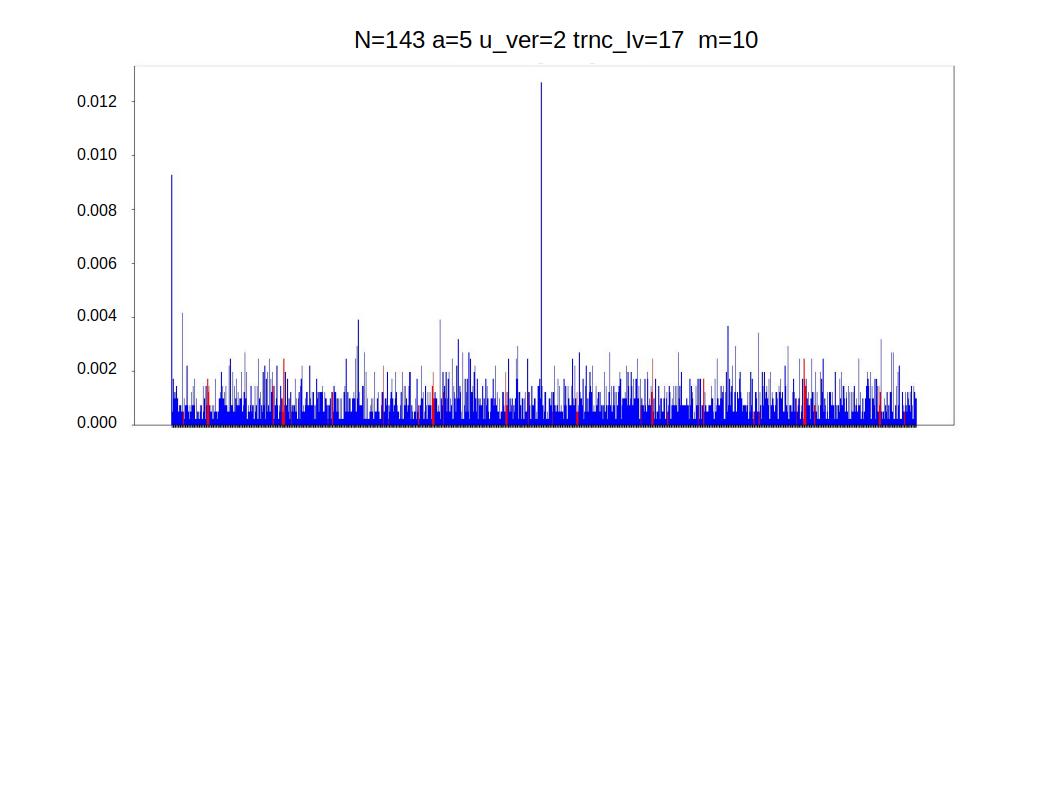}
\end{minipage}
\vskip-3.0cm
\caption{\footnoteskip
Truncation study for $N = 143$, $a =4$, $r = 20$, $n=8$, 
$m =8, 10$. The left panels show $m = 8$, while the right
panels are for $m = 10$. The truncation levels are 
${\tt trnc\_lv} = 0, 11, 15, 16, 17$. Note that we can extract 
signals at deeper levels for higher values of $m$.
}
\label{fig_truncate_N143}
\end{figure}
\begin{figure}[h!]
\includegraphics[scale=0.55]{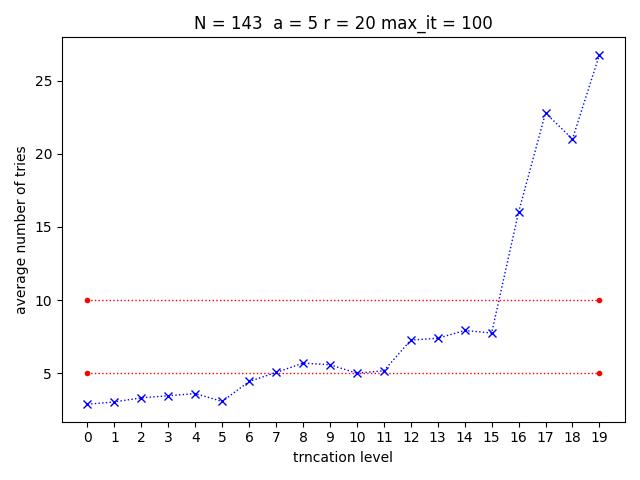} 
\vskip-0.8cm 
\caption{\footnoteskip  
 $N=143$, $a=5$, $r=20$, $n = 8$, $m=8, 10$: 
  Average number of tries vs. truncation level. 
}
\label{fig_tries_N134}
\end{figure}

The action of the operators $U, U^2, U^4, U^{16}, U^{32}, 
U^{64}, U^{128}$, $U^{256}$, $U^{512}$ are given below:
\begin{eqnarray}
  U_{5, 143} && ~:~~~ [1, 5, 25, 125, 53, 122, 38, 47, 92, 31, 12, 60, 14, 70, 64,
              34, 27, 135, 103, 86, 1]
\nonumber\\
  U^2_{5, 143} && ~:~~~ [1, 25, 53, 38, 92, 12, 14, 64, 27, 103, 1]   
    ~+~ [5, 125, 122, 47, 31, 60, 70, 34, 135, 86, 5]
\nonumber\\
  U^4_{5, 143} && ~:~~~ [1, 53, 92, 14, 27, 1]   ~+~ [5, 122, 31, 70, 135, 5] ~+~ 
                                              [25, 38, 12, 64, 103, 25] ~+~
\nonumber\\[-5pt]
                          && ~~~~~~~  [125, 47, 60, 34, 86, 125] 
\label{eq_Up_N21a2_seq_a}
\\
  U^8_{5, 143} && ~:~~~ [1, 92, 27, 53, 14, 1] ~+~  [5, 31, 135, 122, 70, 5] ~+~ 
                                               [25, 12, 103, 38, 64, 25] ~+~
\nonumber\\[-5pt]
                          && ~~~~~~~  [125, 60, 86, 47, 34, 125] 
\nonumber\\
  U^{16}_{5, 143}&& ~:~~~  [1, 27, 14, 92, 53, 1] ~+~ [5, 135, 70, 31, 122, 5] ~+~
      [25, 103, 64, 12, 38, 25] ~+~
\nonumber\\[-5pt]
                          && ~~~~~~~ [125, 86, 34, 60, 47, 125] 
\nonumber\\
  U^{32}_{5, 143}&& ~:~~~ [1, 14, 53, 27, 92, 1]  ~+~ [5, 70, 122, 135, 31, 5] ~+~
                                                   [25, 64, 38, 103, 12, 25] ~+~
\nonumber\\[-5pt]
                          && ~~~~~~~ [125, 34, 47, 86, 60, 125]
\nonumber\\
  U^{64}_{5, 143}&& ~:~~~  [1, 53, 92, 14, 27, 1]  ~+~ [5, 122, 31, 70, 135, 5] ~+~
                                                    [25, 38, 12, 64, 103, 25] ~+~
\nonumber\\[-5pt]
                          && ~~~~~~~ [125, 47, 60, 34, 86, 125] 
\nonumber\\
  U^{128}_{5, 143}&& ~:~~~ [1, 92, 27, 53, 14, 1]   ~+~ [5, 31, 135, 122, 70, 5] ~+~
                                                     [25, 12, 103, 38, 64, 25] ~+~
\nonumber\\[-5pt]
                          && ~~~~~~~  [125, 60, 86, 47, 34, 125] 
\nonumber\\
  U^{256}_{5, 143}&& ~:~~~  [1, 27, 14, 92, 53, 1] ~+~ [5, 135, 70, 31, 122, 5] ~+~
                                                     [25, 103, 64, 12, 38, 25]
\nonumber\\[-5pt]
                          && ~~~~~~~   [125, 86, 34, 60, 47, 125]
\nonumber\\
  U^{512}_{5, 143}&& ~:~~~ [1, 14, 53, 27, 92, 1] ~+~ [5, 70, 122, 135, 31, 5] ~+~
                                                     [25, 64, 38, 103, 12, 25] ~+~
\nonumber\\[-5pt]
                          && ~~~~~~~  [125, 34, 47, 86, 60, 125]
\nonumber
  \ ,
\end{eqnarray}
where the corresponding circuits are listed in Appendix~\ref{sec_N143}.
Note that $U^{4}_{5, 143} = U^{64}_{5, 143}$, $U^{8}_{5, 143} = 
U^{128}_{5, 143}$, $U^{16}_{5, 143} = U^{256}_{5, 143}$, and
$U^{32}_{5, 143} = U^{512}_{5, 143}$ (which can be obtained
by squaring the first relation). 
Figure~\ref{fig_fxN143a5m8_9_10_hist} illustrates the phase 
histograms for ${\tt u\_ver} = 1$  for the control 
resolutions of $m = 8, 10$. The dominant peaks in red correspond 
to the ME phases $\phi_s = s/20$ with $s \in \{0, 1, \cdots, 19\}$ 
and ${\rm gcd}(s, 20) = 1$. Thus the eight phases $s = 1, 3, 7, 9, 1
1, 13, 17, 19$ provide the factors of 11 and 13. Note, however, 
that the peaks for $s=7, 13$ are missing for $m = 8$.  This is 
because $m=8$ does not provide sufficient resolution, whereas 
all relevant peaks occur for the choice $m = 10$. 
Figure~\ref{fig_truncate_N143} shows a truncation study for
$m = 8, 10$ over the truncation levels ${\tt trnc\_lv} = 0, 11, 
15,16, 17$. Since the period is $r =20$, we can remove more 
than half the levels before we start to see a degraded signal. 
To quantify these results, Fig.~\ref{fig_tries_N134} plots the 
average number of tries that is takes to obtain a factor {\em
vs.}  the truncation level. We see that for ${\tt trnc\_lv} = 0, 1,
\cdots, 10$, we can obtain a factor in 5 tries; and for 
${\tt trnc\_lv} = 11, \cdots, 15$, we can obtain a factor in 
under 10 tries. Again, note the steep cliff after level 15. 
Also note that upon increasing  the phase resolution 
from $m = 8$ to $m = 10$, the signal to noise ratio improves 
substantially. Increasing the number of control qubits improves 
the correlations in the work register, and the acceptable truncation 
level increases slightly for $m = 10$. This trend will turn out
to be more pronounced in the next section when we factor
the larger number $N = 247$.

\vfill
\clearpage
\pagebreak

\subsection{$\bm{N=247 = 13 \times 19}$, $\bm{a=2}$, $\bm{r=36}$}

\begin{figure}[b!]
\vskip-11.0cm
\caption{\footnoteskip
$N=247$,  $a=2$,  $r=36$, $n=8$: The left panel is the modular 
exponential function $f_{2,  247}(x) $$= 2^x ~ ({\rm mod}~247)$,  
while the right panel gives the action of the ME operator 
$U_{2, 247}$. The circuit requires $n = \lceil \log_2 247 \rceil 
= 8$ qubits in the work register. 
}
\label{fig_fxN247a2_blank}
\end{figure}

\begin{figure}[t!]
\includegraphics[scale=0.45]{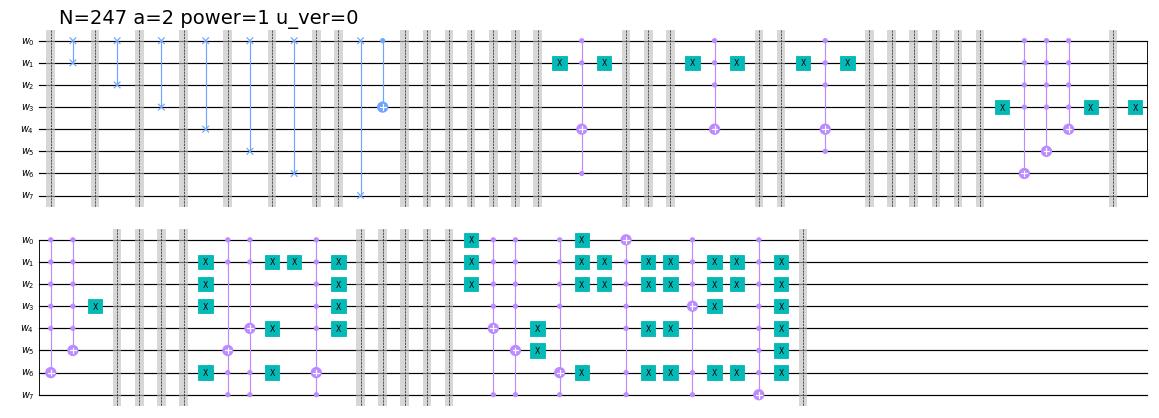} 
\caption{\footnoteskip
$N=247$,  $a=2$,  $r=36$, $n=8$: The modular exponentiation
operator $U_{2, 247}$. Note the large number of automatic
levels. 
}
\label{fig_fxN247a2}
\end{figure}

\begin{minipage}[c]{0.50\linewidth}
\baselineskip 19pt
We now factor the number $N = 247$ into 13 and~19. For 
the base $a =2$,  the left panel of Fig.~\ref{fig_fxN247a2_blank}
shows that the modular exponential function $f_{2,247}(x)$ 
has a period of $r = 36$. Note that we require $n = \lceil 
\log_2 246 \rceil = 8$ work qubits. The action of the ME 
operator $U_{2, 247}$ on the work states $\vert f_{2,247}(x)
\rangle$ is illustrated in right panel of Fig.~\ref{fig_fxN247a2_blank}, 
and its corresponding circuit representation is given in 
Fig.~\ref{fig_fxN247a2}.  In the next paragraph, we shall 
perform a truncation study for $m = 8, 10$ control qubits. 
We therefore require the ME operators $U_{2, 247}^p$ for 
$p =1, 2, 4,\cdots, 512$. As usual, we construct these operators 
using the automated gate construction script, as a period 
of $r = 36$ is so large that manually constructing the 
composite operators $U^p$ would be too time consuming 
and error prone. 

\vskip1.0cm
\includegraphics[scale=0.50]{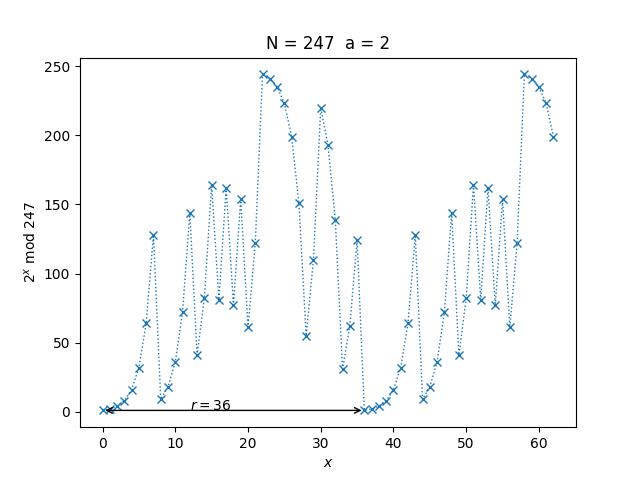}
~~\\
\vskip9.0cm
~~\\

\bodyskip
\end{minipage}
\begin{minipage}[c]{0.1\linewidth}
\end{minipage}
\begin{minipage}[c]{0.0\linewidth}
\begin{tabular}{|c|c|} \hline
 \multicolumn{2}{|c|}{~$U\vert w \rangle = 
\big\vert 2 \times w ~({\rm mod}~247) \big\rangle$~}  \\\hline
$~U\vert 1 \rangle = \vert 2 \rangle$~&
~$U\vert 00000001 \rangle = \vert 00000010 \rangle$~\\[-8pt]
$~U\vert 2 \rangle = \vert 4 \rangle$~&
~$U\vert 00000010 \rangle = \vert 00000100 \rangle$~\\[-8pt]
$~U\vert 4 \rangle = \vert 8 \rangle$~&
~$U\vert 00000100 \rangle = \vert 00001000 \rangle$~\\[-8pt]
$~U\vert 8 \rangle = \vert 16 \rangle$~&
~$U\vert 00001000 \rangle = \vert 00010000 \rangle$~\\[-8pt]
$~U\vert 16 \rangle = \vert 32 \rangle$~&
~$U\vert 00010000 \rangle = \vert 00100000 \rangle$~\\[-8pt]
$~U\vert 32 \rangle = \vert 64 \rangle$~&
~$U\vert 00100000 \rangle = \vert 01000000 \rangle$~\\[-8pt]
$~U\vert 64 \rangle = \vert 128  \rangle$~&
~$U\vert 01000000 \rangle = \vert 10000000 \rangle$~\\[-8pt]
$~U\vert 128 \rangle = \vert 9  \rangle$~&
~$U\vert 10000000 \rangle = \vert 00001001 \rangle$~\\[-8pt]
$~U\vert 9 \rangle ~= \vert 18  \rangle$~&
~$U\vert 00001001 \rangle = \vert 00010010 \rangle$~\\[-8pt]
$~U\vert 18 \rangle ~= \vert 36  \rangle$~&
~$U\vert 00010010 \rangle = \vert 00100100 \rangle$~\\[-8pt]
$~U\vert 36 \rangle ~= \vert 72  \rangle$~&
~$U\vert 00100100 \rangle = \vert 01001000 \rangle$~\\[-8pt]
$~U\vert 72 \rangle = \vert 144 \rangle$~&
~$U\vert 01001000 \rangle = \vert 10010000 \rangle$~\\[-8pt]
$~U\vert 144 \rangle = \vert 41 \rangle$~&
~$U\vert 10010000 \rangle = \vert 00101001 \rangle$~\\[-8pt]
$~U\vert 41 \rangle = \vert 82 \rangle$~&
~$U\vert 00101001 \rangle = \vert 01010010 \rangle$~\\[-8pt]
$~U\vert 82 \rangle = \vert 164 \rangle$~&
~$U\vert 01010010 \rangle = \vert 10100100 \rangle$~\\[-8pt]
$~U\vert 164 \rangle = \vert 81 \rangle$~&
~$U\vert 10100100 \rangle = \vert 01010001 \rangle$~\\[-8pt]
$~U\vert 81 \rangle = \vert 162  \rangle$~&
~$U\vert 01010001 \rangle = \vert 10100010 \rangle$~\\[-8pt]
$~U\vert 162 \rangle ~= \vert 77  \rangle$~&
~$U\vert 10100010 \rangle = \vert 01001101 \rangle$~\\[-8pt]
$~U\vert 77 \rangle ~= \vert 154  \rangle$~&
~$U\vert 01001101 \rangle = \vert 10011010 \rangle$~\\[-8pt]
$~U\vert 154 \rangle ~= \vert 61  \rangle$~&
~$U\vert 10011010 \rangle = \vert 10011010 \rangle$~\\[-8pt]
$~U\vert 61 \rangle ~= \vert 122  \rangle$~&
~$U\vert 10011010 \rangle = \vert 01111010 \rangle$~\\[-8pt]
$~U\vert 122 \rangle = \vert 244 \rangle$~&
~$U\vert 01111010 \rangle = \vert 11110100 \rangle$~\\[-8pt]
$~U\vert 244 \rangle = \vert 241 \rangle$~&
~$U\vert 11110100 \rangle = \vert 11110001 \rangle$~\\[-8pt]
$~U\vert 241 \rangle = \vert 235\rangle$~&
~$U\vert 11110001 \rangle = \vert 11101011 \rangle$~\\[-8pt]
$~U\vert 235 \rangle = \vert 223 \rangle$~&
~$U\vert 11101011 \rangle = \vert 11011111 \rangle$~\\[-8pt]
$~U\vert 223 \rangle = \vert 199 \rangle$~&
~$U\vert 11011111 \rangle = \vert 11000111 \rangle$~\\[-8pt]
$~U\vert 199 \rangle = \vert 151 \rangle$~&
~$U\vert 11000111 \rangle = \vert 10010111 \rangle$~\\[-8pt]
$~U\vert 151 \rangle = \vert 55  \rangle$~&
~$U\vert 10010111 \rangle = \vert 00110111 \rangle$~\\[-8pt]
$~U\vert 55 \rangle ~= \vert 110  \rangle$~&
~$U\vert 00110111 \rangle = \vert 01101110 \rangle$~\\[-8pt]
$~U\vert 110 \rangle ~= \vert 220  \rangle$~&
~$U\vert 01101110 \rangle = \vert 11011100 \rangle$~\\[-8pt]
$~U\vert 220 \rangle ~= \vert 193  \rangle$~&
~$U\vert 11011100 \rangle = \vert 11000001 \rangle$~\\[-8pt]
$~U\vert 193 \rangle ~= \vert 139  \rangle$~&
~$U\vert 11000001 \rangle = \vert 10001011 \rangle$~\\[-8pt]
$~U\vert 139 \rangle = \vert 31 \rangle$~&
~$U\vert 10001011 \rangle = \vert 00011111 \rangle$~\\[-8pt]
$~U\vert 31 \rangle = \vert 62 \rangle$~&
~$U\vert 00011111 \rangle = \vert 00111110 \rangle$~\\[-8pt]
$~U\vert 62 \rangle = \vert 124 \rangle$~&
~$U\vert 00111110 \rangle = \vert 01111100 \rangle$~\\[-8pt]
$~U\vert 124 \rangle = \vert 1 \rangle$~&
~$U\vert 01111100 \rangle = \vert 00000001 \rangle$~\\\hline
\end{tabular} 
~~\\
\vskip9.0cm
~~\\
\end{minipage}
~~

\begin{figure}[t!]
\includegraphics[scale=0.45, center]{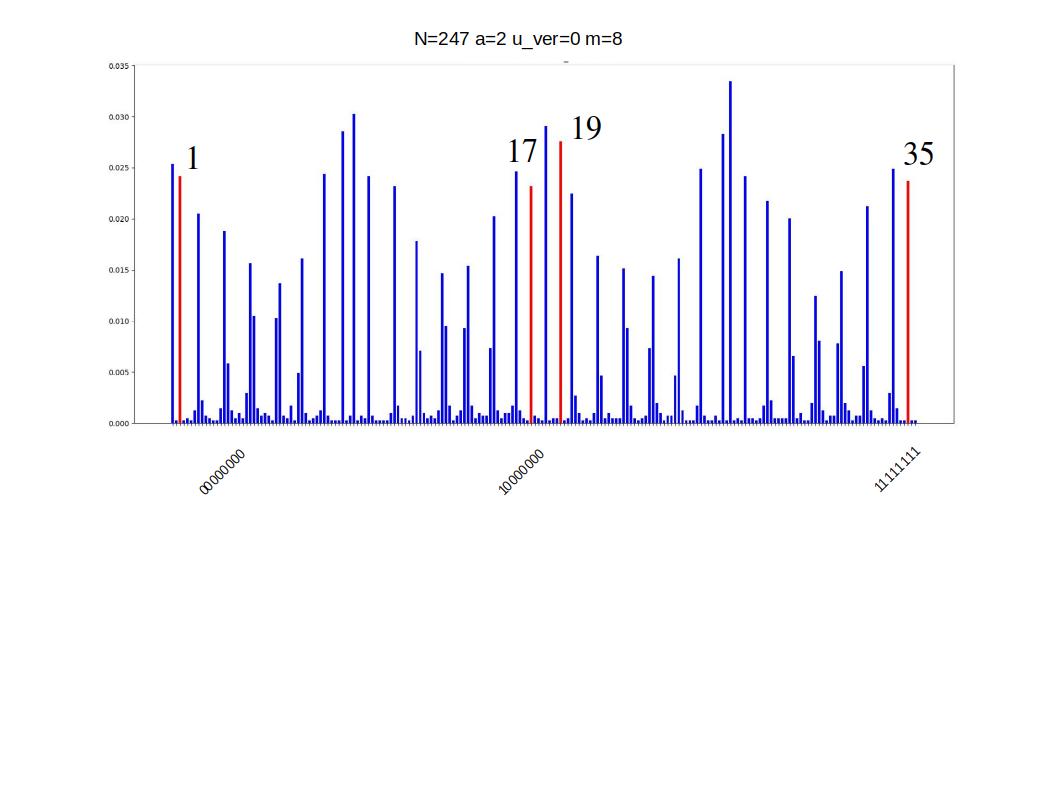}
\vskip-4.3cm
\includegraphics[scale=0.45, center]{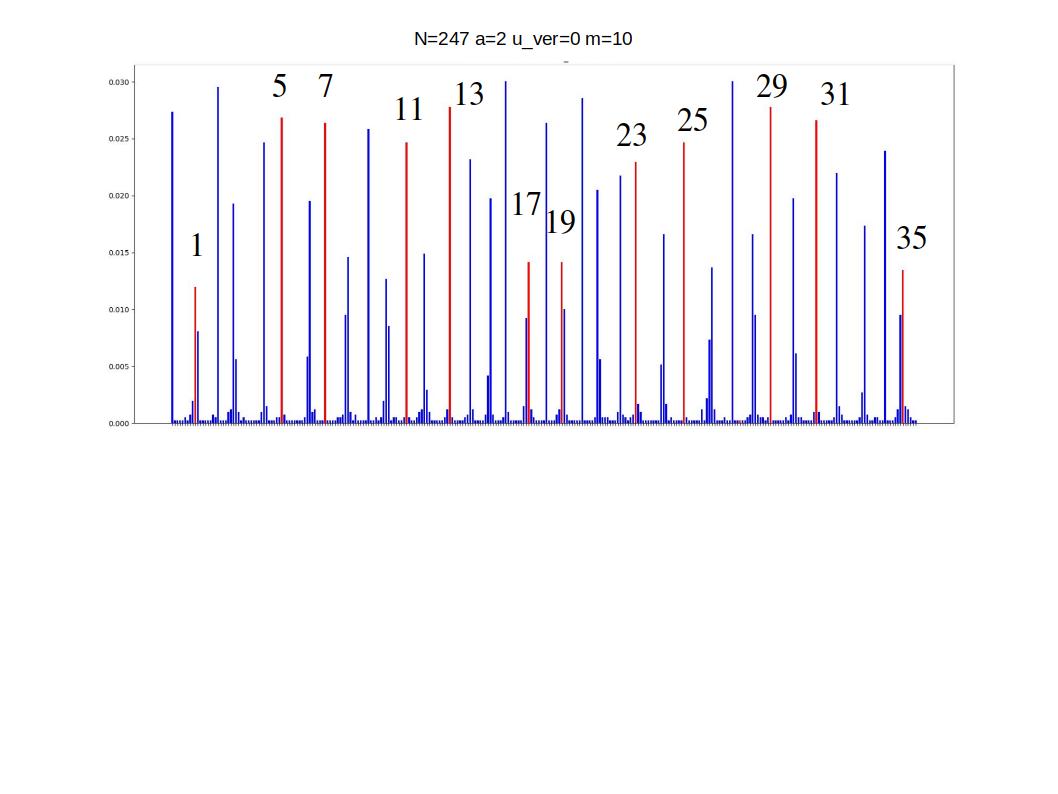}
\vskip-4.2cm
\caption{\footnoteskip
$N=247$,  $a=2$,  $r=36$, $n=8$, $m=8, 10$: Phase histograms 
for ME operator version $\tt{u\_ver}=1$. The top panel shows 
$m = 8$, and the bottom panel illustrates the higher resolution 
$m = 10$. The resolution for $m = 8$ only permits four factorization 
peaks, while $m = 10$ allows for all twelve. 
}
\label{fig_N247a2m8_10_hist}
\end{figure}

The ME operators $U^p$ for $p =  1, 2, 4, \cdots,512$ 
have the following closed cycles:
\begin{eqnarray}
  U_{2, 247} && ~:~
  [1, 2, 4, 8, 16, 32, 64, 128, 9, 18, 36, 72,
 144, 41, 82, 164, 81, 162, 77, 154, 
\nonumber
\\[-5pt] && ~~~~\,
  61, 122, 244, 241, 235, 223, 199, 151, 
  55, 110, 220,193, 139, 31,   62, 124, 1]
\label{eq_Up1_N247a2} 
\\
  U^{2}_{2, 247}   && ~:~
     [1, 4, 16, 64, 9, 36, 144, 82, 81, 77, 61, 244,
     235, 199, 55, 220, 139, 62, 1]
      ~~+~
\nonumber\\[-5pt] && ~~~~\, 
  [2, 8, 32, 128, 18, 72, 41, 164, 162, 154, 122,
   241, 223, 151, 110, 193, 31, 124, 2]
\nonumber\\
   U^4_{2, 247}  && ~:~
  [1, 16, 9, 144, 81, 61, 235, 55, 139, 1] 
    ~+~
  [2, 32, 18, 41, 162, 122, 223, 110, 31, 2]
  \nonumber \\ [-5pt] &&  ~~~~\, 
   [4, 64, 36, 82, 77, 244, 199, 220, 62, 4]
  ~+~
  [8, 128, 72, 164, 154, 241, 151, 193, 124, 8]
\nonumber
\nonumber\\
  U^8_{2, 247}  && ~:~ [1, 9, 81, 235, 139, 16, 144, 61, 55, 1]
   ~+~
  [2, 18, 162, 223, 31, 32, 41, 122, 110, 2]
  \nonumber \\ [-5pt] &&  ~~~~\, 
  [4, 36, 77, 199, 62, 64, 82, 244, 220, 4]
   ~+~
   [8, 72, 154, 151, 124, 128, 164, 241, 193, 8]
\nonumber\\
  U^{16}_{2, 247}   && ~:~
   [1, 81, 139, 144, 55, 9, 235, 16, 61, 1]
   ~+~
   [2, 162, 31, 41, 110, 18, 223, 32, 122, 2]
  \nonumber \\ [-5pt] &&  ~~~~\, 
     [4, 77, 62, 82, 220, 36, 199, 64, 244, 4]
   ~+~
    [8, 154, 124, 164, 193, 72, 151, 128, 241, 8]
\nonumber\\
  U^{32}_{2, 247}   && ~:~
   [1, 139, 55, 235, 61, 81, 144, 9, 16, 1]
   ~+~
  [2, 31, 110, 223, 122, 162, 41, 18, 32, 2]
  \nonumber \\ [-5pt] &&  ~~~~\, 
       [8, 124, 193, 151, 241, 154, 164, 72, 128, 8]
   ~+~
    [4, 62, 220, 199, 244, 77, 82, 36, 64, 4]
\nonumber
\nonumber\\
  U^{64}_{2, 247}  && ~:~  
         [1, 55, 61, 144, 16, 139, 235, 81, 9, 1]
   ~+~
    [2, 110, 122, 41, 32, 31, 223, 162, 18, 2] 
  \nonumber \\ [-5pt] &&  ~~~~\, 
     [8, 193, 241, 164, 128, 124, 151, 154, 72, 8]
     ~+~
    [4, 220, 244, 82, 64, 62, 199, 77, 36, 4] 
\nonumber\\
  U^{128}_{2, 247}   && ~:~
       [1, 61, 16, 235, 9, 55, 144, 139, 81, 1]
   ~+~
         [2, 122, 32, 223, 18, 110, 41, 31, 162, 2]
\nonumber \\ [-5pt] &&  ~~~~\, 
         [4, 244, 64, 199, 36, 220, 82, 62, 77, 4]
   ~+~
        [8, 241, 128, 151, 72, 193, 164, 124, 154, 8]
  \nonumber\\
  U^{256}_{2, 247}   && ~:~
       [1, 16, 9, 144, 81, 61, 235, 55, 139, 1]
   ~+~
        [2, 32, 18, 41, 162, 122, 223, 110, 31, 2]
\nonumber \\ [-5pt] &&  ~~~~\, 
      [4, 64, 36, 82, 77, 244, 199, 220, 62, 4]
   ~+~
        [8, 128, 72, 164, 154, 241, 151, 193, 124, 8]
  \nonumber\\
  U^{512}_{2, 247}   && ~:~
       [1, 9, 81, 235, 139, 16, 144, 61, 55, 1]
   ~+~
        [2, 18, 162, 223, 31, 32, 41, 122, 110, 2]
\nonumber \\ [-5pt] &&  ~~~~\, 
       [4, 36, 77, 199, 62, 64, 82, 244, 220, 4]
   ~+~
      [8, 72, 154, 151, 124, 128, 164, 241, 193, 8]
\nonumber
  \ .
\end{eqnarray}
The corresponding circuits are listed in Appendix~\ref{sec_N247}.
Note that $U_{2, 247}^{4} = U_{2, 247}^{256}$ and
$U_{2, 247}^{8} = U_{2, 247}^{512}$ (the second relation
is given by squaring the first). The phases of the $U_{2, 247}$ 
operator that provide factors occur at $\phi_s = s/36$ for 
$s \in \{ 0,  1,  \cdots, 35\}$, where $r=36$ and $s$ have 
no non-trivial common factors. This gives 12 possible phases:
$s=$ 1, 5, 7,  11, 13, 17, 19,23, 25, 29, 31, 35. The phase 
histograms for $m =8, 10$ are given in 
Fig.~\ref{fig_N247a2m8_10_hist} for ${\tt u\_ver} = 1$. We
see that $m = 8$ only possesses four of these phases, although 
this is enough to extract factors from Shor's algorithm. However, 
if we increase the phase resolution to $m = 10$ control qubits, 
then all 12 phases appear. 

\begin{figure}[h!]
\hskip-3.0cm
\begin{minipage}[c]{0.5\linewidth}
\vskip0.2cm
\includegraphics[scale=0.33]{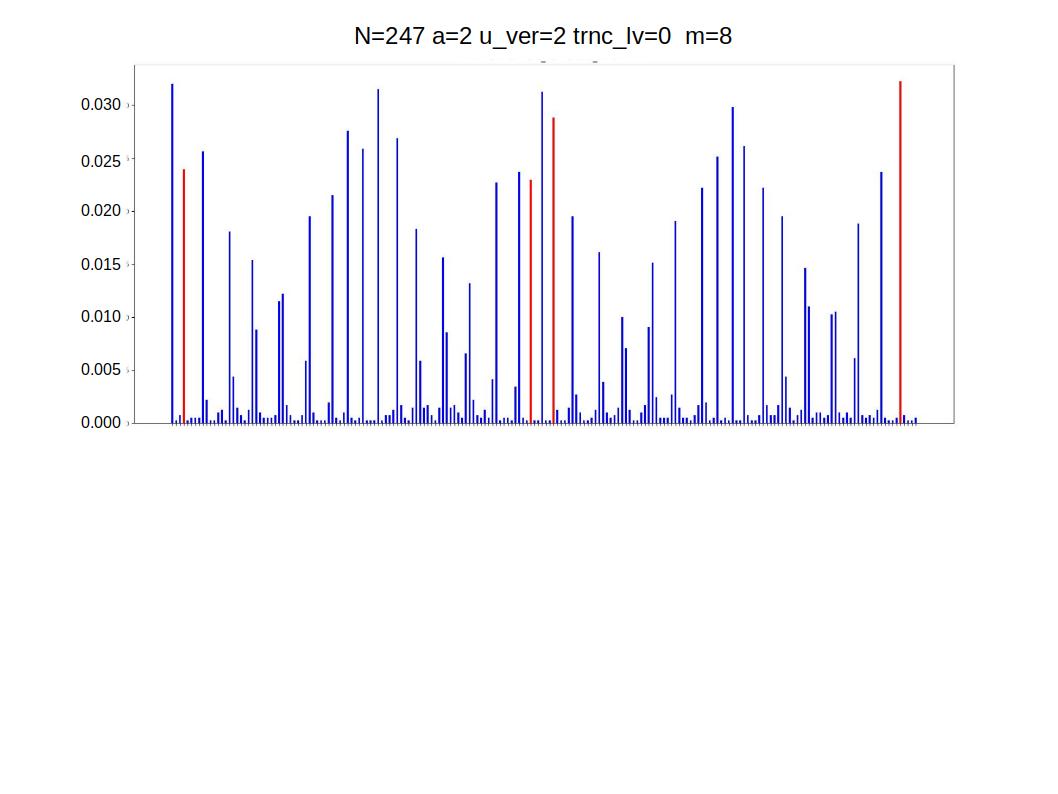} 
\vskip-2.8cm
\includegraphics[scale=0.33]{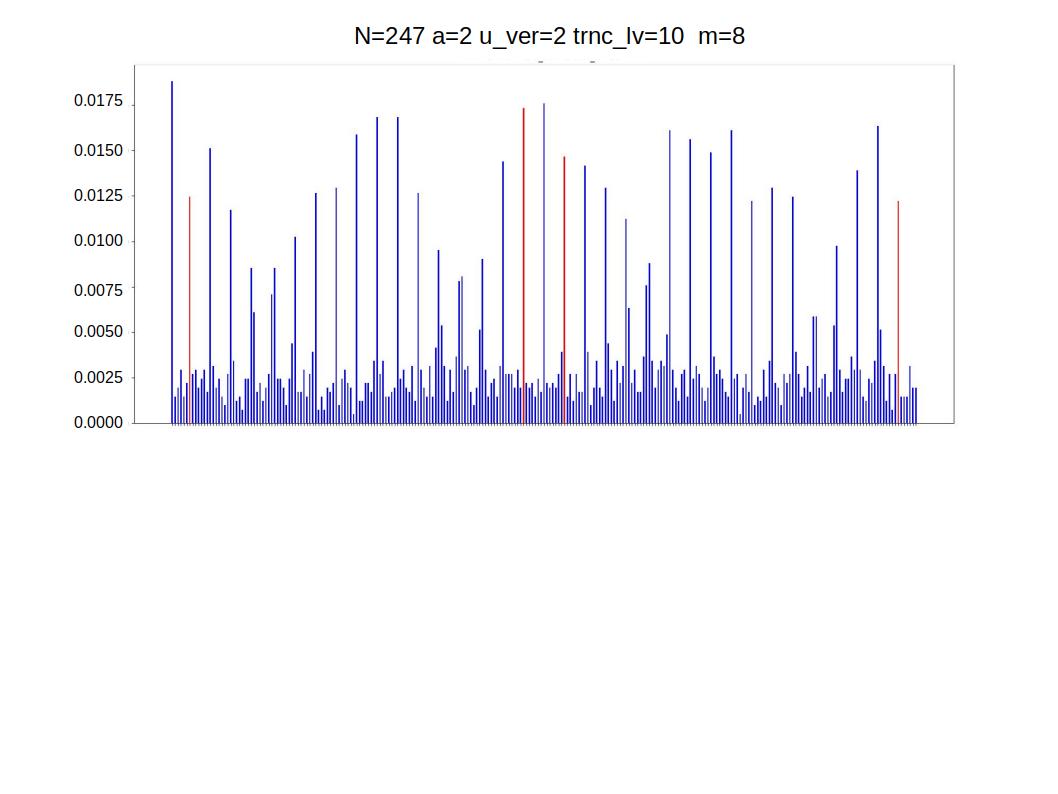} 
\vskip-2.8cm
\includegraphics[scale=0.33]{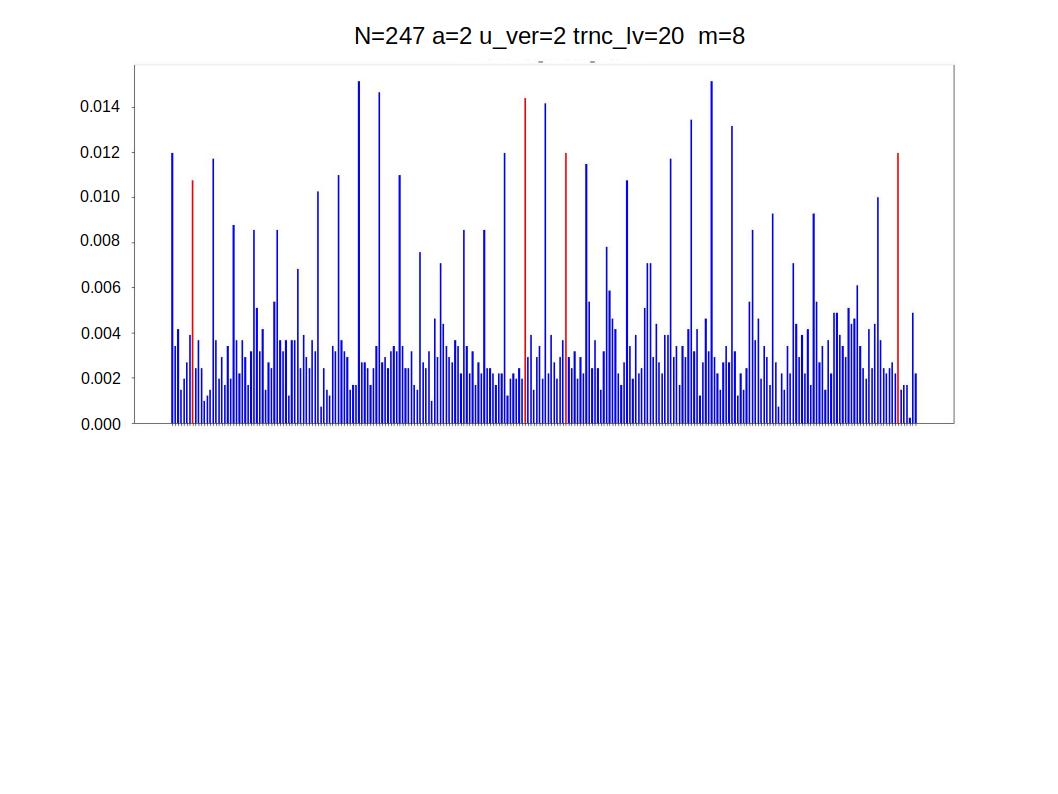} 
\vskip-2.8cm
\includegraphics[scale=0.33]{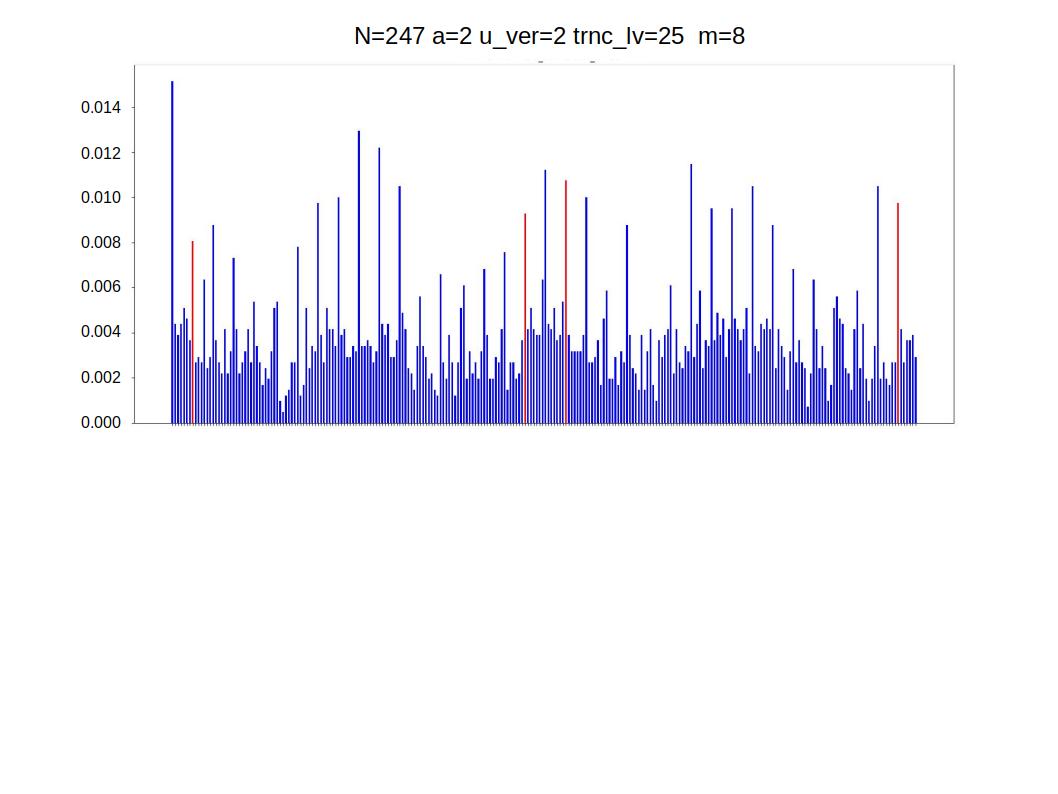} 
\vskip-2.8cm
\includegraphics[scale=0.33]{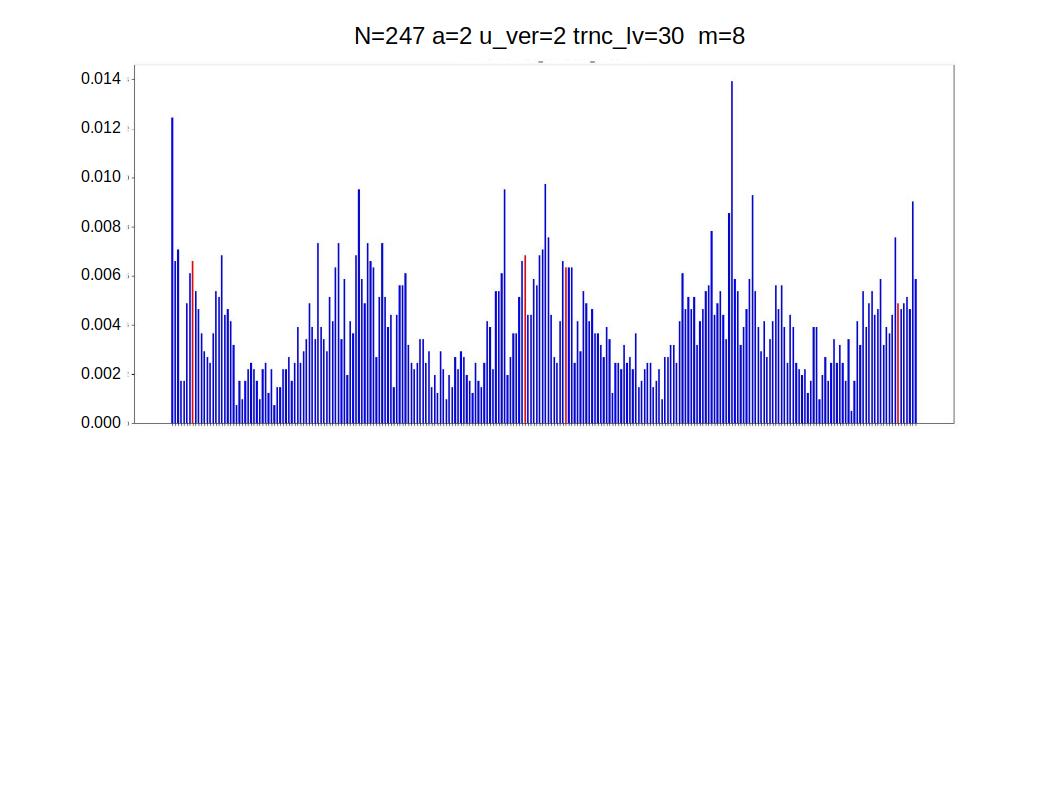} 
\end{minipage}
\begin{minipage}[c]{0.4\linewidth}
\vskip0.4cm
\includegraphics[scale=0.33]{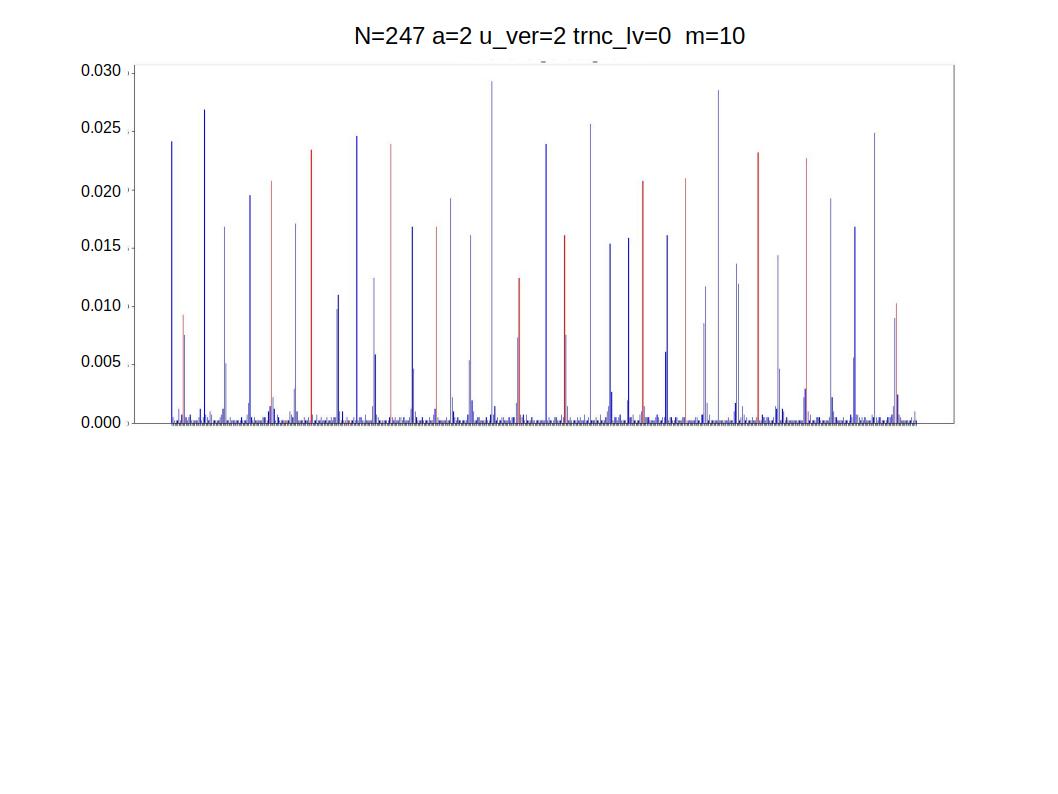}
\vskip-2.8cm
\includegraphics[scale=0.33]{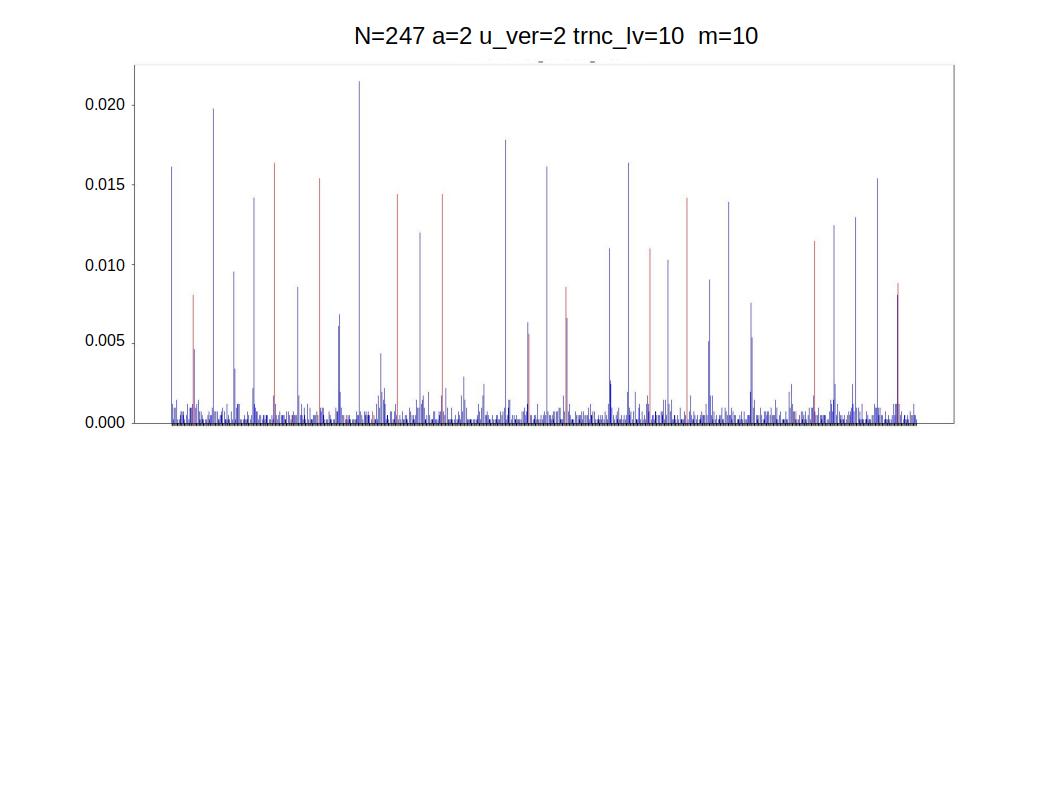} 
\vskip-2.8cm
\includegraphics[scale=0.33]{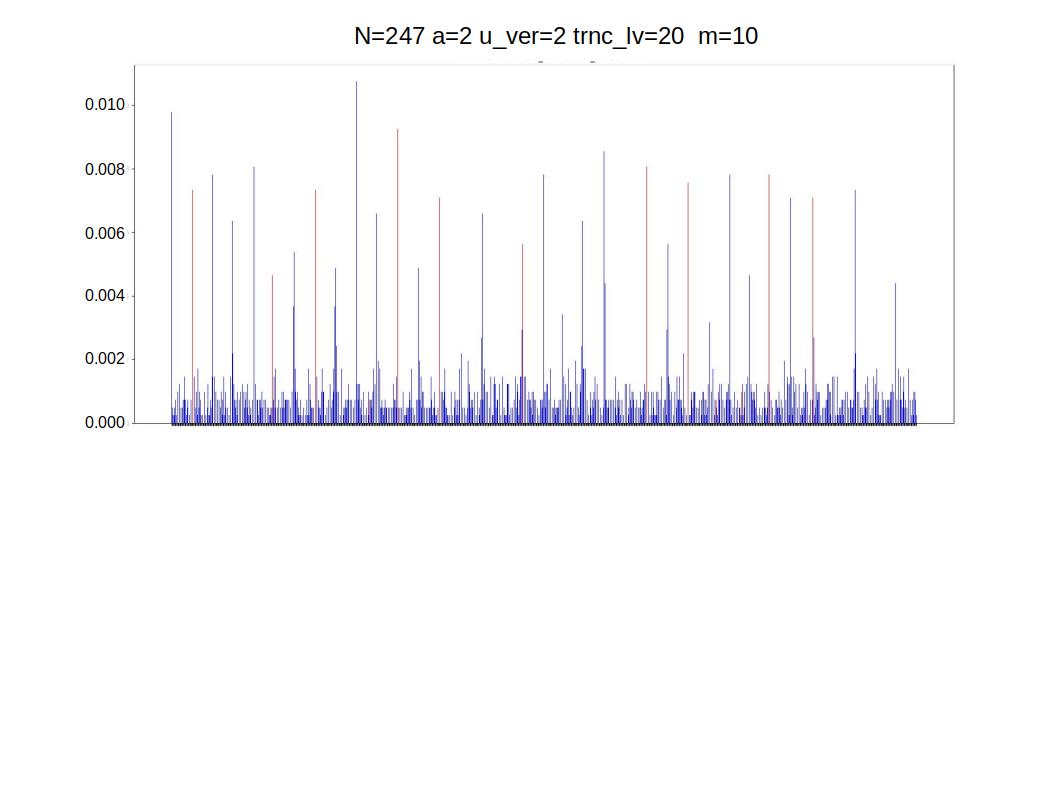} 
\vskip-2.8cm
\includegraphics[scale=0.34]{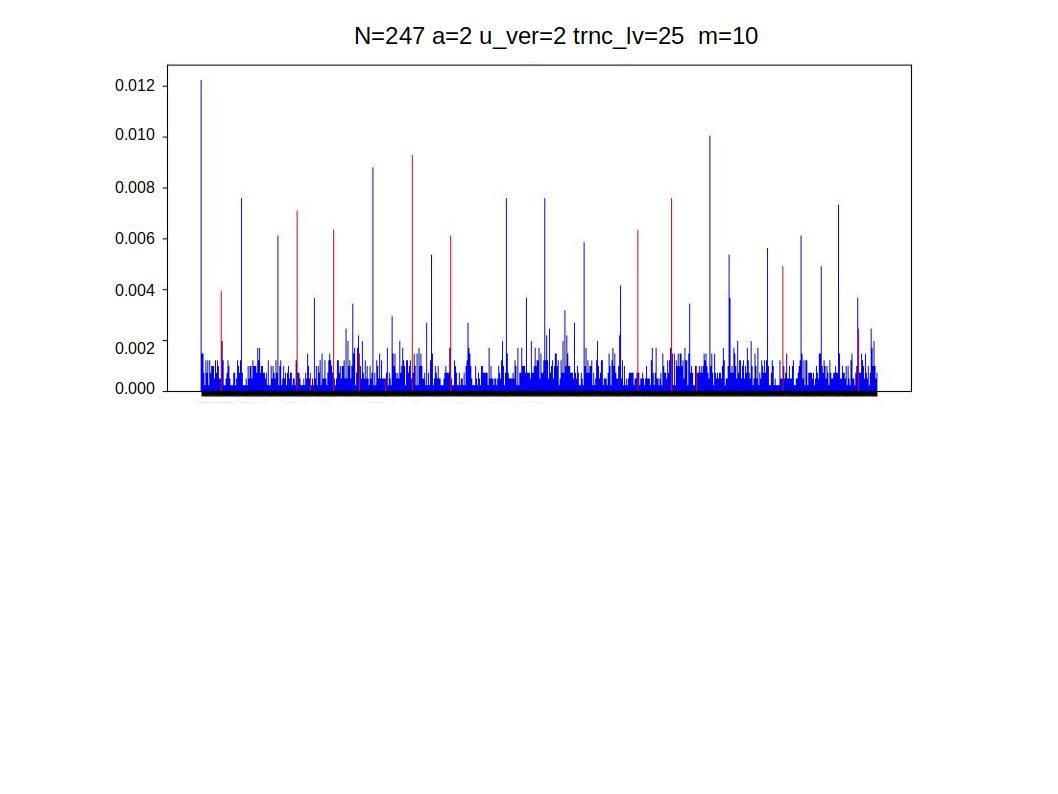} 
\vskip-2.8cm
\includegraphics[scale=0.34]{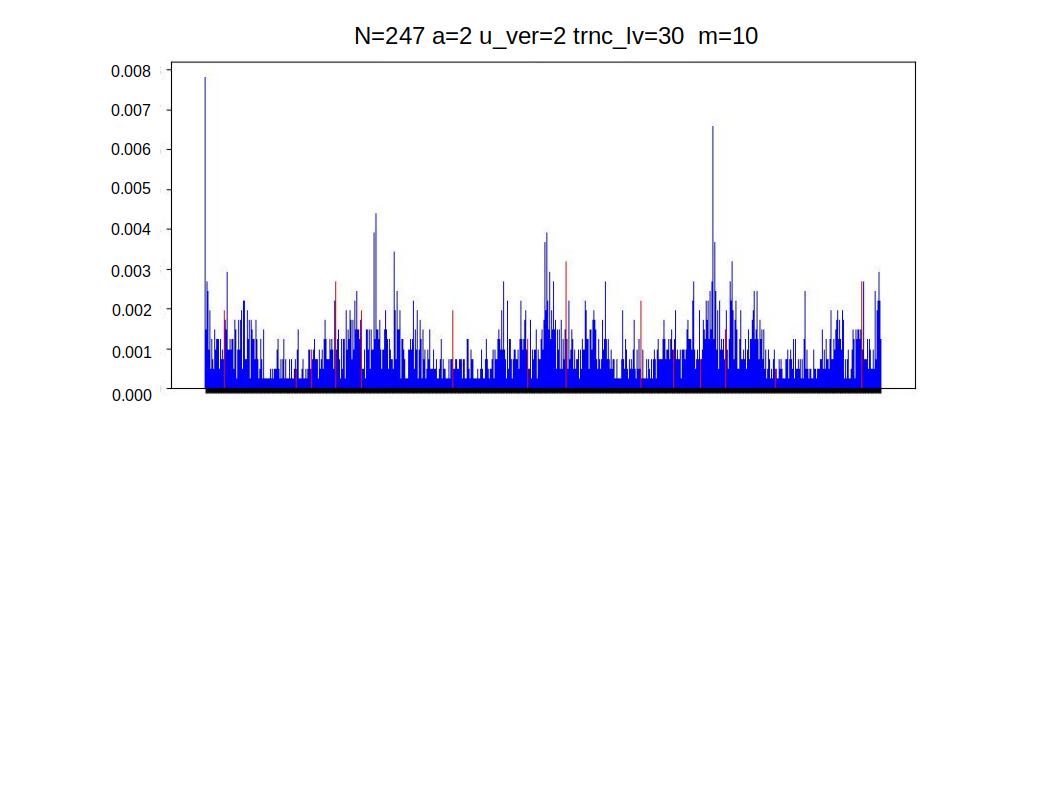} 
\end{minipage}
\vskip-3.0cm
\caption{\footnoteskip
Truncation study for $N = 247$, $a =2$, $r = 36$, $n=8$,
$m =8, 10$ over the range ${\tt trnc\_lv} = 0, 10, 20, 25, 30$,
from  top to bottom. The left panels show $m = 8$, while 
the right panels are for $m = 10$. Note that $m = 8$ starts
becoming noisy shortly after ${\tt trnc\_lv} = 20$, but that 
$m = 10$ retains a strong signal as low as ${\tt trnc\_lv} = 25$. 
In general. the signal is much strong for $m = 10$. 
}
\label{fig_truncate_N247}
\end{figure}

\begin{figure}[h!]
\includegraphics[scale=0.55]{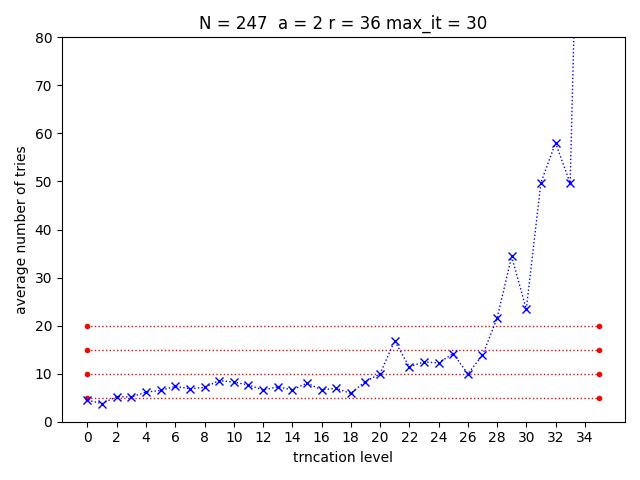} 
\vskip-0.8cm 
\caption{\footnoteskip  
$N=247$,  $a=2$,  $r=36$, $n=8$, $m=10$: 
  Average number of tries vs. truncation level. 
}
\label{fig_tries_N247}
\end{figure}

Finally, Fig.~\ref{fig_truncate_N247} presents a truncation 
study for $m = 8, 10$ over the range of truncation levels
${\tt trnc\_lv} = 0, 10, 20, 25, 30$. The left panels in the Figure 
show the $m = 8$ results, while the right panels illustrate the
corresponding phase histograms for $m = 10$. Note that $m 
= 8$ starts becoming noisy between ${\tt trnc\_lv} = 10$ to 20, 
but that $m =10$ retains a strong signal as low as ${\tt trnc\_lv} 
= 25$. This means that we only need to keep $36 - 25 = 11$
levels to extract a factorization signal. 
Finally, Fig.~\ref{fig_tries_N247} plots the average number
of tries required to find a factor {\em vs.} the truncation
level. This a much more 
dramatic improvement than in the previous case of $N = 143$,
and it will be interesting to see if this trend continues for
even larger numbers. However, these studies push the
limit of the simulator. Therefore, qubit reduction strategies 
like qubit recycling\,\cite{parker1} must be employed 
to increase the value of $N$. 

%

\vfill
\clearpage
\pagebreak

\section{Conclusions}
\label{sec_conclusions}

In this manuscript and in Ref.~\cite{sing1}, I have 
presented a method for constructing modular 
exponentiation (ME) operators for Shor's algorithm. 
Because the work register starts in state $\vert 1 \rangle$,
we do not have to construct a general ME operator 
acting on the entire \hbox{$2^n$-dimensional} Hilbert space 
${\cal W}_n$ of the work register, as in Ref.~\cite{preskill1}. 
Rather, we only need to construct an operator that 
acts on the $r$-dimensional subspace ${\cal U}_r$ 
spanned by the states $\vert f(x) \rangle$ for $x \in 
\{0,1, \cdots,  r-1\}$. This makes the task far easier, 
and drastically reduces the qubit count and gate depth 
as compared to other methods. This approach leads 
to the following strategy for factoring with Shor's 
algorithm. Each composite ME operator $U^p$ for 
$p \in\{2^0, 2^1, 
\cdots, 2^{m-1}\}$ creates closed subsequence of the 
states in the subspace ${\cal U}_r$.  For example, the 
operator $U$ performs the transition from the state
$\vert f(x) \rangle = \vert w_n \cdots w_1 w_0 \rangle$  
to the state $\vert f(x+1) \rangle = \vert w_n^\prime 
\cdots w_1^\prime w_0^\prime \rangle$.  Transforming
a binary number $w = w_n \cdots w_1 w_0$ to another
binary number $w^\prime = w_n^\prime \cdots 
w_1^\prime w_0^\prime$, without altering the 
preceding states, can be turned into a set of formal 
rules. Therefore, this procedure can be automated 
using multi-controlled NOT gates and single-qubit 
NOT gates. Indeed, this is how the ME operators in 
this manuscript, and in Ref.~\cite{sing1}, were constructed. 
The operator $U$ therefore consists of $r$ levels, 
where level $x \in \{0, 1, \cdots, r-1\}$ performs the 
transition $U\,\vert f(x) \rangle = \vert f(x+1) \rangle$. 

The problem with constructing $U$ in this way is that 
we must know the period $r$ of the ME function $f(x)$ 
in advance, and consequently there is no need for Shor's 
algorithm. However, as I have show in the text, the full
sequence of $r$ levels is not required. I perform a 
number of systematic truncation studies, and find that 
factorization is still possible even when well over half 
the levels are omitted. This is  because 
the method of continued fractions, which is used to 
extract the period $r$, still provides an adequate 
approximation to the phase of the $U$ operator. 
The strategy is then to build the operators $U^p$ 
one level at a time, checking for factors at each 
level. The method works because there are enough 
correlations between the $U^p$ operators  to maintain 
a sufficiently correlated control register that permits 
factorization.

For intermediate-scale testing, truncation is of clear 
benefit. However, if the period $r$ becomes exponentially 
large, this method would seem to require an exponential 
number of gates. This is likely to happen for an exponentially 
large number $N$, which is the case of most interest.
We can alleviate this problem by increasing the number 
of qubits $m$ in the control register, which permits
the ME operators to probe deeper. This is because
there are more operators $U^p$ for $p \in \{2^0, 2^2, 
\cdots, 2^{m-1}\}$, and therefore more correlations 
are being tracked. In the coming age of plentiful high 
quality qubits, increasing $m$ would be of no great 
concern. I am also investigating other 
truncation strategies which might alleviate this 
problem. 

Future work will entail factoring larger numbers. 
Qubit recycling, in which the multi-qubit 
control register is replaced by a single-qubit register, 
is an obvious path forward. I am also looking at methods 
to generalize the construction of the ME operators 
$U^p$. By employing templates for the ME operators,
we can perhaps find the optimal approximate operator 
$U_{\rm approx}^p$ for any given $U^p$.  One approach 
would be to employ a variational quantum method to
this end.  Finally, it would be interesting to perform a 
noise study on a simulator to better gauge the effects 
of real quantum gates on this factoring strategy. 
This methodology, or generalizations thereof, could 
help bring Shor's algorithm within closer reach.  It 
remains to be seen just how few levels one can use.

\vfill
\clearpage

\begin{acknowledgments}
I would like to thank Don Shirk and David Ostby
for reading the manuscript for clarity, and for a
number of useful discussions. 
\end{acknowledgments}

\appendix

\section{Modular Exponentiation Operators}
\label{sec_meo_rel_prime}

\subsection{Modular Exponentiation Operators for $\bm {N = 143}$}
\label{sec_N143}

This appendix provides the modular exponentiation (ME) operators
$U_{5, 143}^p$ for the powers $p \in \{1, 2, 4, \cdots, 512\}$. Quantum 
circuits are often called {\em scores} because of their resemblance 
to written musical scores. The quantum score can still be read even 
when the gates are not quite resolvable to the eye. The green boxes 
are single-qubit NOT gates $X$, the blue lines ending with a plus 
sign in a circle indicate controlled-NOT gates $CX$, and purple 
lines with a dots and a circle-plus are multi-controlled-NOT gates 
$CC \cdots CX$. 

\begin{figure}[h!]
\includegraphics[scale=0.45, center]{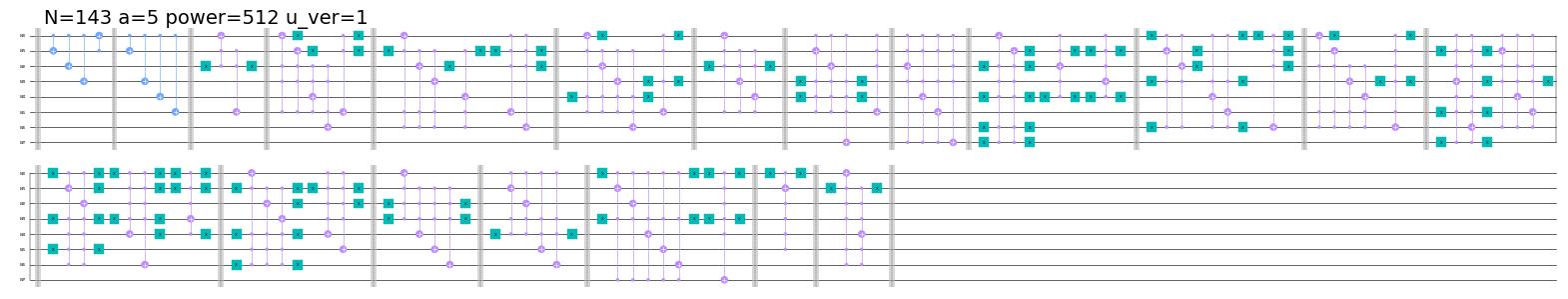} 
\includegraphics[scale=0.45, center]{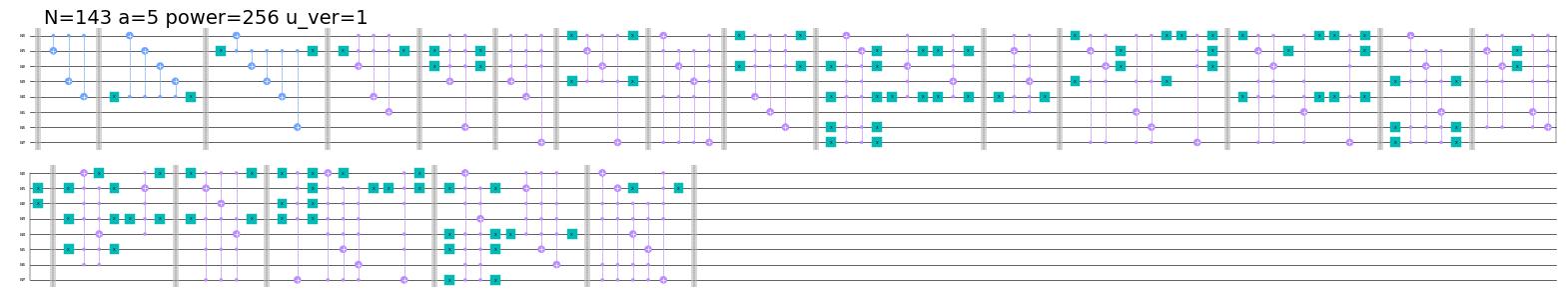} 
\includegraphics[scale=0.38, center]{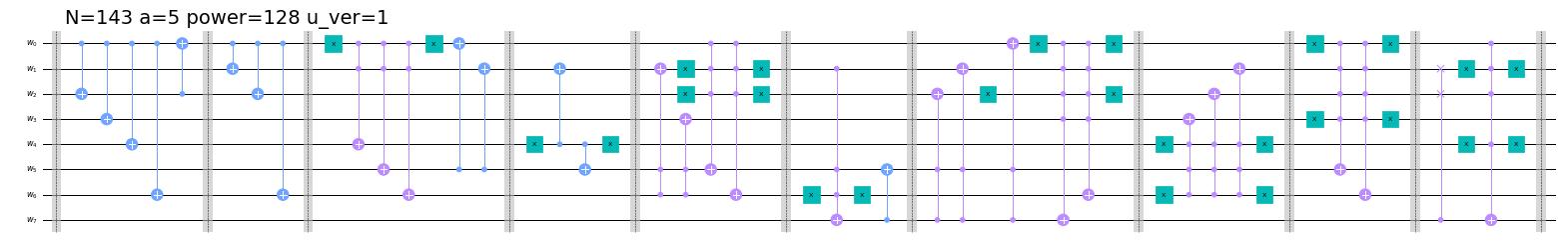} 
\includegraphics[scale=0.40, center]{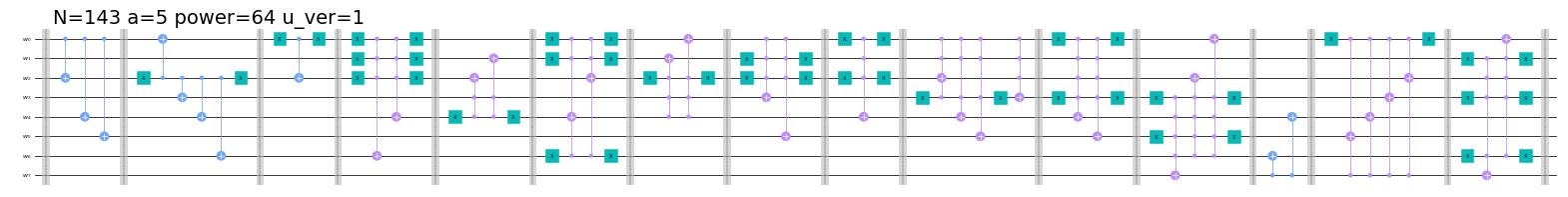} 
\caption{\footnoteskip  
  $N = 143$, $a = 5$, $r = 10$, $n = 8$: The ME operator $U^p$ for the 
  powers $p \in \{64, 128, 256, 512\}$.
}
\label{fig_N143_a}
\end{figure}

\begin{figure}[h!]
\includegraphics[scale=0.40, center]{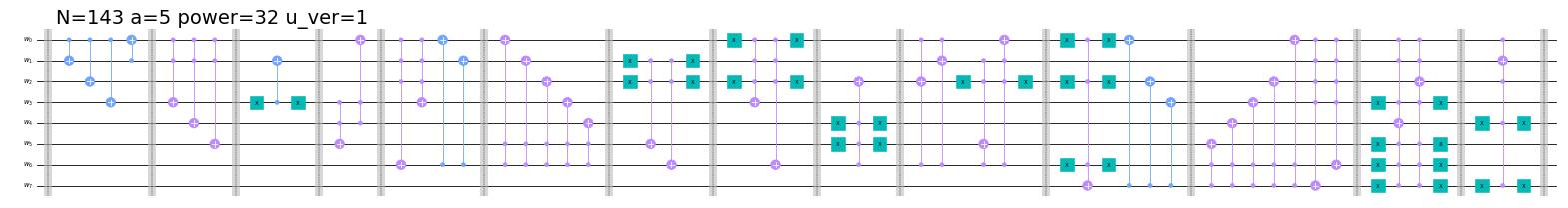} 
\includegraphics[scale=0.40, center]{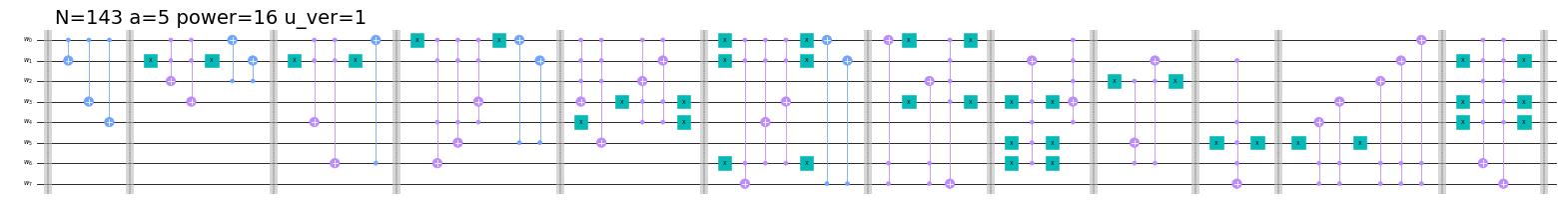} 
\includegraphics[scale=0.40, center]{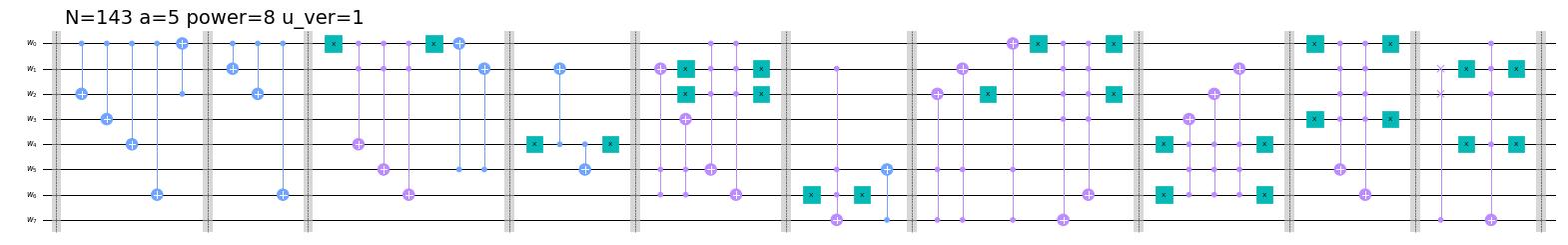} 
\includegraphics[scale=0.40, center]{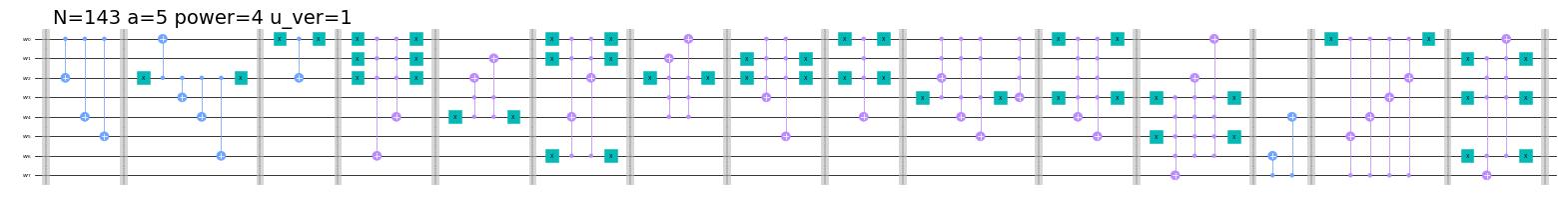} 
\includegraphics[scale=0.40, center]{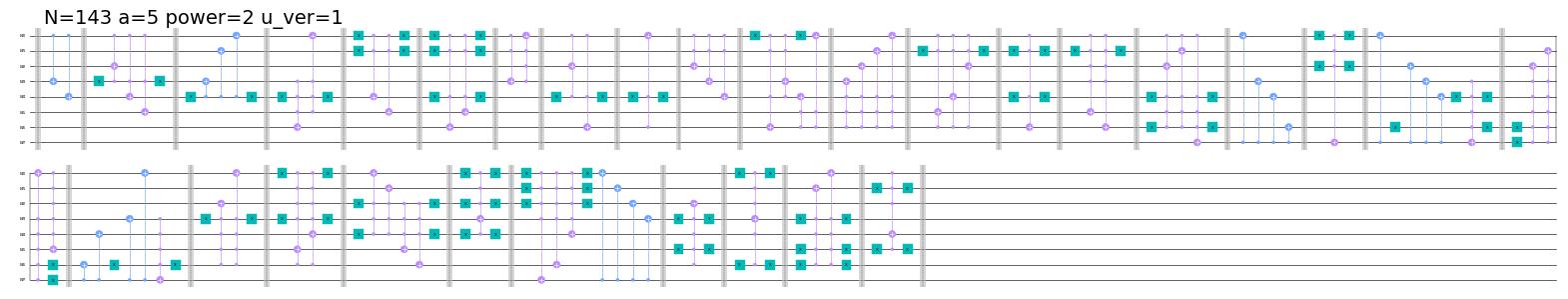} 
\includegraphics[scale=0.40, center]{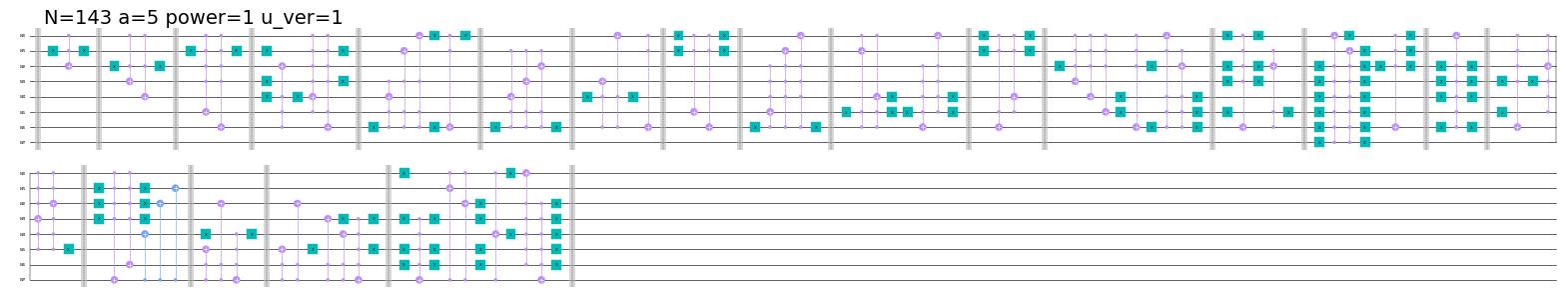} 
\caption{\footnoteskip  
  $N = 143$, $a = 5$, $r = 10$, $n = 8$: The ME operator $U^p$ for the
  powers $p \in \{1, 2, 4, 8, 16, 32\}$.
}
\label{fig_N143_b}
\end{figure}

\vfill
\clearpage
\subsection{Modular Exponentiation Operators for $\bm {N = 247}$}
\label{sec_N247}

This appendix provides the modular exponentiation (ME) operators
$U_{2, 247}^p$ for the powers $p \in \{1, 2, 4, \cdots, 512\}$. Quantum 
circuits are often called {\em scores} because of their resemblance 
to music. As emphasized in the previous appendix, it is not necessary 
to resolve every ``note'' in the {\em score} of the circuit, as the score 
can still be read. The green boxes are single-qubit NOT gates $X$, 
the blue lines ending with a plus sign in a circle indicate controlled-NOT 
gates $CX$, and purple lines with dots and a circle-plus are 
multi-controlled-NOT gates $CC \cdots CX$. 
\begin{figure}[h!]
\includegraphics[scale=0.45, center]{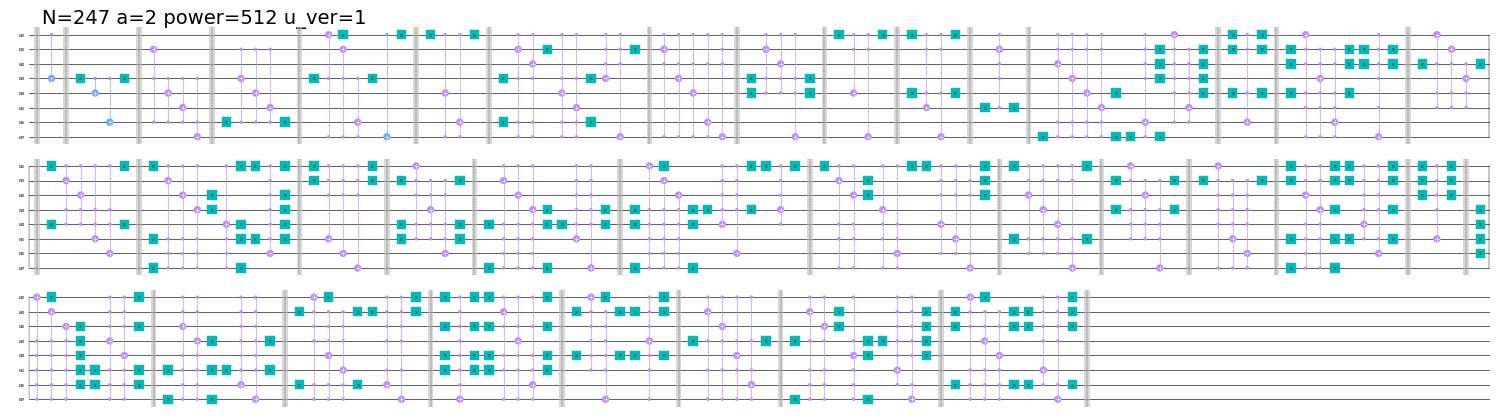} 
\includegraphics[scale=0.45, center]{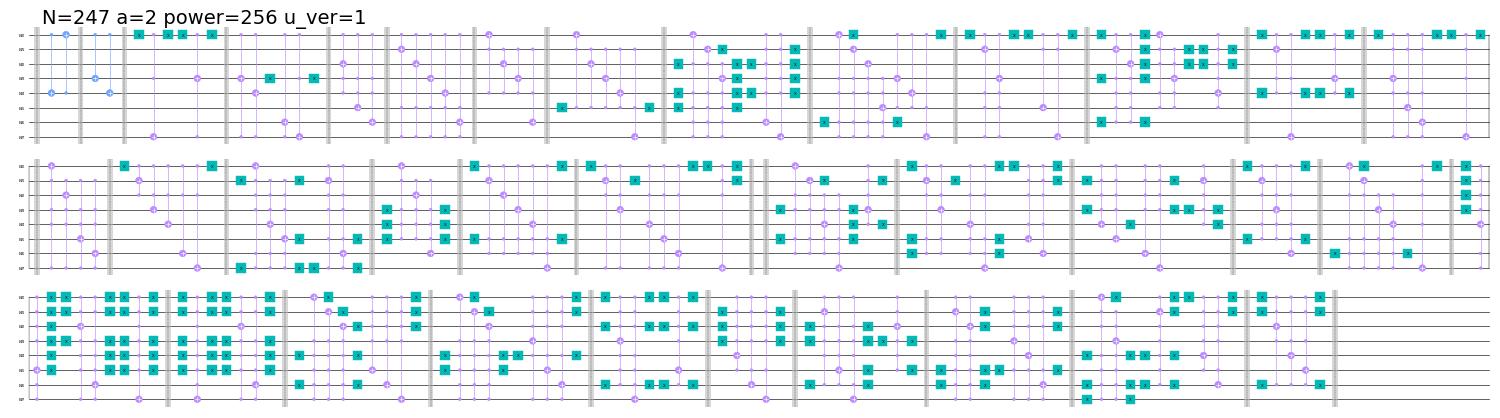} 
\includegraphics[scale=0.45, center]{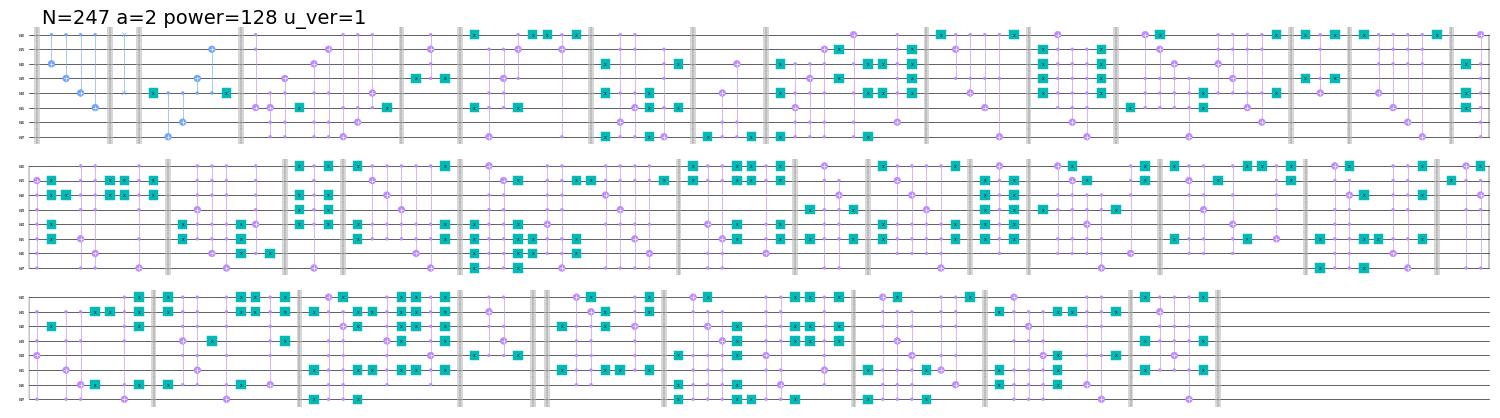} 
\caption{\footnoteskip  
  $N = 247$, $a = 2$, $r = 36$, $n = 8$: The ME operator $U^p$ for the
  powers $p \in \{128, 256, 512\}$.
}
\label{fig_N247_a}
\end{figure}

\begin{figure}[h!]
\includegraphics[scale=0.60, center]{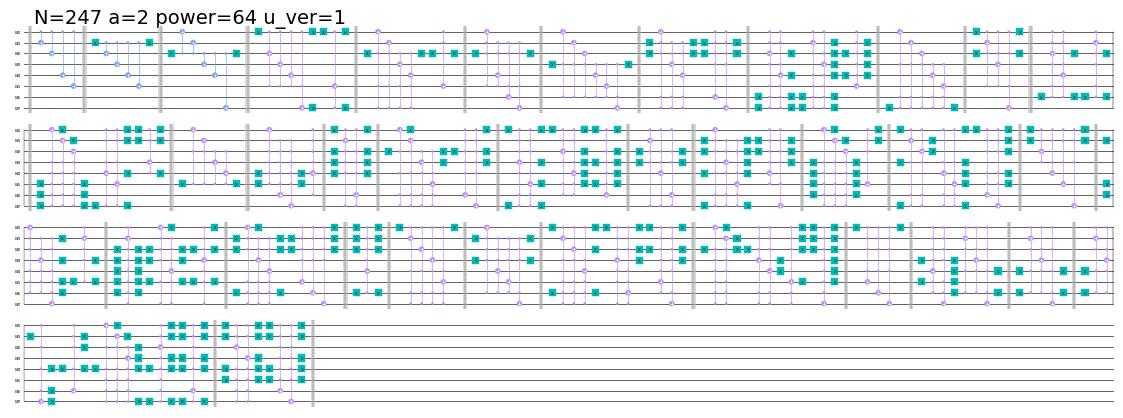} 
\includegraphics[scale=0.45, center]{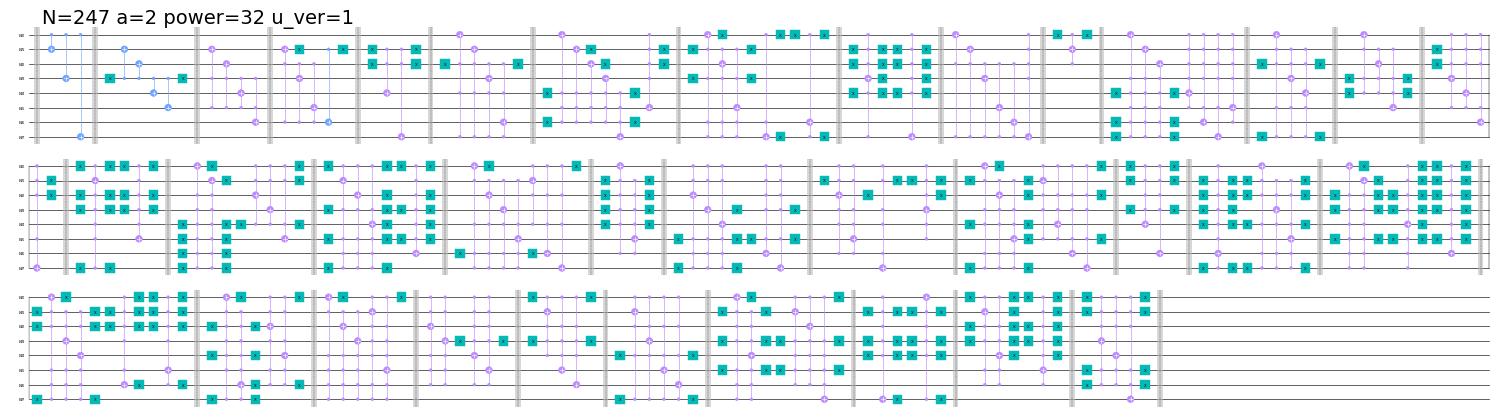} 
\includegraphics[scale=0.45, center]{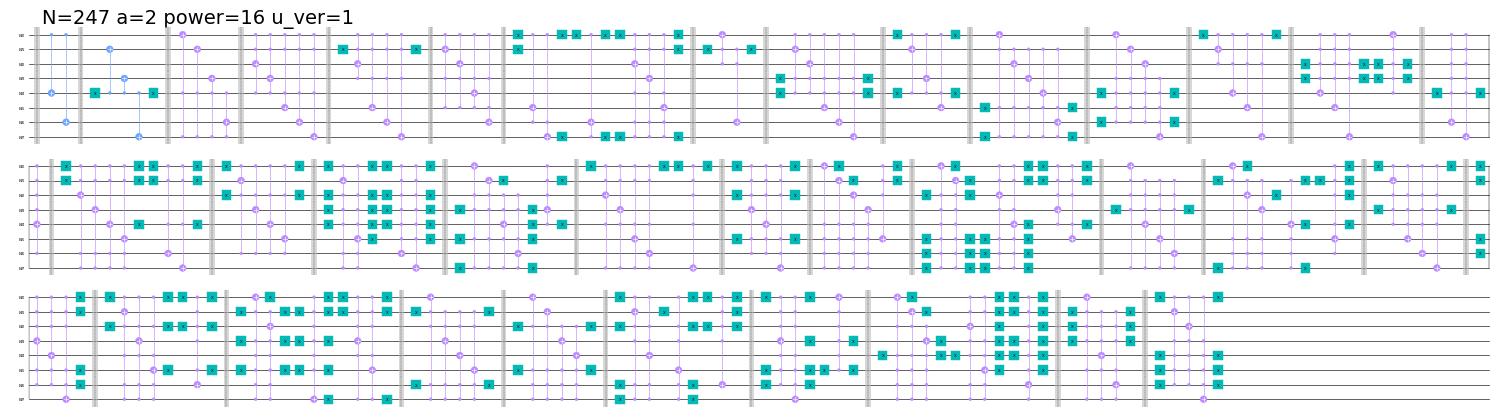} 
\includegraphics[scale=0.45, center]{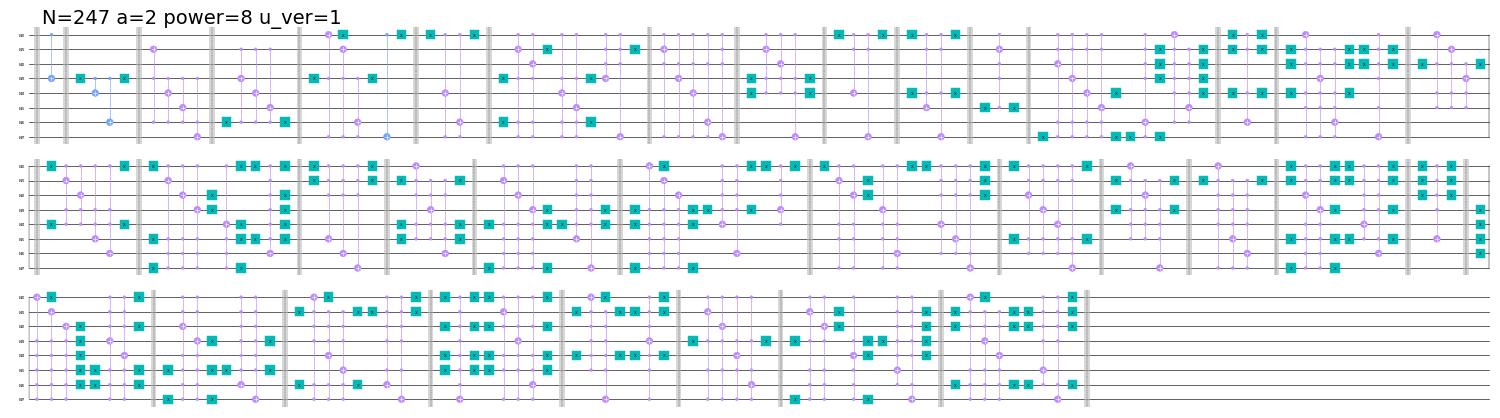} 
\caption{\footnoteskip  
  $N = 247$, $a = 2$, $r = 36$, $n = 8$: The ME operator $U^p$ for the
  powers $p \in \{8, 16, 32, 46\}$.
}
\label{fig_N247_b}
\end{figure}
 
\begin{figure}[h!]
\includegraphics[scale=0.45, center]{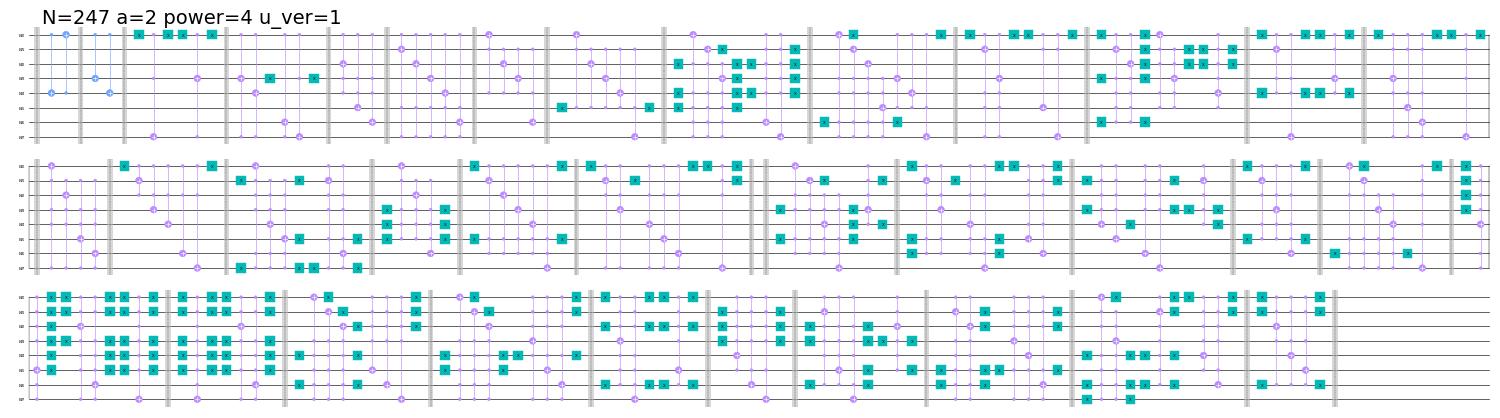} 
\includegraphics[scale=0.45, center]{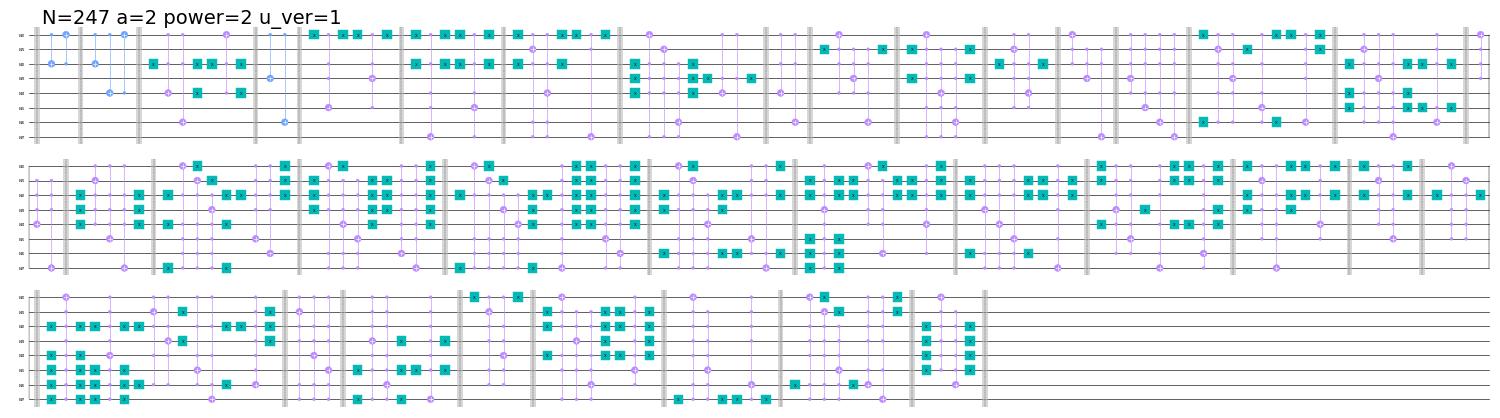} 
\includegraphics[scale=0.43, center]{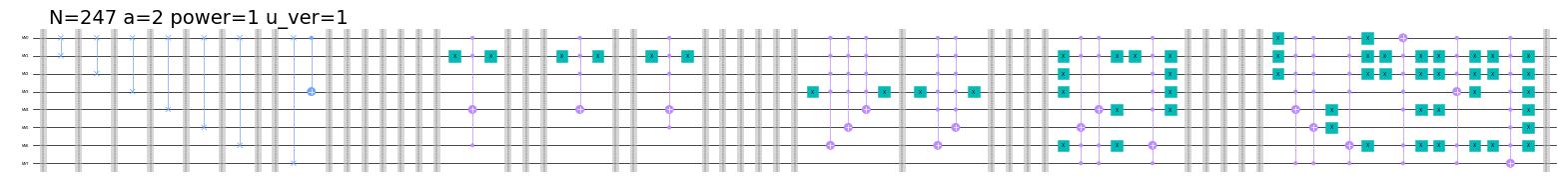} 
\caption{\footnoteskip  
  $N = 247$, $a = 2$, $r = 36$, $n = 8$: The ME operator $U^p$ for the
  powers $p \in \{1, 2, 4\}$.
}
\label{fig_N247_c}
\end{figure}

\vfill
\pagebreak

\end{document}